%% file: Revision_HIT_CAMCOS.tex
\documentclass{amsart}

\input{ex_shared}

\ifpdf
\hypersetup{
  pdftitle={Investigation of finite-volume methods to capture shocks and turbulence spectra in compressible flows},
  pdfauthor={Emmanuel Motheau and John Wakefield}
}
\fi

\begin{document}

\maketitle

\begin{abstract}
The aim of the present paper is to provide a comparison between several finite-volume methods of different numerical accuracy: second-order Godunov method with PPM interpolation and high-order finite-volume WENO method. The results show that while on a smooth problem the high-order method perform better than the second-order one, when the solution contains a shock all the methods collapse to first-order accuracy. In the context of the decay of compressible homogeneous isotropic turbulence with shocklets, the actual overall order of accuracy of the methods reduces to second-order, despite the use of fifth-order reconstruction schemes at cell interfaces. Most important, results in terms of turbulent spectra are similar regardless of the numerical methods employed, except that the PPM method fails to provide an accurate representation in the high-frequency range of the spectra. It is found that this specific issue comes from the slope-limiting procedure and a novel hybrid PPM/WENO method is developed that has the ability to capture the turbulent spectra with the accuracy of a high-order method, but at the cost of the second-order Godunov method. Overall, it is shown that virtually the same physical solution can be obtained much faster by refining a simulation with the second-order method and carefully chosen numerical procedures, rather than running a coarse high-order simulation. Our results demonstrate the importance of evaluating the accuracy of a numerical method in terms of its actual spectral dissipation and dispersion properties on mixed smooth/shock cases, rather than by the theoretical formal order of convergence rate.
\end{abstract}

\section{Introduction}

The utility of high-order accurate numerical methods has been a subject of discussion within the Computational Fluid Dynamics (CFD) community for several decades. As suggested in the review paper of \cite{Wang:2013}, one of the myths in the debate over low versus high-order numerical methods is the ability the get an accurate solution at a reduced computational cost. High-order methods are more costly on a per point basis but can potentially obtain a solution of the desired accuracy on a coarse mesh. Low-order methods are easier to implement, less costly per point but require a finer mesh to obtain accuracy equivalent to a high-order method.

The theoretical order of accuracy $k$ of a numerical method describes the order of the truncation error made when approximating the derivative of a function via a numerical discretization. In practice, the order of accuracy can be quantified by the asymptotic rate of convergence of the solution error $\epsilon$ respect to the mesh size $h$, namely $\epsilon \propto h^k$. This type of theoretical asymptotic estimates argues for the utility of high-order methods (\cite{Wang:2013} defines high-order for $k>3$). However, realizing this type of convergence depends on the smoothness of the solution.

In most CFD applications, particularly those involving turbulent flow, the solution is adequately resolved well before reaching the asymptotic regime of the numerical method (see discussion in \cite{Almgren:2013}). This issue is exacerbated for compressible flow. The solution can include shock waves that require local dissipation to prevent the appearance of spurious nonphysical oscillations in the solution, reducing the order of accuracy of the numerical method employed. 

A more realistic way to assess a numerical method is to determine the cost needed to obtain a desired accuracy. In the context of viscous compressible turbulent flow, we can frame the question in terms of the resolution required to resolve the spectrum of the turbulent flow. Indeed, it is emphasized that the performance of a numerical method should not be defined only by the order of the convergence of the error for smooth solutions. A better measure for the actual accuracy is the ability of the numerical method to adequately resolve both the inertial range and the dissipation range of the turbulent
energy spectrum. 

Unfortunately the literature on the development of numerical methods often provides tests and comparisons based on canonical  cases, which consist on the propagation of smooth solutions or very specific cases with discontinuities. As an alternative we propose  investigating the performance of numerical schemes for resolving the spectrum of a complex turbulent flow, especially in a context where both shocks and a wide range of turbulent scales interact in the flow field. To the author's knowledge, only a few papers \cite{Aspden:2009,Johnsen:2010} deal with such a complete study. However, \cite{Aspden:2009} only investigates incompressible flow, while in \cite{Johnsen:2010} the impact of the mesh resolution is not investigated and simulations are only performed on a coarse mesh. As it will be shown in the present paper, refinement of the mesh allows spurious small structures to develop and may lead to inaccurate spectra in the high-frequency range. We advocate that one of the most important features of a  numerical method should be its robustness to any discretization size.

Many different numerical methods exist to solve partial differential equations, and each of them present pros and cons depending on the problem investigated. For example compact finite difference schemes \cite{Lele:1992} are very efficient to capture accurately turbulent
energy spectrum, but their performances quickly degrade when applied to geometries more complicated than a triple periodic cubic box, and/or if the solution is not smooth enough. In the context of the simulation of flows in engineering applications, complex geometries are often involved and multiphysics phenomena can occur. See for example \cite{Motheau:2014b} where simulations of flames are performed in a realistic gas turbine combustion chamber. For such complicated applications, finite-volume methods are often preferred because they are intrinsically conservative, robust, and flexible enough to handle both unstructured and structured meshes. Moreover, finite-volume methods fit naturally within the paradigm of  Adaptive Mesh Refinement (AMR) using the concept of re-fluxing across multi-grids to achieve conservation properties.

The goal of the present paper is to compare and investigate the performance of several popular finite-volume methods for the compressible Navier-Stokes equations. Let's recall that a finite-volume method seeks to reconstruct data at the interface between cells, and then to solve a Riemann problem so as to evaluate the fluxes that cross the cells.  As explained above, a flow may contain shocks. In typical compressible Navier-Stokes the associated shock profiles are so thin that they cannot for all practical purposes be represented by the points of a numerical mesh. Because from one cell to another there is a strong difference in the states of the flow, a specialized treatment is made to reconstruct fluxes that capture the discontinuities without introducing spurious oscillations. Several techniques have been proposed in the literature but a complete review is beyond the scope of the present paper and can be found in reference textbooks \cite{Leveque:2002,Toro:2013}. 

In the present paper, two techniques are considered. First, in the asymptotic second-order Godunov method, the classical PPM interpolation procedure \cite{Colella:1984,Miller:2002} considers several limiters to enforce the monotonicity, for example starting with the van Leer method \cite{vanleer:1979}. Second, the present paper also investigates the high-order finite-volume method developed by \cite{Titarev:2004}, which is based on the Weighted Essentially Non-Oscillatory (WENO) schemes. There is an extensive literature on different variants of WENO schemes and a complete description is beyond the scope of the present paper, a review can be found in \cite{SHU:2016}. The basic idea of WENO schemes is to provide a high-order non-linear reconstruction method, which effectively captures discontinuities but can also be dissipative on smooth solutions. Note that several different WENO variants were tested during the present study and it has been found that, in overall, they provide similar results despite exhibiting some robustness discrepancies. Thus, for clarity purpose, only one WENO variant is employed in this paper, but we provide in \cref{sec:appendix_weno_comparisons} more comprehensive results to highlight the performance and robustness issues that we encountered while testing the different WENO variants. 

Three test cases of increasing complexity are investigated in the present paper. First, the convection of a smooth vortex is considered, followed by the simulation of a classical shock-driven Shu-Osher problem. It is emphasized that these test cases are chosen here because they are commonly employed in the literature to assess performance of numerical schemes. Here the results show that while on a smooth problem the high-order method perform better than the second-order one, when the solution contains a shock all the methods collapse to first-order accuracy. Finally, the decay of compressible homogeneous isotropic turbulence (HIT) with shocklets is investigated. Comparisons reveal that a second-order Godunov method with the classical PPM interpolation provides essentially the same results as a fourth-order finite-volume WENO scheme but at a significant lower cost. It is emphasized that virtually the same physical solution can be
obtained much faster by refining a simulation with the second-order method, rather than running a coarse high-order simulation. However, the results also show that the refinement of the mesh presents some limit when using the second-order Godunov procedure with the classical PPM interpolation. Indeed, it is found that when the mesh is fine enough, a non-physical pile-up of energy appears in the high-frequency range of the turbulent spectra. After an intensive trial and error process, it has been found that the limiting procedures employed by the PPM to ensure monotonicity are responsible to this pile-up of energy in the high-frequency range of the spectra. 

One of the most significant innovation of the present paper is to propose to replace the interpolation and limiting procedures at cell interfaces in the classical PPM algorithm by a WENO interpolation. It is shown that the novel proposed hybrid PPM/WENO method has the ability to capture the turbulent spectra with the accuracy of a high-order method, but at the cost of the second-order Godunov method.

This study makes use of CFD software developed in the Center for Computational Sciences and Engineering (CCSE) group\footnote{\url{https://ccse.lbl.gov/index.html}} at Lawrence Berkeley National Laboratory in the USA. The codes are implemented in the \textbf{AMReX} framework\footnote{\url{https://amrex-codes.github.io/amrex/}} that facilitates the development of a generic post-processing chain as well as the assessment of computing costs via embedded profiling functionality. Note that while the \textbf{AMReX} library supports AMR applications, only single level grids are employed in the present paper. Two codes are being compared:
\begin{itemize}
\item \textbf{PeleC}, which is based on a second-order Godunov procedure. Interpolation to evaluate data at cell faces is performed either with the orignial unsplit PPM \cite{Miller:2002} method, or with the hybrid PPM/WENO developed in the present paper. The diffusion operators are evaluated with a second-order finite-volume discretization.
 
\item \textbf{RNS}, which is based on a fourth-order finite-volume WENO method \cite{Titarev:2004} in space. Note that \textbf{RNS} was originally built for the development of the Adaptive Multi-Level Spectral Deferred Correction (AMLSDC), which is a fourth-order time integration method  \cite{Emmett:2018}, but in the present paper the classical Runge-Kutta algorithm is employed instead. Note that the diffusion terms are discretized with a fourth-order conservative finite-volume technique. First, the cell-averaged conserved variables are used to compute fourth-order approximations to
point values at cell centers using the procedure outlined in McCorquodale and Colella \cite{mccorquodale:2011} and then explicit formulae are used to compute derivatives needed to compute the diffusive fluxes at Gauss points on the cell-faces directly.
\end{itemize}

The remainder of the present paper is organized as follows. In \cref{sec:governing_equations}, the set of equations solved by the codes are presented. In \cref{sec:RNS} the \textbf{RNS} code is presented, as well as a short description of the high-order finite-volume WENO scheme that is employed for the spatial discretization. Next, in \cref{sec:PeleC} the \textbf{PeleC} code together with the original PPM algorithm are presented, followed in \cref{subsec:hybrid_ppm_weno} by the novel hybrid PPM/WENO method developed in the present paper that captures the turbulent spectra with the accuracy of a high-order method at the cost of a second-order Godunov method. Results are then presented in \cref{sec:results}. The convection of a smooth vortex and the Shu-Osher problem are investigated in \cref{subsec:COVO} and \cref{subsec:Shu_Osher}, respectively, while the decay of compressible homogeneous isotropic turbulence with shocklets is investigated in \cref{subsec:compressible_HIT}.

\section{Governing equations}
\label{sec:governing_equations}

The software employed in the present study was initially developed for the simulation of combustion problems and the codes solve the multicomponent reacting Navier-Stokes equations. However, only non-reacting problems with no specific mixture are investigated in the present study. Consequently, the set of equations solved are significantly simplified and are given by

\begin{align}
\frac{\partial \rho}{\partial t} + \nabla \cdot (\rho
    \vec{u})= { } & 0, \label{eq:NS:mass} \\
\frac{\partial \rho \vec{u}}{\partial t} + \nabla \cdot (\rho
    \vec{u} \otimes \vec{u}) + \nabla p= { } & \nabla \cdot \tensor{\tau}, \label{eqs:NS:momentum} \\
\frac{\partial \rho E}{\partial t} + \nabla \cdot [(\rho E + p)
  \vec{u}] = { } & \nabla \cdot (\lambda \nabla T) + \nabla \cdot
  \left(\tensor{\tau} \cdot \vec{u}\right), \label{eqs:NS:energy}
\end{align}
where $\rho$ is the density, $\vec{u}$ is the velocity, $p$ is the pressure, $E = e + \mathbf{u} \cdot \mathbf{u} / 2$ is the total energy, $T$ is the temperature and $\lambda$ is the thermal conductivity. The viscous stress tensor is given by
\begin{equation}
\tensor{\tau} = \eta (\nabla \vec{u} + (\nabla \vec{u})^T) + (\varsigma - \frac{2}{3} \eta ) (\nabla \cdot \vec{u}) \mathbf{I},
\end{equation}
where $\eta$ and $\varsigma$ are the shear and bulk viscosities.

The system is closed by an equation of state (EOS) that specifies $p$ as a function of $\rho$ and $T$.  An ideal gas mixture for the EOS is assumed:
\begin{equation}
 p = \rho T \mathfrak{R},
 \label{eqn:eos}
\end{equation} 
where $\mathfrak{R}$ is the specific gas constant. Here we set $C_p$ and $C_v$ the heat capacity at constant pressure and volume, respectively, to follow an ideal gas law proportional to the ratio of the specific heats $\gamma$ so that \cref{eqn:eos} is equivalent to the following relation:
\begin{equation}
e = p /  \left( \gamma - 1 \right) \rho 
\end{equation}
where $e$ is the specific internal energy and $\gamma$ is set to $\gamma=1.4$.

Note that for the ease of simplicity, the system presented at \cref{eq:NS:mass,eqs:NS:momentum,eqs:NS:energy} is recast in the form of
\begin{equation}
\frac{\partial \mathbf{U}}{\partial t} + \nabla \cdot \mathbf{F} = \mathbf{S},
\label{eqn:generic_conservative}
\end{equation}
where $\mathbf{U}$ is the vector of conservative variables, while $\mathbf{F}$ represents the convective flux vector and $\mathbf{S}$ contains the diffusive terms, respectively.

\section{RNS: a high-order WENO-based finite-volume solver}
\label{sec:RNS}

The \textbf{RNS} code implements high-order  temporal and spatial AMR integration methods for combustion applications. The major innovative feature of this code is the development of the Adaptive Multi-Level Spectral Deferred Correction (AMLSDC) method, which is fourth-order in time \cite{Emmett:2018}. The Runge-Kutta method for AMR applications as presented in \cite{mccorquodale:2011} is also implemented. In the present paper, second-order explicit midpoint Runge-Kutta method is used. Note that although not shown in the present paper, the results were compared to the fourth-order Runge-Kutta and AMLSDC approaches, and results were virtually the same without impacting the spatial solutions, which is attributed to the fact that the time-steps involved are small, and the spatial errors introduced at shocks dominate the solution.

The diffusion terms are discretized using standard finite volume techniques.  First, the cell-averaged conserved variables are used to compute fourth-order approximations to point values at cell centers using the procedure outlined in 
McCorquodale and Colella \cite{mccorquodale:2011}. These point values of conserved quantities are then used to compute primitive variables, and explicit formulae are then used to compute derivatives needed to evaluate the diffusive fluxes at Gauss points on the cell-faces directly.  Similarly, diffusion coefficients are computed at cell centers using point values and are then interpolated to Gauss points.

The spatial discretization of the advection terms in the algorithm uses the conservative finite-volume WENO reconstruction presented in \cite{Titarev:2004}. The following approach is repeated for each stage of the Runge-Kutta integration scheme: 

\begin{enumerate}
\item For each cell, the conservative  \cref{eqn:generic_conservative} is rewritten in terms of primitive variables.
\item The primitive variables are reconstructed at the cell interfaces with a fifth-order WENO scheme in order to provide a left and right state for each face. For 2D and 3D cases, the variables are first reconstructed to Gauss quadrature nodes to evaluate their average value in the direction normal to the faces. This procedure is obviously computationally expensive, but as shown by \cite{Zhang_zhang_shu:2011}, a midpoint rule for integrating fluxes is not sufficiently accurate to obtain fourth-order convergence. Note that although the solution is reconstructed at cell interfaces with fifth-order WENO procedures, the method is formally fourth-order accurate because a fourth-order quadrature rule is employed to integrate the flux over faces.
\item The HLLC algorithm \cite{Toro:1994} is employed to reconstruct the fluxes through the faces.  
\end{enumerate}

In the present study, several different WENO schemes were investigated and results show that the so-called WENO-Z variant \cite{Borges:2008} performs the best. Some elements of the comparisons results are presented in \cref{sec:appendix_weno_comparisons}. As the conservative finite-volume WENO method presented by \cite{Titarev:2004} is based on the so-called WENO-JS scheme generalized by Jiang and Shu \cite{WENO_JS}, we first review the basic principles, followed by a short description of the WENO-Z variant.

\subsection{The WENO-JS method}
\label{subsec:WENO_JS}

For a given cell $i$, the principle of a WENO method is to provide a high-order approximation of the variable $q$ interpolated on the left and the right side of a face, denoted $\hat{q}^L_{i+\frac{1}{2}}$ and $\hat{q}^R_{i-\frac{1}{2}}$. In the remainder of this section, the procedures to evaluate $\hat{q}^L_{i+\frac{1}{2}}$ are provided. $\hat{q}^R_{i-\frac{1}{2}}$ is evaluated analogously.

In the WENO-JS method proposed by \cite{WENO_JS}, a fifth-order polynomial approximation of $\hat{q}^L_{i+\frac{1}{2}}$ is constructed through a convex combination of the  values $\hat{q}^k_{i+\frac{1}{2}}$ interpolated with a third degree polynomial on a three point stencil $k$, such that:
\begin{equation}
\hat{q}^L_{i+\frac{1}{2}} = \sum_{k=0}^2 \omega_k \hat{q}^k_{i+\frac{1}{2}}
\end{equation}
with
\begin{align}
	\hat{q}^0_{i+\frac{1}{2}} &= \frac{1}{6}\left(2q_{i-2} - 7 q_{i-1} + 11q_{i} \right),   \\ 
	\hat{q}^1_{i+\frac{1}{2}} &=  \frac{1}{6} \left(-q_{i-1} + 5 q_{i} + 2q_{i+1} \right), \\
	\hat{q}^2_{i+\frac{1}{2}} &=  \frac{1}{6} \left( 2q_{i}  + 5 q_{i+1} -q_{i+2} \right).
\end{align}
Here, $\omega_k$ are non-linear weights balancing the contribution of each stencil, and the challenge is to find the best values to capture shocks the most accurately while preserving the resolution of the spectrum of a solution.

The weights $\omega_k$ are defined as
\begin{equation}
\omega_k = \frac{\alpha_k}{\sum^2_{l=0}\alpha_l},\hspace{1cm} \alpha_k = \frac{d_k}{\left( \beta_k + \epsilon \right)^p},
\label{eqn:weights_WENO}
\end{equation}
where $d_k$ are the so-called optimal weights because they reconstruct the fifth-order upstream central scheme for the $5$-points stencil, $\beta_k$ are the smoothness indicators, $\alpha_k$ are referred as the unnormalized weights and $\epsilon$ is a parameter set to avoid a division by zero. The parameter $p$ controls the adaption rate. According to \cite{Arshed:2013}, a large value of $p$ leads to unnecessarily high dissipation in smooth regions of the flow. In the present study, the parameter is set to $p=1$ for all the test cases. Moreover, as suggested by \cite{Arshed:2013}, $\epsilon$ is set to $\epsilon=10^{-40}$.

The smoothness indicators $\beta_k$ are given by
\begin{align}
&\beta_0 = \frac{13}{12} \left(q_{i-2} - 2 q_{i-1} + q_i\right)^2 + \frac{1}{4}\left( q_{i-2} - 4 q_{i-1} + 3 q_i \right)^2, \\
&\beta_1 = \frac{13}{12} \left(q_{i-1} - 2 q_{i} + q_{i+1}\right)^2 + \frac{1}{4}\left(   q_{i-1} -  q_{i+1} \right)^2, \\
&\beta_2 = \frac{13}{12} \left(q_{i} - 2 q_{i+1} + q_{i+2}\right)^2 + \frac{1}{4}\left( 3 q_{i} - 4 q_{i+1} +  q_{i+2} \right)^2.
\end{align}

One of the feature of the conservative finite-volume WENO method is that the optimal weights as well as the formulae for the reconstructed values differ if the interpolation is performed in the normal direction at faces or at the Gauss integration points $\xi = \xi_i \pm \Delta\xi /(2\sqrt{3})$ (see \cite{Titarev:2004}).
\begin{itemize}
\item For the normal direction through a face, the optimal weights are:
\begin{equation}
d_0 = \frac{1}{10}, \hspace{1cm} d_1=\frac{6}{10}, \hspace{1cm} d_2=\frac{3}{10},
\label{eqn:optimal_weights}
\end{equation}
and $\hat{q}^L_{i+\frac{1}{2}}$ is given by:
\begin{multline}
\hat{q}^L_{i+\frac{1}{2}} = \frac{1}{6}\omega_0\left(2q_{i-2} - 7 q_{i-1} + 11q_{i} \right)   \\ + \frac{1}{6}\omega_1\left(-q_{i-1} + 5 q_{i} + 2q_{i+1} \right) + \frac{1}{6}\omega_2\left( 2q_{i}  + 5 q_{i+1} -q_{i+2} \right).
\label{eqn:reconstructed_forumla}
\end{multline}
\item For the first Gaussian integration point $\xi = \xi_i + \Delta\xi /(2\sqrt{3})$, the optimal weights are:
\begin{equation}
d_0 = \frac{210-\sqrt{3}}{1080}, \hspace{1cm} d_1=\frac{11}{18}, \hspace{1cm} d_2=\frac{210+\sqrt{3}}{1080},
\label{eqn:optimal_weights_gauss}
\end{equation}
and $q\left(\xi_i + \frac{\Delta\xi}{2\sqrt{3}}\right)$ is given by:
\begin{multline}
q\left(\xi_i + \frac{\Delta\xi}{2\sqrt{3}}\right)=\omega_0\left[q_{i} -\left(-3 q_{i} + 4 q_{i-1} - q_{i-2}\right) \frac{\sqrt{3}}{12} \right] \\ +\omega_1\left[q_{i} -\left( q_{i-1} - q_{i+1} \right) \frac{\sqrt{3}}{12} \right] + \omega_2\left[q_{i} -\left(3 q_{i} - 4 q_{i+1} + q_{i+2}\right) \frac{\sqrt{3}}{12} \right] .
\label{eqn:reconstructed_forumla_gauss} 
\end{multline}
\end{itemize}
Recall here that a simple mirror-symmetric change to the coefficients and the formulae will provide $\hat{q}^R_{i-\frac{1}{2}}$ and $q\left(\xi_i - \frac{\Delta\xi}{2\sqrt{3}}\right)$.

\subsection{The WENO-Z method}
\label{subsec:WENO_Z}

A well-known issue with the original WENO-JS method is that the smoothness indicators $\beta_k$ employed to compute the weights $\omega_k$ fail to recover the maximum order of the scheme at critical points when the derivatives of flux function vanish. Borges \textit{et al.} \cite{Borges:2008} propose a different approach to overcome the issues of the WENO-JS method by acting directly on the smoothness indicator $\beta_k$ with a very simple formulation. The so-called WENO-Z method is given by:
\begin{equation}
\omega_k^{(\rm z)} = \frac{\alpha_k^{(\rm z)}}{\sum_{i=0}^2\alpha_i^{(\rm z)}}, \hspace{.5cm} \text{with} \hspace{.5cm} \alpha_k^{(\rm z)} = d_k\left(1 + \frac{\tau_5}{\beta_k + \epsilon} \right)^p,
\end{equation} 
where
\begin{equation}
\tau_5 = |\beta_0 - \beta_2|.
\end{equation}
Similarly to the WENO-JS method, the parameter $p$ controls the detection of the smoothness of the solution. In the present study, the parameter is set to $p=1$ to reduce as much as possible the dissipation of the numerical scheme. Note that the WENO-Z method simply provides a new way to compute the non-linear weights $\omega_k$ and can be directly implemented in the conservative finite-volume WENO method, regardless if the interpolation is performed in the normal direction at faces or at the Gauss integration points.

\section{PeleC: the second-order Godunov-based finite-volume solver}
\label{sec:PeleC}

The \textbf{PeleC} code is a second-order AMR finite-volume solver for reacting and non-reacting fluid simulations with complex geometry and support for Lagrangian spray particles. The simulations performed in the present paper only uses a fraction of the capability of the software, namely the Godunov-based integration procedure on a single level mesh grid. Note also that \textbf{PeleC} is part of the \textbf{Pele Suite} of codes, which are publicly available and may be freely downloaded\footnote{\url{https://amrex-combustion.github.io/}}, and that all the test cases investigated in the present paper are available from the \textbf{PeleC} distribution and can be reproduced.

The solution is advanced from time $n$ to time $n+1$ with the following second-order Godunov method:
\begin{align}
\mathbf{U}^{*} &= \mathbf{U}^{n} - \Delta t \nabla \cdot \mathbf{F}^{n + 1/2} + \Delta t ~\mathbf{S}^n , \label{eqn:PeleC_1st_step} \\
\mathbf{U}^{n+1} &= \mathbf{U}^{*} + \frac{1}{2}\Delta t \left(\mathbf{S}^* - \mathbf{S}^n \right),
\label{eqn:PeleC_2nd_step}
\end{align}
where $\Delta t=t^{n+1} - t^n$ is the time step. The second step at \cref{eqn:PeleC_2nd_step} is a correction of the solution to ensure second-order accuracy by effectively time-centering the diffusion source terms. The conserved state vector $\mathbf{U}$ is stored at cell centers and the flux vectors are computed on cell edges. 

The convective flux vector $\mathbf{F}$ that appears in \cref{eqn:PeleC_1st_step} is constructed from time-centered edge states computed with a conservative, shock-capturing, unsplit Godunov method, which makes use of the Piecewise Parabolic Method (PPM) \cite{Colella:1984}, characteristic tracing and full corner 
coupling \cite{Almgren:2010a,Miller:2002}. As the present paper proposes a modification of the PPM method, for ease of exposition the whole algorithm will be detailed in 1D for the Euler equations. It is emphasized that the algorithm can be extended to multi-dimensional problems and multi-component flows. Moreover, since the publication of the original paper \cite{Colella:1984} presenting the PPM method, several modifications have been proposed in the literature (see \cite{Miller:2002,Colella:2008,Colella:2011b}). Consequently, the algorithm implemented in the code \textbf{PeleC} incorporates some of the variants, but it is emphasized that these changes only slightly differ from the original PPM method. Many variants have been tested through this study, and while not reported in the present paper, none change fundamentally the results.

\subsection{System of primitive variables}
\label{subsec:primitive_var_system}

The conservative \cref{eqn:generic_conservative} is rewritten in terms of primitive variables, such that:
\begin{equation}
  \frac{\partial \mathbf{Q}}{\partial t} + \mathbf{A} \frac{\partial \mathbf{Q}}{\partial x} = \mathbf{S}_{\mathbf{Q}}. \label{eqn:prim_var_eq}
\end{equation}
Here $\mathbf{Q}$ is the primitive state vector, $\mathbf{A}=\partial \mathbf{F}/\partial \mathbf{Q}$ and $\mathbf{S}_{\mathbf{Q}}$ are the viscous source terms reformulated in terms of the primitive variables.

In one dimension, this comes:

\begin{equation}
\left(\begin{array}{c}
\rho \\
u \\
p \\
\rho e
\end{array}\right)_t
+
\left(\begin{array}{cccc}
u & \rho &  0 & 0  \\
0 & u &  \frac{1}{\rho} & 0  \\
0 & \rho c^2 & u & 0 \\
0 & \rho e + p & 0 & u 
\end{array}\right)
\left(\begin{array}{c}
\rho \\
u \\
p \\
\rho e 
\end{array}\right)_x
=
\mathbf{S}_{\mathbf{Q}}
\end{equation}

Note that here, the system of primitive variables has been extended to include an additional equation for the internal energy, denoted $e$. This avoids several calls to the equation of state, especially in the Riemann solver step. 

The eigenvalues of the matrix $\mathbf{A}_x$ are given by:

\begin{equation}
\mathbf{\Lambda}\left(\mathbf{A}_x\right) = \{u-c,u,u,u+c\}.
\end{equation}
The right column eigenvectors are:
\begin{equation}
\mathbf{r}_x =
\left(\begin{array}{ccccc}
1 & 1 &  0  & 1 \\
-\frac{c}{\rho} &  0 & 0 & \frac{c}{\rho} \\
c^2 & 0  & 0 & c^2 \\
h & 0 &  1  & h
\end{array}\right). \label{eqn:matrix_lx}
\end{equation}
The left row eigenvectors, normalized so that $\mathbf{l}_x\cdot\mathbf{r}_x = \mathbf{I}$ are:
\begin{equation}
\mathbf{l}_x =
\left(\begin{array}{ccccc}
0 & -\frac{\rho}{2c} &  \frac{1}{2c^2}  & 0 \\
1 & 0  & -\frac{1}{c^2}  & 0 \\
0 & 0 &  -\frac{h}{c^2}  & 0 \\
0 & \frac{\rho}{2c} & \frac{1}{2c^2}  & 0
\end{array}\right). \label{eqn:matrix_rx}
\end{equation}
Note that here, $c$ and $h$ are the sound speed and the enthalpy, respectively.

\subsection{Edge state prediction}
\label{subsec:edge_state_prediction}

As discussed at the beginning of \cref{sec:PeleC}, the fluxes are reconstructed from time-centered edge state values. Thus, the primitive variables are first interpolated in space with the PPM method, then a characteristic tracing operation is performed to extrapolate in time their values at $n+1/2$.

\subsubsection{Interpolation and slope limiting}
\label{subsubsec:interpolation_limiting}

Basically the goal of the algorithm is to compute a left and a right state of the primitive variables at each edge in order to provide inputs for the Riemann problem to solve. 

First, the average cross-cell difference is computed for each primitive variable with a quadratic interpolation as follows:
\begin{equation}
\delta q_i = \frac{1}{2} \left(q_{i+1} - q_{i-1}\right).
\end{equation} 
In order to enforce monotonicity, $\delta q_i$ is limited with the van Leer \cite{vanleer:1979} method:
\begin{equation}
\delta q_i^* = \min \left(|\delta q_i|,2|q_{i+1}-q_i|,2|q_i - q_{i-1}|\right)\text{sgn}\left(\delta q_i\right),
\end{equation}
and the interpolation of the primitive values to the cell face $q_{i+\frac{1}{2}}$ is estimated with:
\begin{equation}
q_{i+\frac{1}{2}} = q_i + \frac{1}{2}\left(q_{i+1}-q_i \right)-\frac{1}{6}\left(\delta q_{i+1}^* - \delta q_i^* \right).
\end{equation}

In order to enforce that $q_{i+\frac{1}{2}}$ lies between the adjacent cell averages, the following constraint is imposed:
\begin{equation}
\min\left(q_i,q_{i+1} \right) \leqslant q_{i+\frac{1}{2}} \leqslant \max\left(q_i,q_{i+1} \right).
\end{equation}

The next step is to set the values of $q_{R,i-\frac{1}{2}}$ and $q_{L,i+\frac{1}{2}}$, which are the right and left state at the edges bounding a computational cell. Here, a quartic limiter is employed in order to enforce that the interpolated parabolic profile is monotone. The procedure proposed by \cite{Miller:2002} is adopted, which slightly differs from the original one proposed in \cite{Colella:1984}. In \cite{Miller:2002}, this specific procedure is followed by the imposition of another limiter based on a flattening parameter to prevent artificial extrema in the reconstructed values. In the present paper, the order of imposition of the different limiting procedures is reversed.

First, the edge state values are defined as:
\begin{align}
q_{L,i+\frac{1}{2}} = q_{i+\frac{1}{2}}, \\
q_{R,i-\frac{1}{2}} = q_{i-\frac{1}{2}}.
\end{align}
Then the flattening limiter is imposed as follows:
\begin{align}
q_{L,i+\frac{1}{2}} \leftarrow \chi_i q_{L,i+\frac{1}{2}} + \left(1+\chi_i\right) q_i, \label{eqn:flattening_eq_1} \\
q_{R,i-\frac{1}{2}} \leftarrow \chi_i q_{R,i-\frac{1}{2}} + \left(1+\chi_i\right) q_i, \label{eqn:flattening_eq_2}
\end{align}
where $\chi_i$ is a flattening coefficient computed from the local pressure, and its evaluation is presented in \cref{sec:appendix_flattening}.

Finally, the monotonization is performed with the following procedure:

\begin{align}
q_{L,i+\frac{1}{2}} = q_{R,i-\frac{1}{2}} = q_i \hspace{0.8cm} &\text{if}  \hspace{0.2cm}   \left(q_{L,i+\frac{1}{2}} - q_i \right)\left(q_i - q_{R,i-\frac{1}{2}}\right) > 0, \\
q_{L,i+\frac{1}{2}} = 3 q_i - 2 q_{R,i-\frac{1}{2}} \hspace{0.8cm} &\text{if}  \hspace{0.2cm} |q_{L,i+\frac{1}{2}}-q_i| \geqslant 2|q_{R,i-\frac{1}{2}}-q_i|, \\ 
q_{R,i-\frac{1}{2}} = 3 q_i - 2 q_{L,i+\frac{1}{2}} \hspace{0.8cm} &\text{if}  \hspace{0.2cm} |q_{R,i-\frac{1}{2}}-q_i| \geqslant 2|q_{L,i+\frac{1}{2}}-q_i|.
\end{align}

\subsubsection{Piecewise Parabolic Reconstruction}
\label{subsubsec:PPM}

Once the limited values $q_{R,i-\frac{1}{2}}$ and $q_{L,i+\frac{1}{2}}$ are known, the limited piecewise parabolic reconstruction in each cell is done by computing the average value swept out by parabola profile across a face, assuming that it moves at the speed of a characteristic wave $\lambda_k$. The average is defined by the following integrals:
\begin{align}
\mathcal{I}^{(k)}_{+} \left(q_i \right) &= \frac{1}{\sigma_k \Delta x}\int^{(i+1/2)\Delta x}_{((i+1/2)-\sigma_k)\Delta x} q_i^I\left(x\right){\rm d}x, \label{eqn:int_parab_1}\\
\mathcal{I}^{(k)}_{-} \left(q_i \right) &= \frac{1}{\sigma_k \Delta x}\int^{((i-1/2)+\sigma_k)\Delta x}_{(i-1/2)\Delta x} q_i^I\left(x\right){\rm d}x, \label{eqn:int_parab_2}
\end{align}
with $\sigma_k = |\lambda_k|\Delta t / \Delta x$, where $\lambda_k=\{u-c,u,u,u+c\}$, while $\Delta t$ and $\Delta x$ are the discretization step in time and space, respectively, with the assumption that $\Delta x$ is constant in the computational domain.

The parabolic profile is defined by
\begin{equation}
q_i^I \left(x\right) = q_{R,i-\frac{1}{2}} + \xi\left(x\right)\left[q_{L,i+\frac{1}{2}} - q_{R,i-\frac{1}{2}} + q_{i,6}\left(1-\xi\left(x\right)\right)\right]
\end{equation}
with 
\begin{equation}
q_{i,6} = 6 q_i - 3\left(q_{R,i-\frac{1}{2}} + q_{L,i+\frac{1}{2}} \right).\label{eqn:parabolic_profile}
\end{equation}
and
\begin{equation}
\xi \left(x\right) = \frac{x-x_{i-\frac{1}{2}}}{\Delta x}, \hspace{0.8cm} x_{i-\frac{1}{2}} \leqslant x \leqslant x_{i+\frac{1}{2}}
\end{equation}

Substituting \cref{eqn:parabolic_profile} in \cref{eqn:int_parab_1,eqn:int_parab_2}
leads to the following explicit formulations:
\begin{align}
\mathcal{I}^{(k)}_{+} \left(q_i \right) &= q_{L,i+\frac{1}{2}} - \frac{\sigma_k}{2}\left[q_{L,i+\frac{1}{2}} - q_{L,i+\frac{1}{2}} - \left(1-\frac{2}{3}\sigma_k \right) q_{i,6} \right], \\
\mathcal{I}^{(k)}_{-} \left(q_i \right) &= q_{R,i-\frac{1}{2}} + \frac{\sigma_k}{2}\left[q_{L,i+\frac{1}{2}} - q_{L,i+\frac{1}{2}} + \left(1-\frac{2}{3}\sigma_k \right) q_{i,6} \right].
\end{align}

\subsubsection{Characteristic tracing and flux reconstruction}
\label{subsubsec:characteristic_tracing}

The next step is to extrapolate in time the integrals $\mathcal{I}^{(k)}_{\pm}$ to get the left and right edge states at time $n+1/2$. This procedure is complex, especially in multi-dimensions where transverse terms are taken into account; the complete detailed procedure can be found in \cite{Miller:2002}. In 1D, the left and right edge states are computed as follows:
\begin{align}
q_{L,i+\frac{1}{2}}^{n+\frac{1}{2}} &= \mathcal{I}^{(k=u+c)}_{+} - \sum_{k:\lambda_k \geqslant 0} \beta_k \mathbf{l}_k \cdot \left[\mathcal{I}^{(k=u+c)}_{+}-\mathcal{I}^{(k)}_{+}  \right] \mathbf{r}_k + \frac{\Delta t}{2} S_i^n, \\
q_{R,i-\frac{1}{2}}^{n+\frac{1}{2}} &= \mathcal{I}^{(k=u-c)}_{-} - \sum_{k:\lambda_k \leqslant 0} \beta_k \mathbf{l}_k \cdot \left[\mathcal{I}^{(k=u-c)}_{-}-\mathcal{I}^{(k)}_{-}  \right] \mathbf{r}_k + \frac{\Delta t}{2} S_i^n. \\
\end{align}
where 
\begin{equation}
    \beta_k = \begin{cases}
        \frac{1}{2}, & \text{if}\;\lambda_k = 0,  \\
        1, & \text{otherwise},
    \end{cases}
\end{equation}
and $\mathbf{l}_k$ and $\mathbf{r}_k$ are the left row and right column of the matrices defined at \cref{eqn:matrix_lx,eqn:matrix_rx} for each eigenvalue $k$. Note that here, $S_i^n$ represents any source terms at time $n$ to include in the characteristic tracing operation.

Finally, the time-centered fluxes are computed using an approximate Riemann problem solver. Here the HLLC algorithm \cite{Toro:1994} is employed. At the end of this procedure the primitive variables are centered in time at $n+1/2$, and in space at the edges of a cell. This is the so-called \emph{Godunov state} and the convective fluxes can be computed to advance \cref{eqn:PeleC_1st_step}.

\subsection{The hybrid PPM/WENO method}
\label{subsec:hybrid_ppm_weno}

As it will be shown in the results \cref{sec:results}, the PPM method presented above gives good results for a small computational time compared to the fourth-order finite-volume WENO strategy that is costly. However, for fine meshes, the PPM method exhibits a significant pile-up of energy in the high-frequency range of the spectra, which is undesirable and limits mesh refinement. It has been found that the pile-up of energy at the high-frequencies was sensitive to the slope-limiting procedure presented at \cref{subsubsec:interpolation_limiting}. As many variants can be found in the literature, an attempt to tweak this procedure was tried, for example by playing with the numerical parameters (see \cref{sec:appendix_flattening}) or by removing the slope limiting operation completely. Also, the procedure given in \cite{Colella:2008} was tested. For all cases, the results were very similar and the impact on the pile-up of energy was modest and not satisfying.  

After an intensive trial and error process, it became apparent that the interpolation and slope-limiting procedure described in \cref{subsubsec:interpolation_limiting} was not robust, leading to poor results in the high-frequency range. Here we consider replacing this whole procedure by a WENO interpolation. 

Basically, the purpose of the hybrid PPM/WENO method is only to replace the procedure in \cref{subsubsec:interpolation_limiting}, and $q^L_{i+\frac{1}{2}}$ and $q^R_{i-\frac{1}{2}}$ are instead given by \cref{eqn:reconstructed_forumla}. Then the PPM algorithm continues exactly the same as in \cref{subsubsec:PPM}. 

As shown in \cref{sec:appendix_weno_comparisons}, as the WENO-Z \cite{Borges:2008} appears to be the most robust and gives satisfying results for a small computational cost compared to other WENO methods, only the WENO-Z method is employed below, but it is emphasized that any other WENO reconstruction methods can be employed. For the ease of exposition, the hybrid method will be called PPM/WENO in the remainder of the paper, but one has to keep in mind that the WENO-Z method has been used for the reconstruction at faces.

\section{Results}
\label{sec:results}

The numerical methods presented in the previous section are tested and compared on three very different test cases. The first one is the convection of a smooth compressible vortex. This test case is chosen because it highlights the theoretical order of accuracy of the numerical methods. The second test case is the Shu-Osher problem, which represents the extreme opposite of the smooth vortex test case. The Shu-Osher problem is very difficult to solve numerically, because a shock wave is propagating in an oscillating entropy field, and the challenge is to capture the shock while resolving the phase and amplitude of the fluctuating entropy. As will be shown, all the methods perform correctly, but for all of them the rate of convergence collapses to first-order. The last test case is the decay of compressible homogeneous isotropic turbulence in the presence of eddy shocklets. This test case can be viewed as a combination of the two previous test case, because it contains both shocks and discontinuities, as well as smooth turbulence structures that lie in a large-bandwidth turbulent spectrum. More specifically, this test case is representative of flows that are encountered in practical CFD applications (see  \cite{Motheau:2014b} for an example). Note that in the remainder of this section, the initial solution comes from either an analytical solution or a synthetic manufactured solution. It is important to note that there is an averaging process over the volume to provide a consistent initial solution. For the fourth-order method, the procedure is slightly different to preserve a high-order accuracy: the solution is first expressed at the Gauss points, and the integration over the volume is performed via the quadrature rule.

\subsection{2D convection of a smooth compressible vortex}
\label{subsec:COVO}

The following test case consists of the convection of a 2D compressible vortex. This test case has been used frequently in the literature to assess the performance of outflow characteristic boundary conditions \cite{Motheau:2017aa,Poinsot:1992b}. The interest for this test case is that the solution is smooth and presents weak compressibility effects.  Here, the vortex is convected in a periodic domain so as to accumulate numerical errors from the discretization schemes. For each numerical method, the same test case is simulated with increasing mesh resolution. The time-step is computed based on the mesh resolution via a constraint on the CFL number, set to $0.7$. At the end of a simulation, convergence is measured using the $\mathcal{L}^1$-norm of the difference of the $x$-velocity between the final computed solution and the analytical solution:
\begin{equation}
\varepsilon_{u} = \mathcal{L}^1_{u} \left(S_{sol} - S_{init} \right) = \frac{{\sum_1^{N}}\left|u_{sol} - u_{init} \right|}{N },
\label{eqn:L1_norm_error}
\end{equation} 
where subscripts {\it{sol}} and {\it{ref}} identify the numerical solution and the initial solution, and $N$ is the number of computational cells.

The configuration is a single vortex superimposed on a uniform flow field along the $x$-direction. The stream function $\Psi$ of the initial vortex is given by

\begin{equation}
\Psi = \Gamma \exp\left(\frac{-r^2}{2 R_v^2} \right),
\end{equation}
where $r=\sqrt{\left(x-x_v\right)^2 + \left(y-y_v\right)^2}$ is the radial distance from the center of the vortex located at $\left[ x_v, y_v\right]$, while $\Gamma$ and $R_v$ are the vortex strength and radius, respectively. The velocity field is then defined as

\begin{equation}
u = \frac{\partial  \Psi}{\partial y}+ u_{0}, \hspace{1cm} v = -\frac{\partial  \Psi}{\partial x}.
\end{equation}
The initial pressure field is expressed as
\begin{equation}
p\left(r\right) = p_{\rm ref} \exp\left(-\frac{\gamma}{2} \left(\frac{\Gamma}{c R_v} \right)^2 \exp \left( -\frac{r^2}{R_v^2} \right) \right),
\end{equation}
and the corresponding density field is given by
\begin{equation}
\rho\left(r\right) = \frac{p\left(r\right)}{\mathcal{R} T_{\rm ref}}, 
\end{equation}
where $T_{\rm ref}$ is assumed constant. Note that here, $\gamma$ is the ratio of specific heats and is set to $\gamma=1.4$.

The computational domain is a square of dimension $L=0.01$~m. The reference temperature $T_{\rm ref}$ and pressure $p_{\rm ref}$ are set to $300$~K and $101320$~Pa, respectively. The vortex is located at $\left[ x_v, y_v\right]$ = $\left[ 0, 0\right]$, and its parameters are set to $\Gamma = 0.11$~m$^2$/s and $R_v = 0.1 L$. The initial flow velocity is $u_{0} = 100$~m/s. In the present test case, only the Euler equations  are solved. Thus, the transport coefficients  $\eta$, $\varsigma$ and $\lambda$  in  \cref{eq:NS:mass,eqs:NS:momentum,eqs:NS:energy} are set to zero.

The simulations are performed over a physical time of $5$~ms, corresponding
to $5$ flow through times (FTT), in order to accumulate enough numerical errors from
the spatial discretization schemes. 

\begin{figure}[tbhp]
\centering
\includegraphics[width=0.9\textwidth]{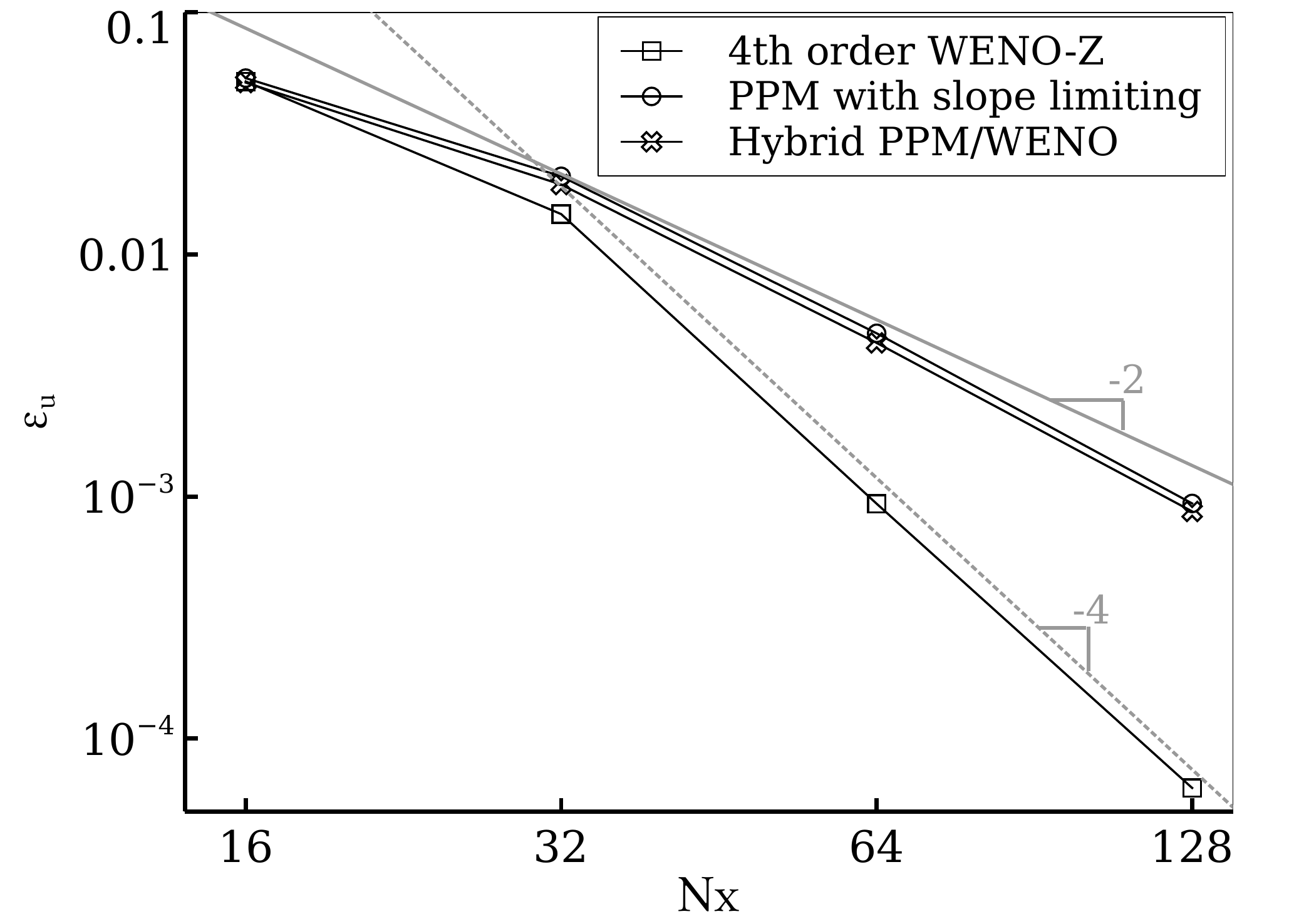}
\caption{Convection of a vortex, evolution of the $\mathcal{L}_1$-norm of the error of the $x-$velocity for different mesh size $N_x$.}
\label{fig:COVO_error}
\end{figure}

Results are shown in \cref{fig:COVO_error}. The solid and dotted grey lines represent  second- and fourth-order slopes, respectively. As expected, because the solution is smooth, all the numerical methods exhibit a convergence rate that follows their  theoretical order of accuracy. The \textbf{PeleC} code (see \cref{sec:PeleC}) using a second-order Godunov method with either the PPM or the hybrid PPM/WENO method for interpolation presents an almost constant second-order convergence rate. The finite-volume WENO method of the \textbf{RNS} code exhibits fourth-order convergence. From the results depicted in \cref{fig:COVO_error}, it is clear that a high-order method is superior to a second-order numerical method, because for the same mesh resolution the numerical error of the solutions is significantly lower. However, this superiority is possible because the solution is smooth, and as it will be shown below, this observation no longer holds when the solution features shocks and high gradients in the flow.

\subsection{Shock-driven test case: the Shu-Osher problem}
\label{subsec:Shu_Osher}

The so-called Shu-Osher test case simulates the one-dimensional propagation of a normal shock wave interacting with a fluctuating entropy wave, generating a flow field containing both small scale structures as well as discontinuities. The initial conditions are given by:
\begin{equation}
    \left( \rho, u, p \right) = \begin{cases}
        \left(3.857143, 2.629369, 10.3333 \right), & \text{if}\;x \leqslant 1,  \\
        \left( 1+ 0.2 \sin \left(5x\right),0,1\right), & \text{otherwise}.
    \end{cases}
\end{equation}

The length of the computational domain is $x \in [0,10]$ and the solution is advanced in time to $t=1.2$. For all numerical methods investigated, the mesh is progressively refined from $N_x=256$ to $N_x = 2048$. The convergence is measured using the $\mathcal{L}^1$-norm (see  \cref{eqn:L1_norm_error}) of the difference in density between the final computed solution and a reference solution defined to be the solution computed with the second-order Godunov method with PPM interpolation and with a very fine mesh $N_x =32768$. In all simulations the CFL number is set to $0.5$.

\begin{figure}[tbhp]
\centering
\subfloat[Full domain]{\label{fig:Shu_Osher_WENO_PPM_256_a}\includegraphics[width=0.5\textwidth]{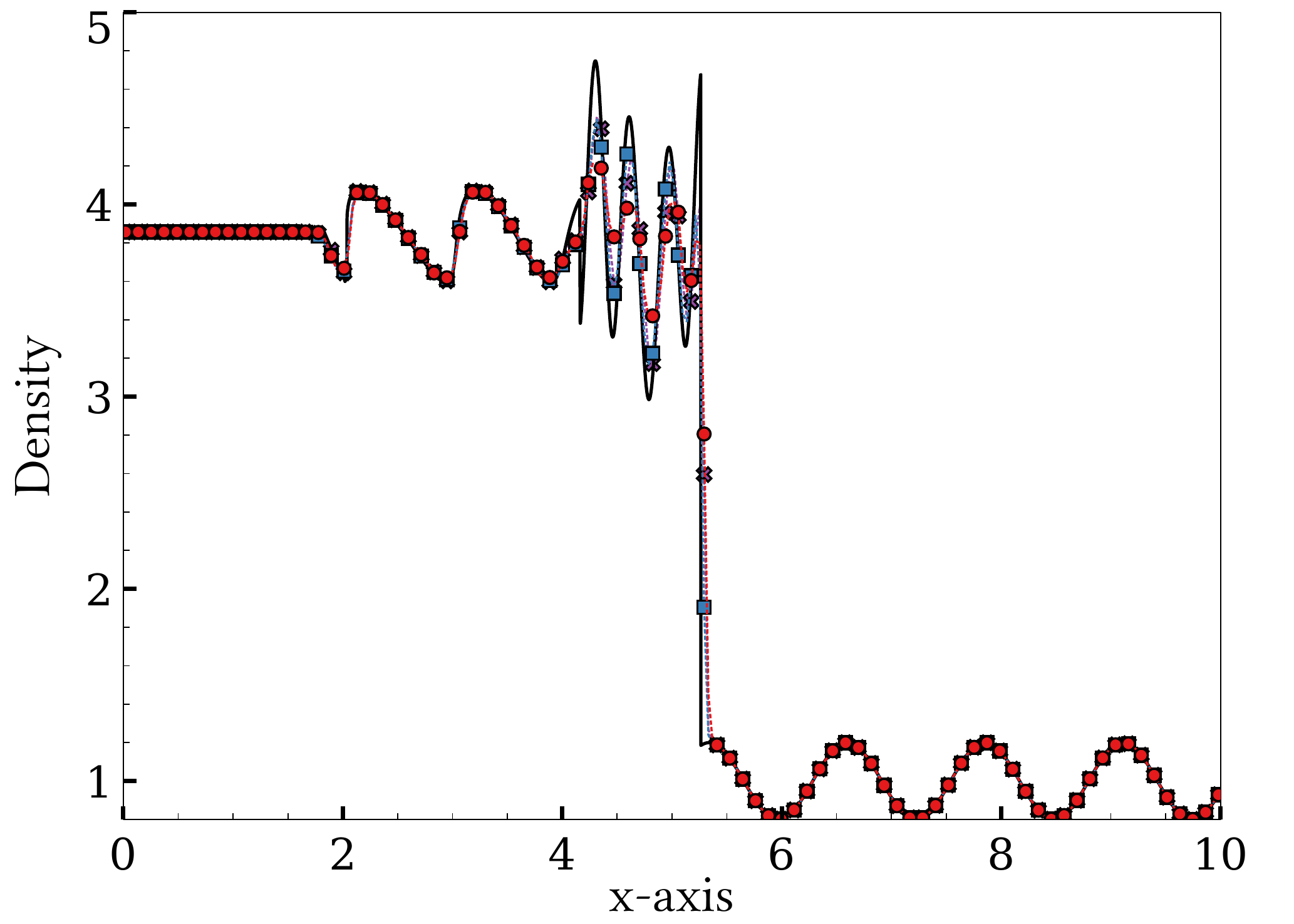}}
\subfloat[Zoom]{\label{fig:Shu_Osher_WENO_PPM_256_b}\includegraphics[width=0.5\textwidth]{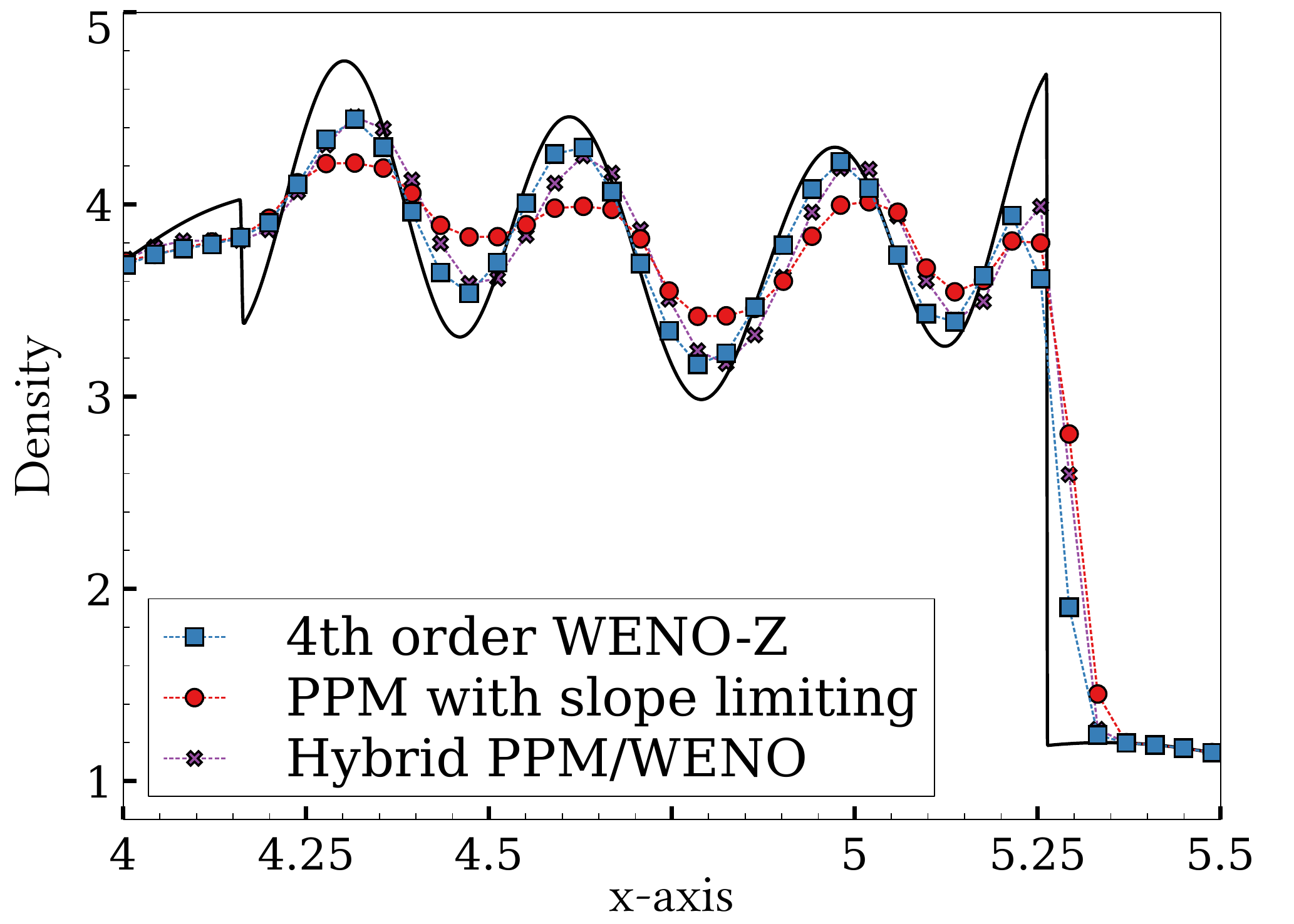}}
\caption{Shu-Osher test case: profile of density for $N_x=256$.}
\label{fig:Shu_Osher_WENO_PPM_256}
\end{figure}

\begin{figure}[tbhp]
\centering
\subfloat[Full domain]{\label{fig:Shu_Osher_WENO_PPM_512_a}\includegraphics[width=0.5\textwidth]{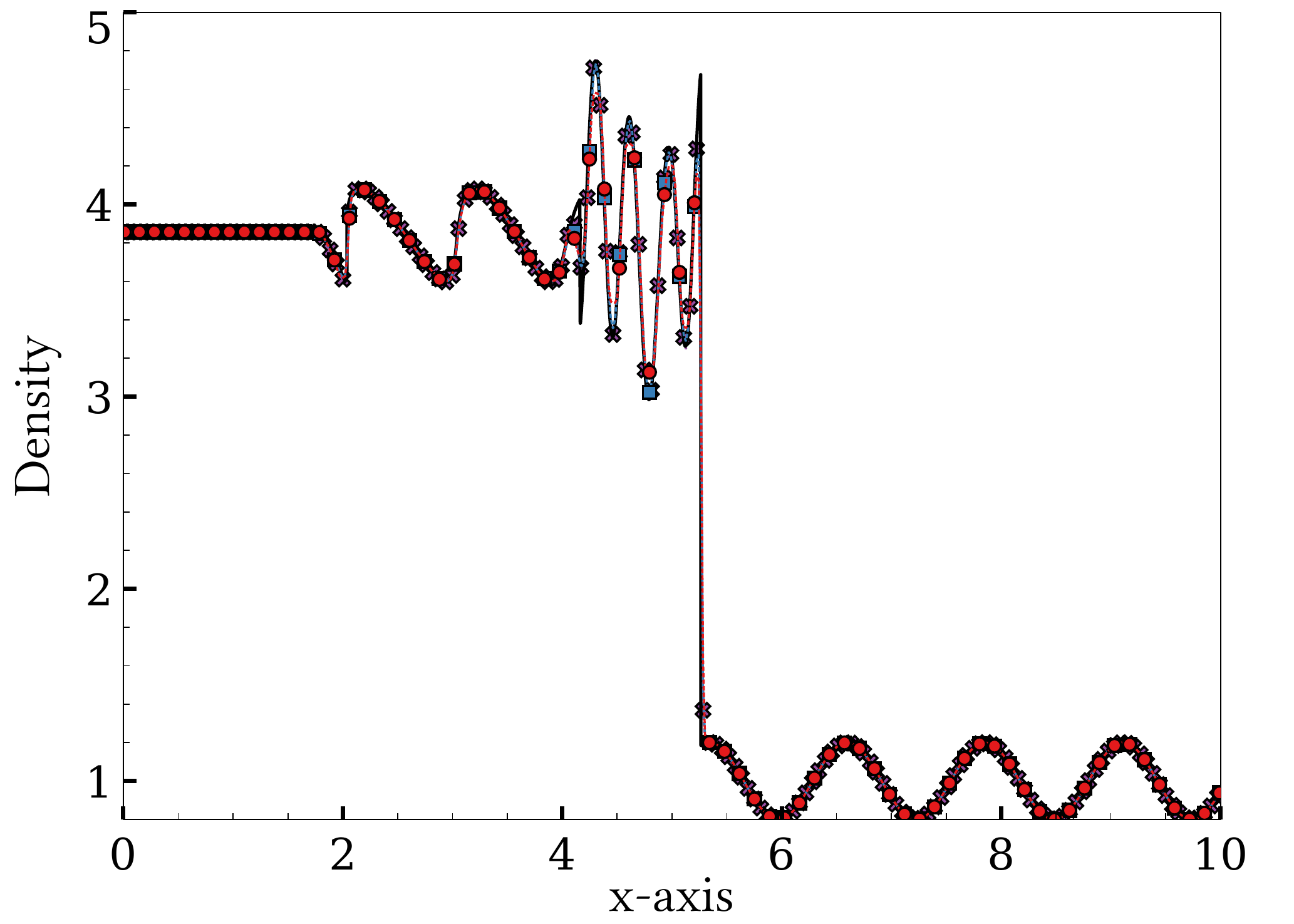}}
\subfloat[Zoom]{\label{fig:Shu_Osher_WENO_PPM_512_b}\includegraphics[width=0.5\textwidth]{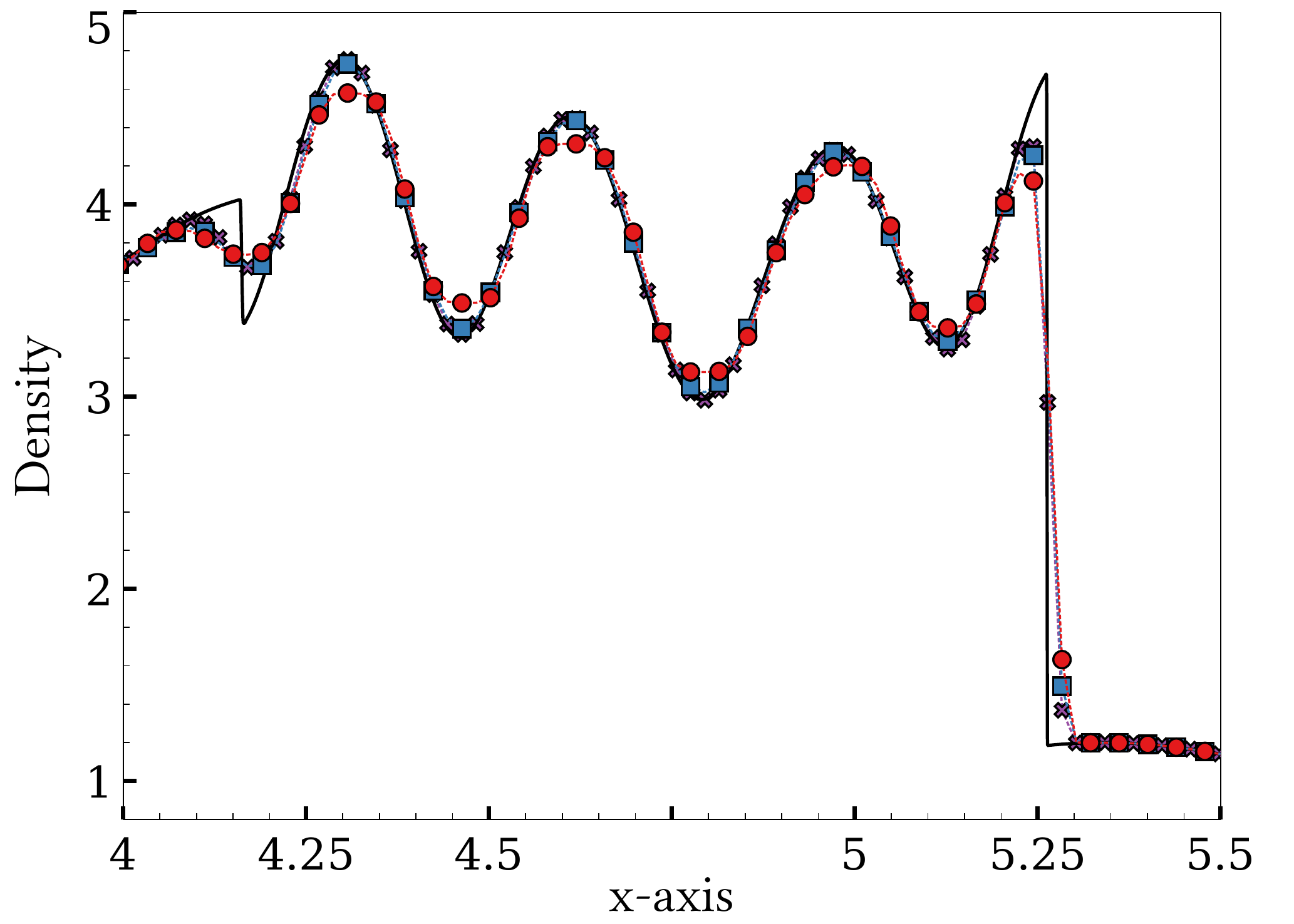}}
\caption{Shu-Osher test case: profile of density for $N_x=512$.}
\label{fig:Shu_Osher_WENO_PPM_512}
\end{figure}

\begin{figure}[tbhp]
\centering
\subfloat[Full domain]{\label{fig:Shu_Osher_WENO_PPM_1024_a}\includegraphics[width=0.5\textwidth]{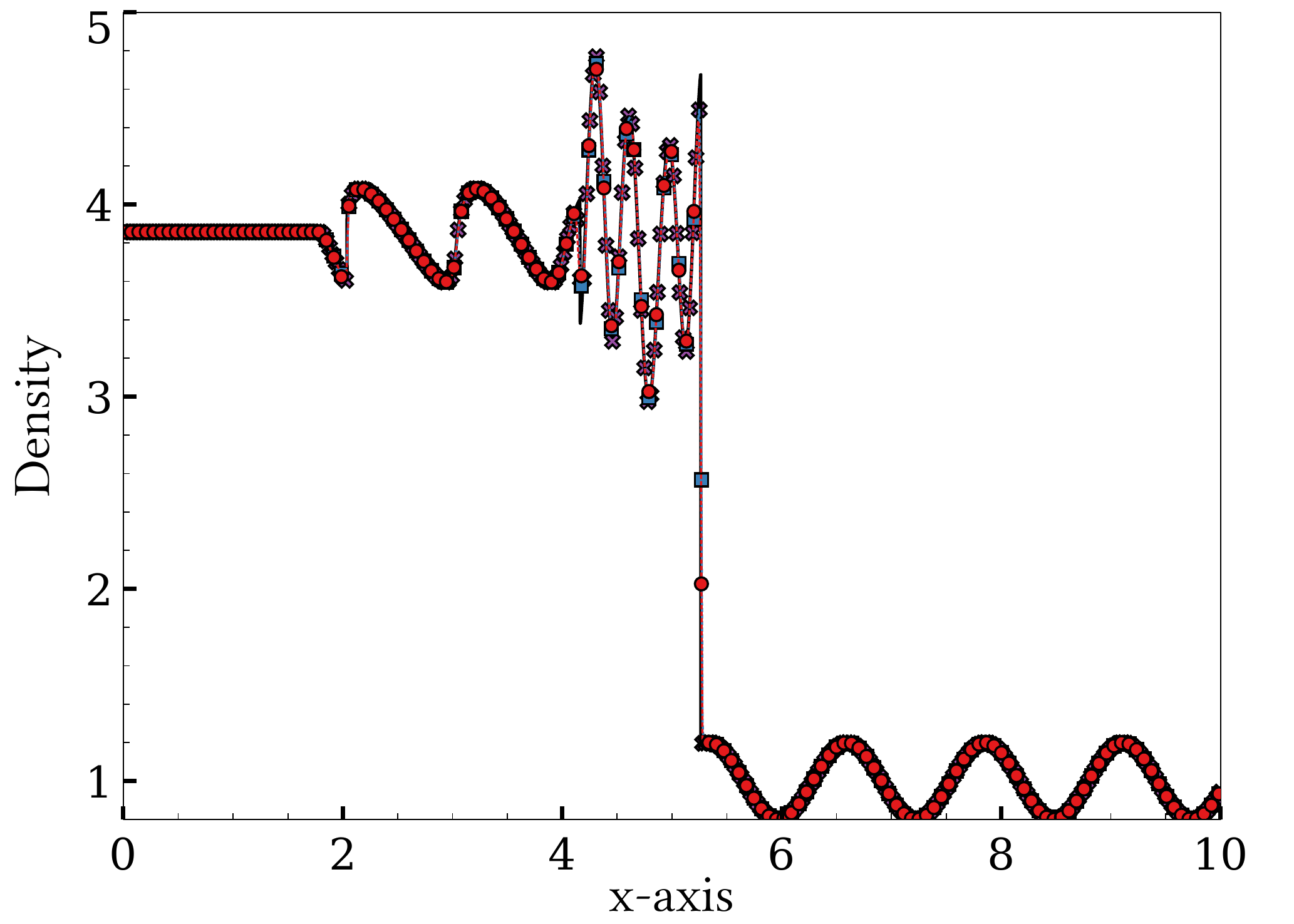}}
\subfloat[Zoom]{\label{fig:Shu_Osher_WENO_PPM_1024_b}\includegraphics[width=0.5\textwidth]{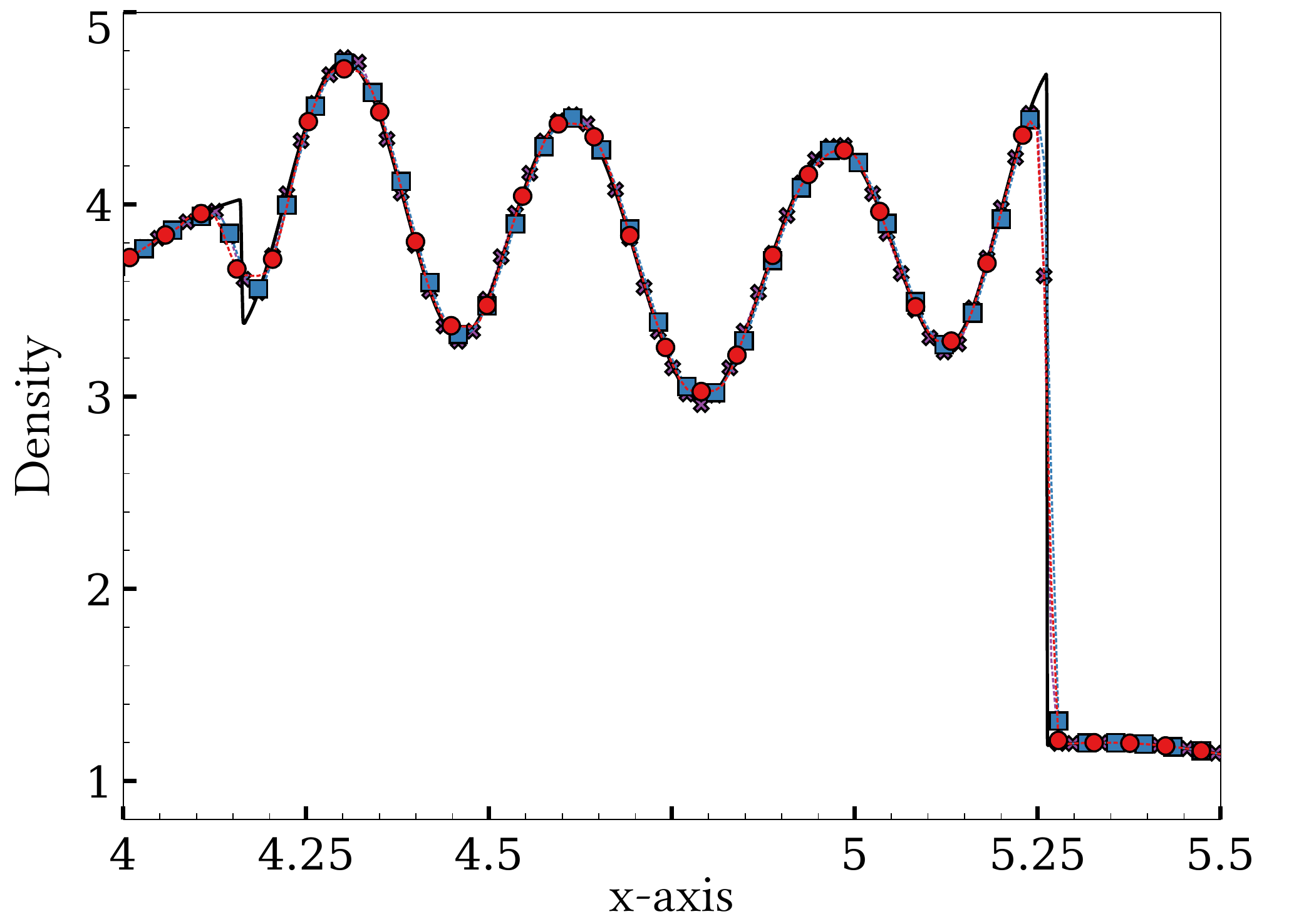}}
\caption{Shu-Osher test case: profile of density for $N_x=1024$.}
\label{fig:Shu_Osher_WENO_PPM_1024}
\end{figure}

\begin{figure}[tbhp]
\centering
\includegraphics[width=0.9\textwidth]{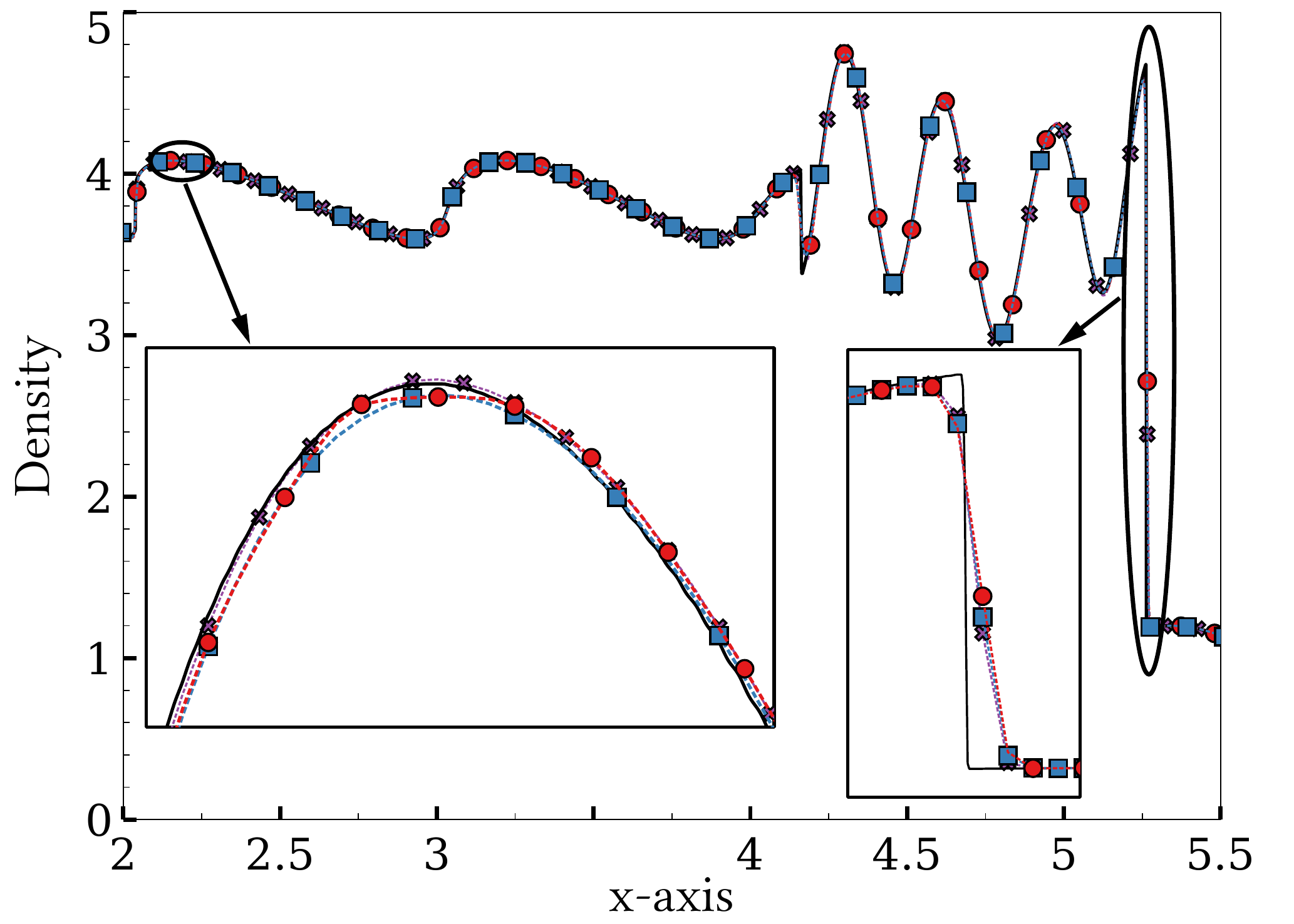}
\caption{Shu-Osher test case: profile of density for $N_x=2048$.}
\label{fig:Shu_Osher_WENO_PPM_2048}
\end{figure}

The density field at $t=1.2$ computed with $N_x = 256, 512, 1024$ and $2048$ is shown in \cref{fig:Shu_Osher_WENO_PPM_256,fig:Shu_Osher_WENO_PPM_512,fig:Shu_Osher_WENO_PPM_1024,fig:Shu_Osher_WENO_PPM_2048}, respectively. In these figures, the blue square, red circle and purple cross represents the fourth-order finite-volume WENO method with the WENO-Z variant, the original PPM method with slope limiting and the hybrid PPM/WENO method developed in the present paper, respectively (see legend in \cref{fig:Shu_Osher_WENO_PPM_256_b}). Note that the panels (a) and (b) in \cref{fig:Shu_Osher_WENO_PPM_256}, \cref{fig:Shu_Osher_WENO_PPM_512} and  \cref{fig:Shu_Osher_WENO_PPM_1024} present the full domain and a zoom in the domain, respectively, while \cref{fig:Shu_Osher_WENO_PPM_2048} is only a zoom in the domain. Note also that there is no relation between the symbols and the number of grid points. Several symbols have been removed from the figures for clarity purpose.

For a coarse mesh ($N_x=256$), a close look at \cref{fig:Shu_Osher_WENO_PPM_256_b} reveals that the fourth-order finite-volume WENO method is able to capture the correct phase of the waves, despite a damping of the amplitude. The second-order Godunov method with the original PPM interpolation and the slope limiting procedure does not accurately capture the correct profile of density. However, the hybrid PPM/WENO method presents a profile very similar to the one captured by the fourth-order finite-volume method. It turns out that changing the slope-limiting procedure in the PPM method to the WENO interpolation makes the second-order Godunov method recover the correct profile of density. This can be explained by the fact that the shock is better resolved by the WENO interpolation and that the slope limiting procedure introduces spurious wiggles in the density waves.

As seen in \cref{fig:Shu_Osher_WENO_PPM_512_b}, a mesh refinement by a factor $2$ makes all the methods to accurately capture the phase of the density waves. However the original PPM method with slope-limiting (red circle symbols) shows a damping of the amplitude, while the hybrid PPM/WENO method solution correctly captures both the phase and the amplitude, and is very close to the solution computed with the fourth-order finite-volume WENO method.

As the mesh is further refined, all the methods tend to collapse to the same solution. However, as can be seen in  \cref{fig:Shu_Osher_WENO_PPM_2048} for a fine mesh ($N_x=2048$), the fourth-order finite-volume WENO method shows a slight damping of the amplitude of the density wave, whereas the second-order Godunov method with PPM interpolation and slope-limiting exhibits some smooth high-frequency oscillations. The best solution is the one computed with the second-order Godunov method and the hybrid PPM/WENO method. The shape and amplitude of the density are closer to the reference solution.

\begin{figure}[tbhp]
\centering
\includegraphics[width=0.9\textwidth]{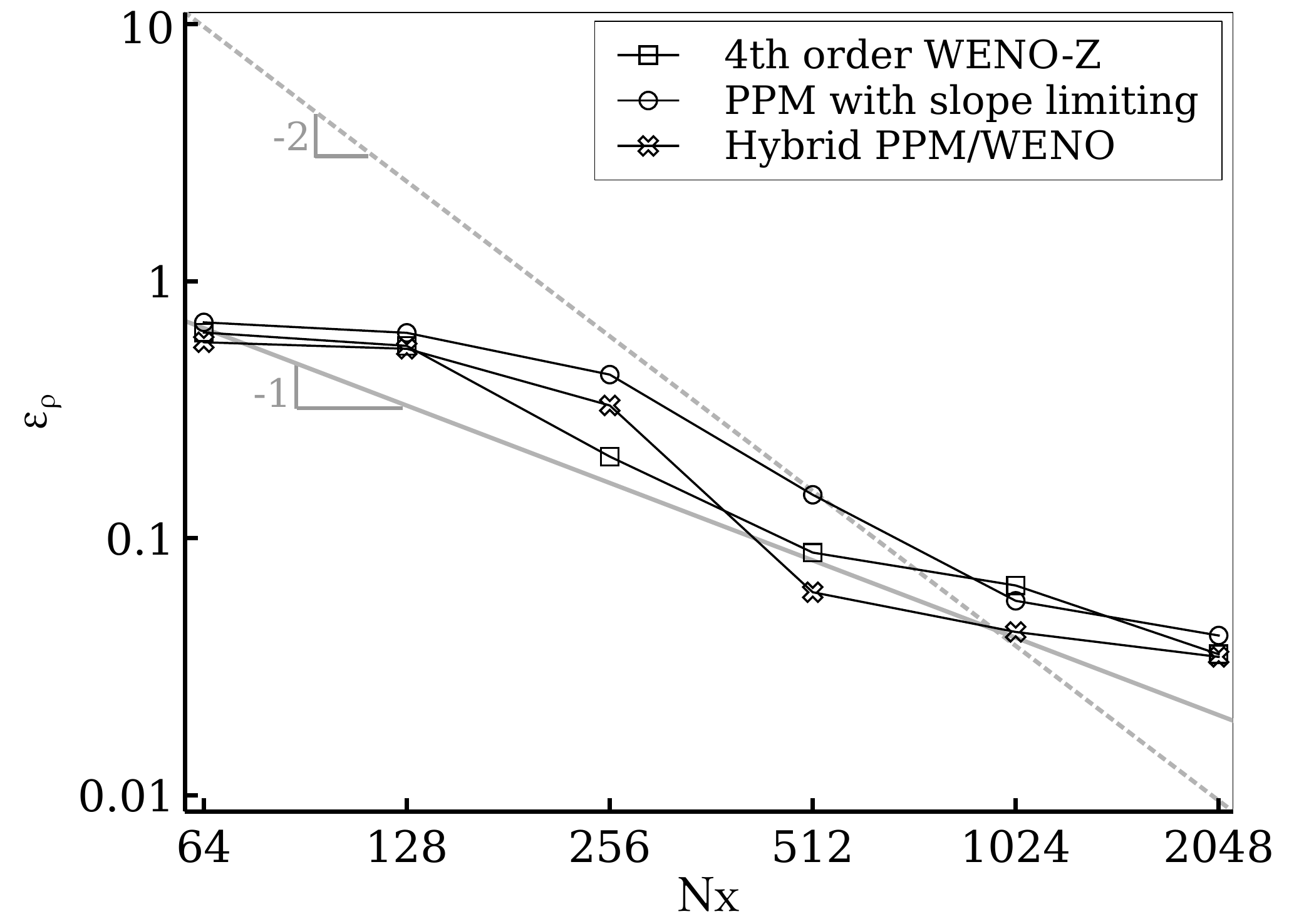}
\caption{Shu-Osher test case: $\mathcal{L}^1$-norm of the error on the density.}
\label{fig:Shu_Osher_convergence_study}
\end{figure}

Overall, it turns out that for this specific test case, the use of high-order methods is questionable. This is highlighted by the study of the convergence rate of the $\mathcal{L}^1$-norm of the error on the density profile. The error $\epsilon_\rho$ is reported in \cref{fig:Shu_Osher_convergence_study} and the convergence rate computed with a best-fitting curve method is reported in \cref{tab:Shu_Osher_convergence_rate}. It is obvious that all the numerical methods, either theoretically second- or fourth-order accurate, collapse to less than first-order accuracy because of the presence of the discontinuity. Overall, the present study suggests that reaching a correct approximation of a flow solution can be achieved by a second-order method and sufficient mesh resolution. In the following section, a more realistic three-dimensional compressible turbulent flow is simulated to investigate the capabilities of the second- and fourth-order numerical methods, as well as their effective cost in terms of mesh resolution, when both shocks and small turbulence structures interact in the same domain.

\begin{table}[p]
{\footnotesize
\caption{Shu-Osher test case: convergence rate of the $\mathcal{L}^1$-norm of the error on the density.}\label{tab:Shu_Osher_convergence_rate}
\begin{center}
\begin{tabular}{|c|c|} \hline
Method & \bf $\mathcal{O}\left( \epsilon_\rho\right)$  \\ \hline
PPM with slope-limiting & $0.92$  \\
$4$th-order WENO-Z &  $0.89$ \\
Hybrid PPM/WENO & $0.96$  \\  \hline
\end{tabular}
\end{center}
}
\end{table}

\subsection{Three-dimensional isotropic compressible turbulence decay}
\label{subsec:compressible_HIT}

The present test case consists on the simulation of the decay of a compressible isotropic turbulent field with the presence of eddy shocklets. Originally a physical study of turbulence in the work of Lee \textit{et al.} \cite{Lee:1991}, these simulations have become a framework to study the properties of numerical schemes to capture turbulence spectra and the decay of physical quantities. Here, the numerical setup described in \cite{Johnsen:2010} is reproduced.

The initial condition is built by generating a solenoidal velocity field $\vec{u}_0$ that satisfies:
\begin{equation}
E\left(k\right) \sim k^4 \exp\left(-2\left(k/k_0 \right)^2 \right), \hspace{.5cm} \frac{3 u^2_{{\rm rms},0}}{2} = \frac{<\vec{u}_0 \cdot \vec{u}_0>}{2} = \int_0^{\infty} E\left( k \right){\rm d} k 
\end{equation}
Here, $k_0$ is the most energetic wavenumber and is set to $k_0=4$.  The simulation is controlled by two non-dimensional parameters: the turbulent Mach number
\begin{equation}
M_{t,0} = \frac{\sqrt{<\vec{u}_0 \cdot \vec{u}_0>}}{c_0}
\label{eq:mach_turbulent}
\end{equation}
where $c_0$ is the sound speed in the initial solution, and the Taylor-scale Reynolds number defined as
\begin{equation}
Re_{\psi,0} = \frac{\rho_0  \psi_0 u_{\rm rms,0}}{\eta_0}
\label{eq:reynolds_number}
\end{equation}
where 
\begin{equation}
u_{{\rm rms},0} = \sqrt{\frac{<\vec{u}_0 \cdot \vec{u}_0>}{3}}, \hspace{.5cm} \psi_0 = \frac{2}{k_0}.
\end{equation}

In the present simulation, $M_{t,0} = 0.6$ and $Re_{\lambda,0}=100$. These values are set such that weak shock waves can develop spontaneously from the turbulent motions \cite{Johnsen:2010}, and allow numerical convergence for relatively coarse mesh grids to keep the computational cost reasonable. Once $M_{t,0}$ and $Re_{\psi,0}$ are set, $u_{{\rm rms},0}$ can be deduced from \cref{eq:mach_turbulent} with the known sound speed, and the viscosity $\eta_0$ can be deduced from  \cref{eq:reynolds_number}. Unlike the simulations presented in \cite{Johnsen:2010}, in the present study the viscosity is held constant throughout the simulation. Moreover, a constant thermal conductivity is set according to 
\begin{equation}
\lambda_0 = \frac{\eta_0 C_p}{Pr}
\end{equation}
where $C_p$ is the specific heat capacity, set to $C_p = 1.173$~kJ/kg.K and the Prandtl number $Pr$ is set to $Pr=0.71$. Moreover, the initial temperature and pressure in the flow are set to $T_0 = 1200$~K and $p_0=1$~atm.

All the simulations are performed over a non-dimensional time set to $t/\tau=4$ where $\tau = \psi_0/u_{{\rm rms},0}$. Several mesh resolutions are investigated: $N_x=64$, $N_x=128$, $N_x=256$ and $N_x=512$, and the CFL number is kept constant at $0.5$. 
Note that the practical procedure to generate the velocity fields $\vec{u}_0$ is detailed in \cite{Johnsen:2010}. It is also important to note that the initial turbulent velocity fields are first generated on a grid of $N_x = 512$ and then integrated over each cell in the mesh. Moreover, the initial solution is exactly the same for all simulations, regardless of the codes, numerical methods or mesh grids employed.

In order to assess the performance of the second-order Godunov methods and the fourth-order finite volume WENO method, a reference solution is generated with the very high-order code \textbf{SMC} \cite{Emmett:2014}  that employs eighth-order accurate centered finite-difference schemes for the spatial discretization, and a fourth-order Runge-Kutta algorithm for the time advancement. A convergence study for the reference solution is presented in \cref{sec:reference_SMC}. This reference solution will be depicted with a black solid line in the remainder of the paper.

\begin{figure}[tbhp]
\centering
\subfloat[Kinetic Energy]{\label{fig:HIT_tseries_WENO_PPM_full_a}\includegraphics[width=0.5\textwidth]{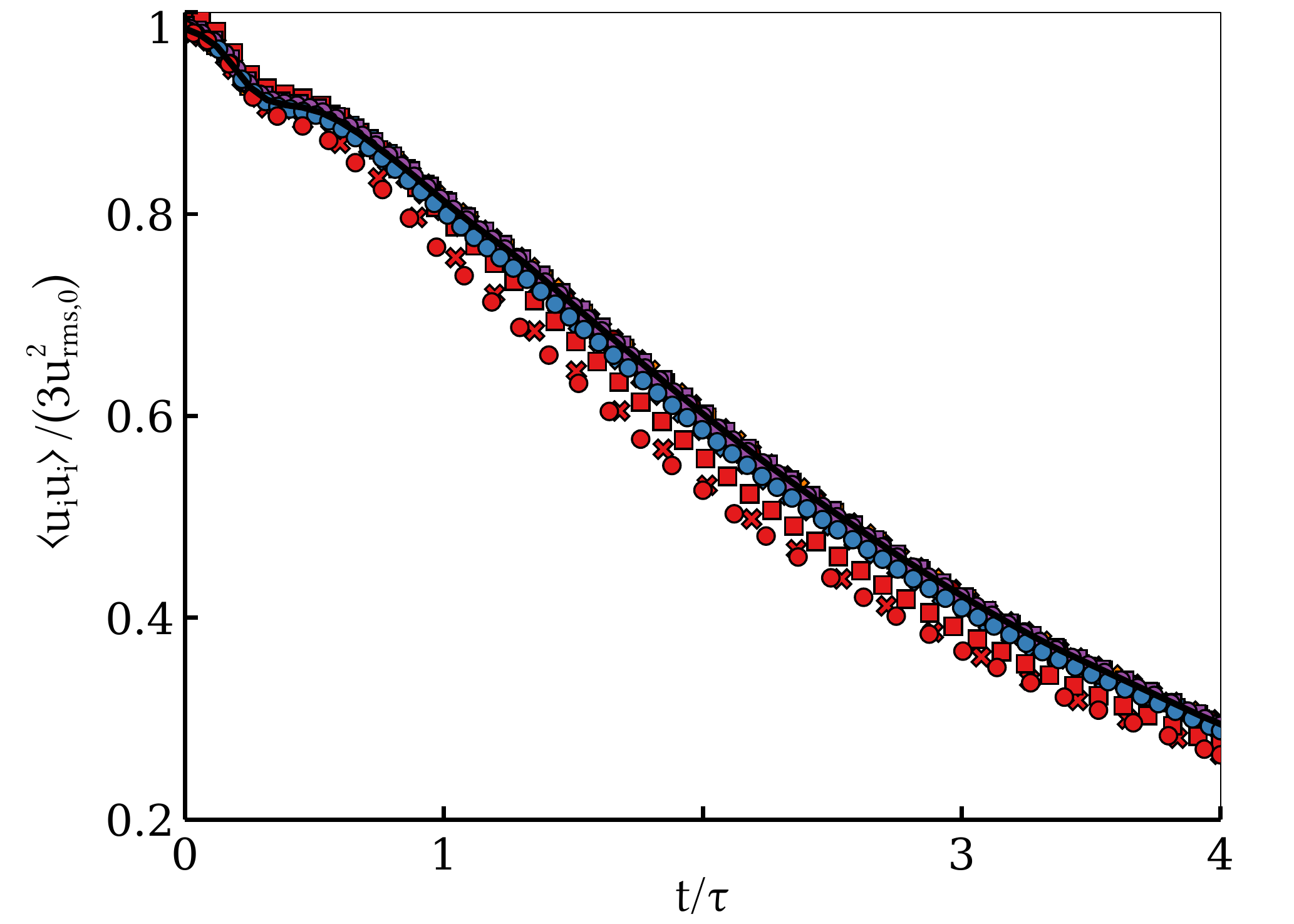}}
\subfloat[Enstrophy]{\label{fig:HIT_tseries_WENO_PPM_full_b}\includegraphics[width=0.5\textwidth]{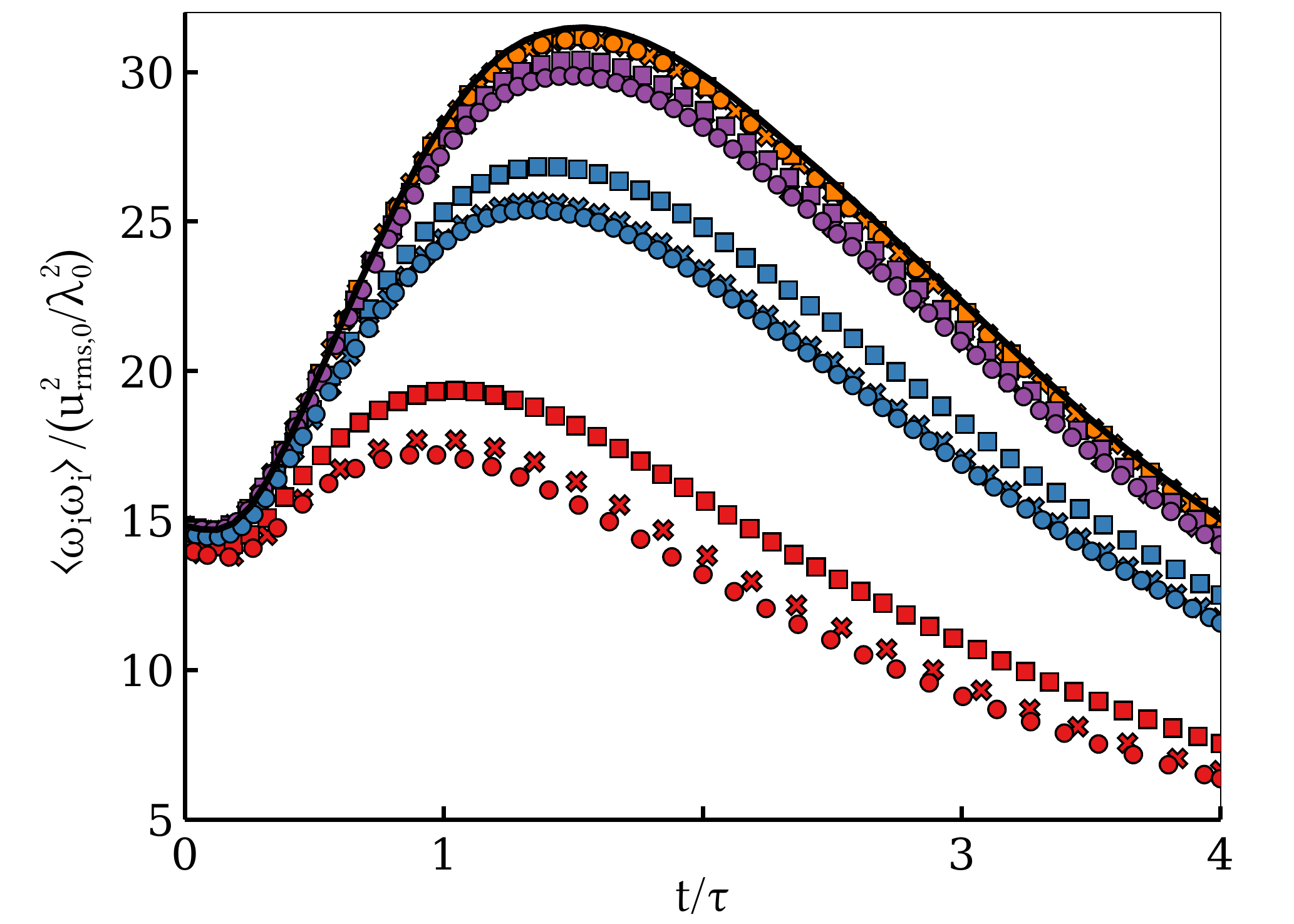}} \\
\subfloat[Temperature]{\label{fig:HIT_tseries_WENO_PPM_full_c}\includegraphics[width=0.5\textwidth]{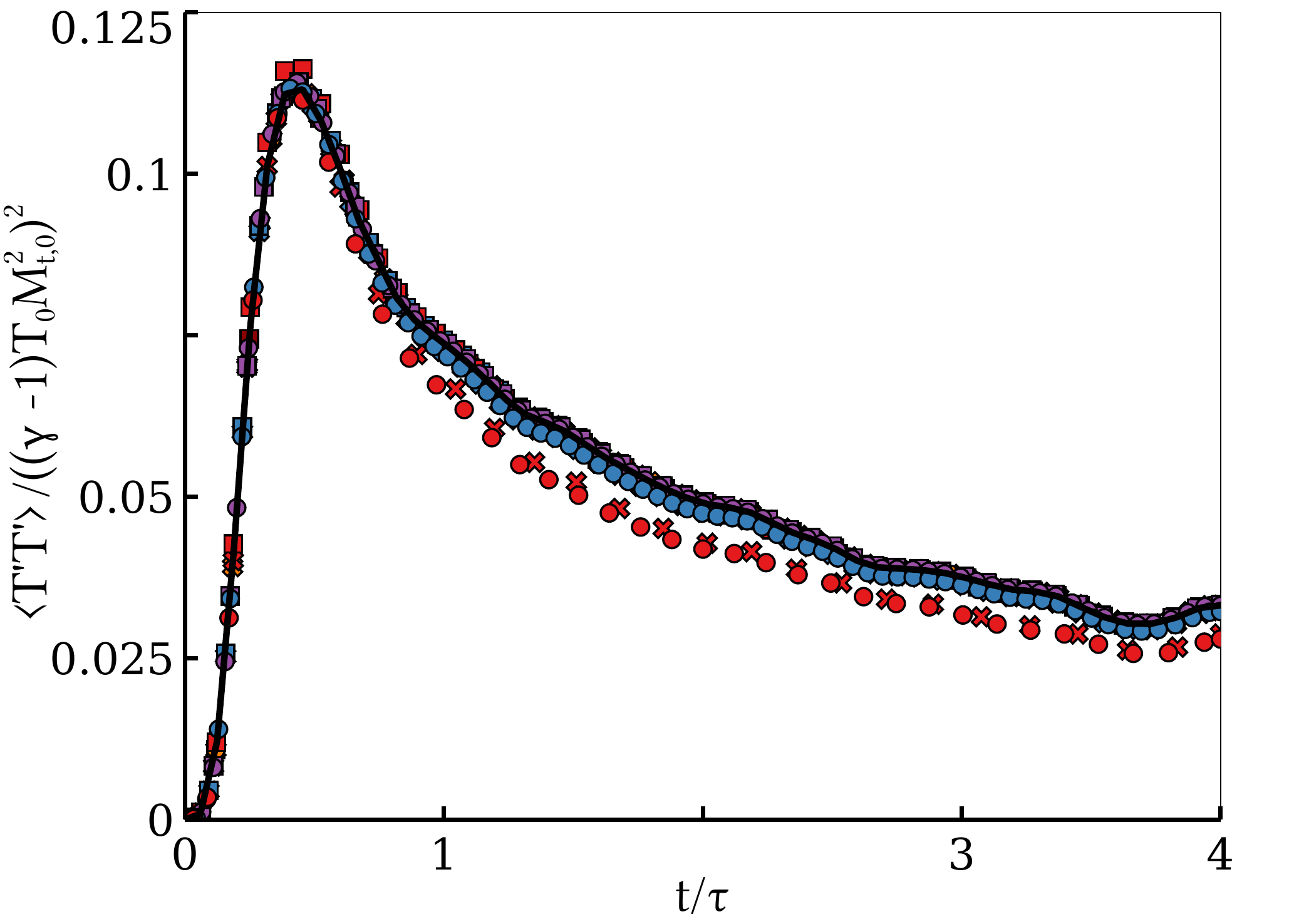}}
\subfloat[Dilatation, $\theta = \partial_j u_j$]{\label{fig:HIT_tseries_WENO_PPM_full_d}\includegraphics[width=0.5\textwidth]{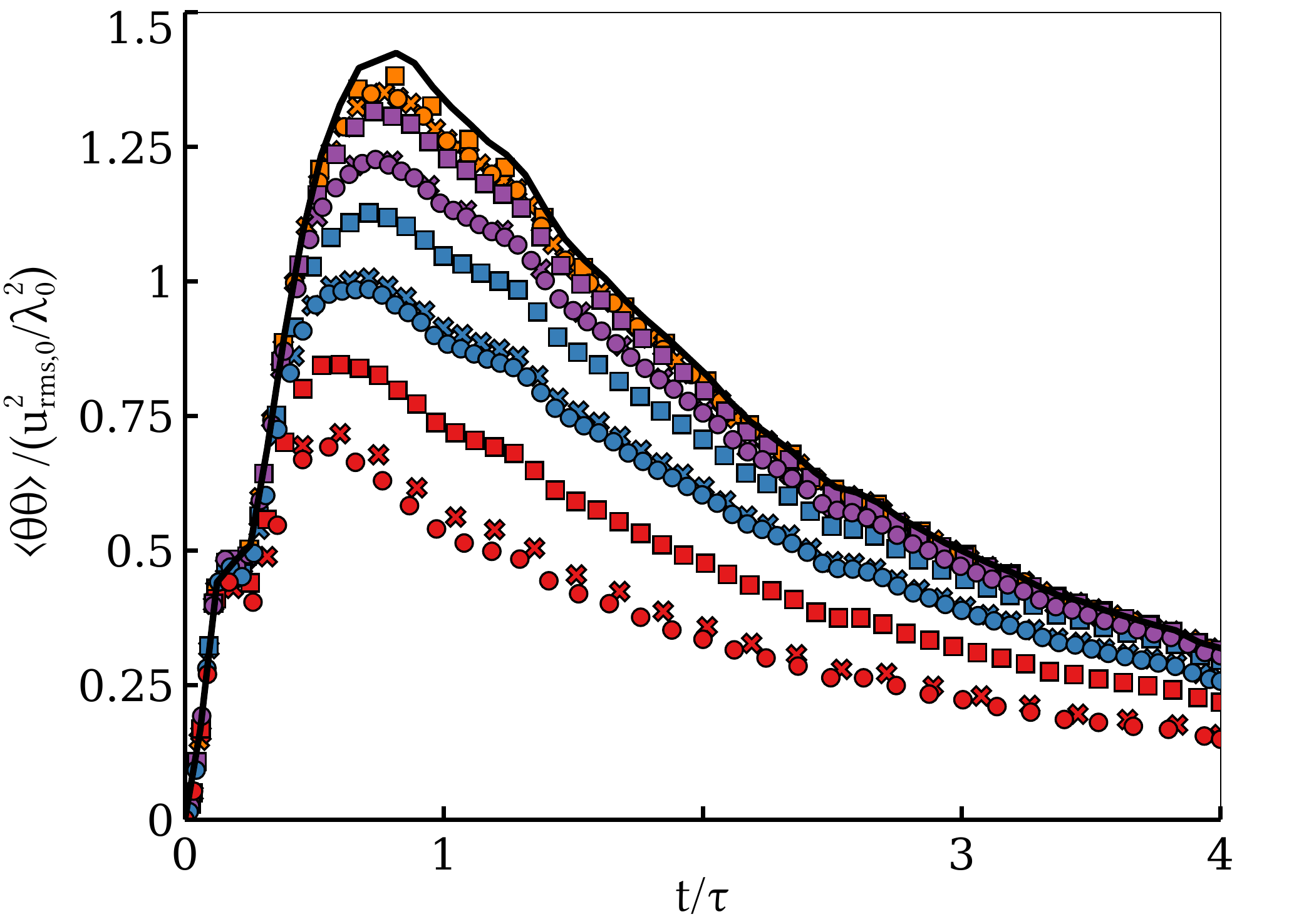}}
\caption{Time series of selected physical quantities for simulations performed with different mesh resolution and numerical methods. Legend is recalled in the text and in \cref{fig:HIT_legend}.}
\label{fig:HIT_tseries_WENO_PPM_full}
\end{figure}

\begin{figure}[tbhp]
\centering
\subfloat[Kinetic Energy]{\label{fig:HIT_spectra_WENO_PPM_full_a}\includegraphics[width=0.5\textwidth]{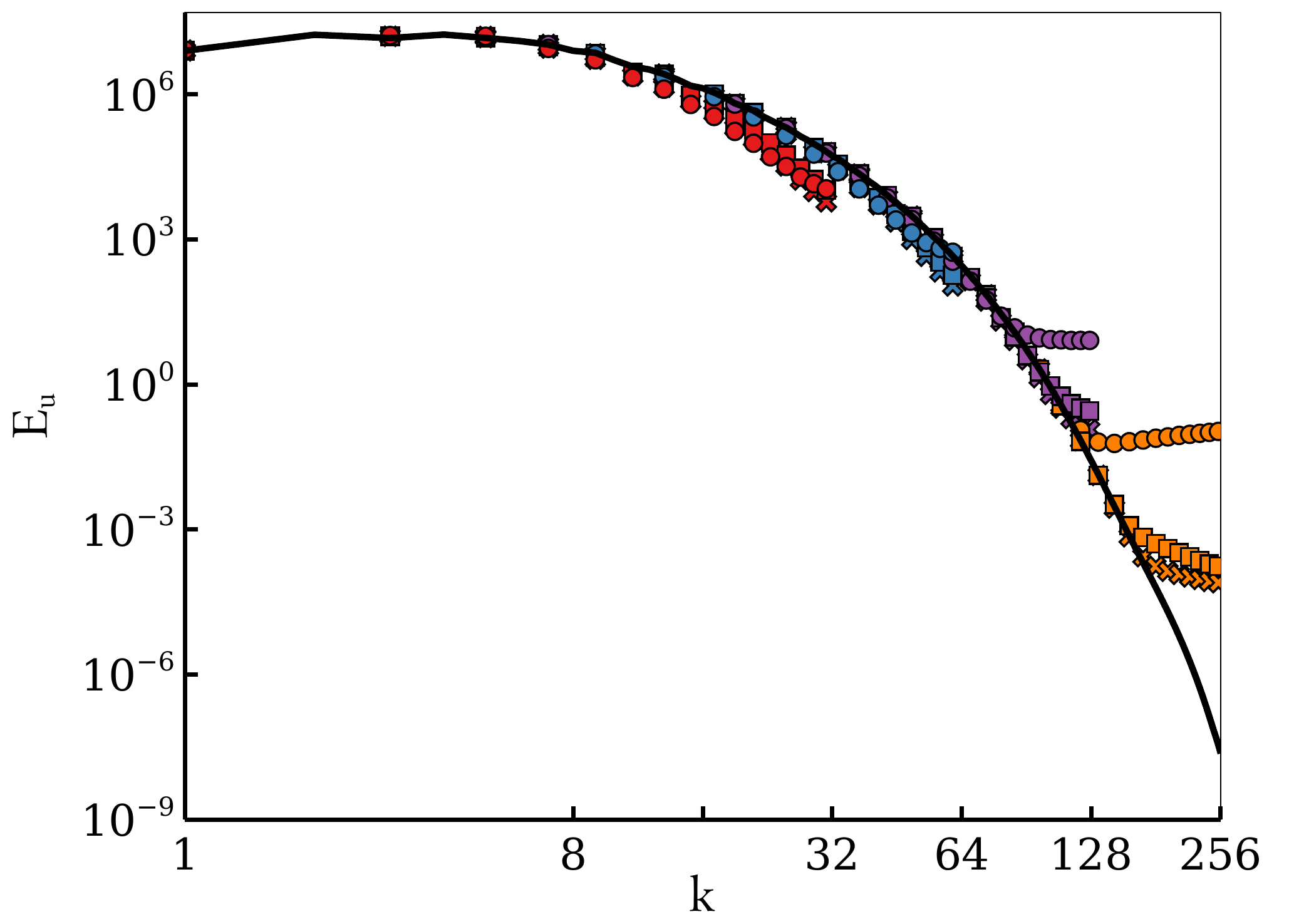}}
\subfloat[Vorticity]{\label{fig:HIT_spectra_WENO_PPM_full_b}\includegraphics[width=0.5\textwidth]{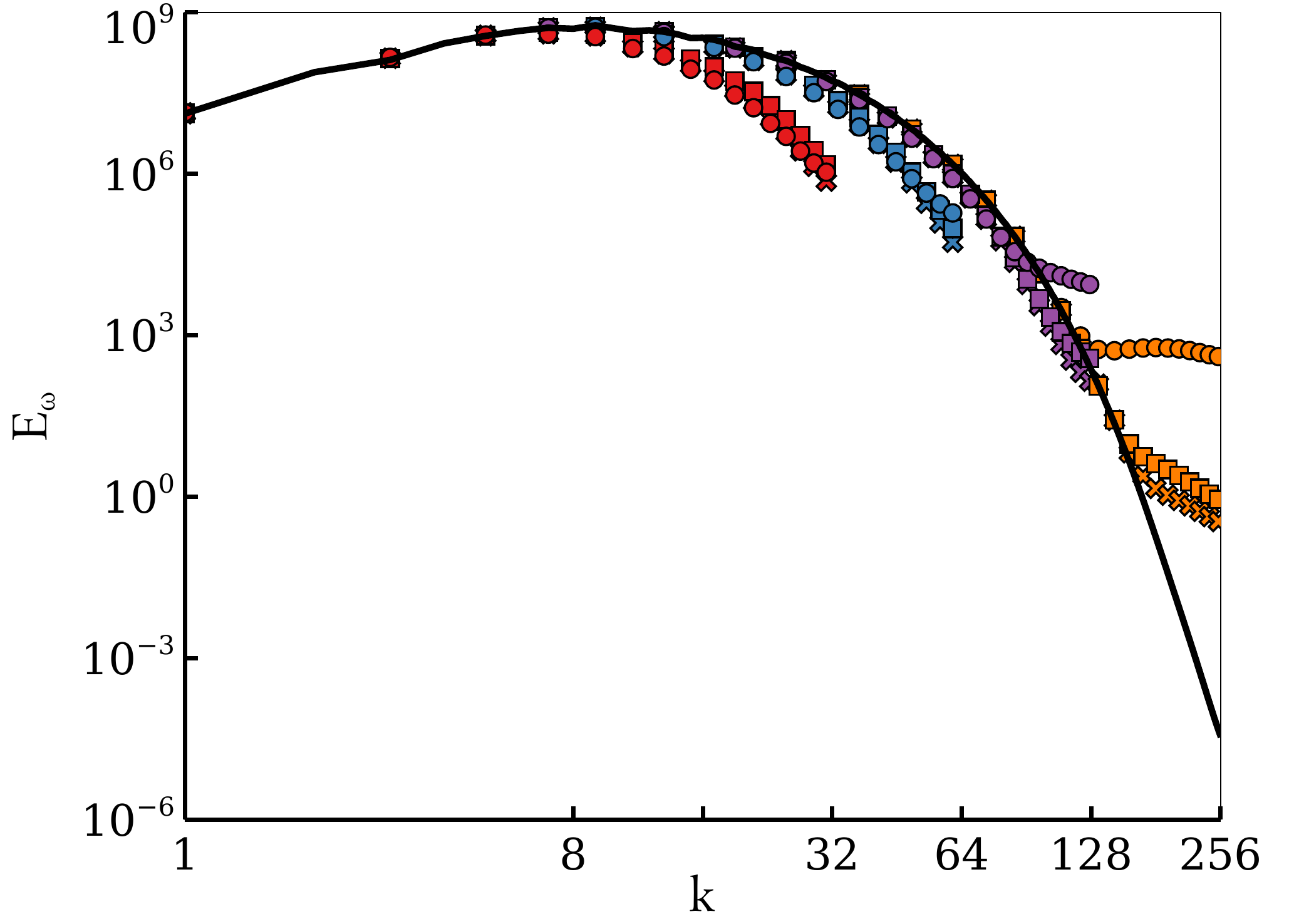}} \\
\subfloat[Dilatation]{\label{fig:HIT_spectra_WENO_PPM_full_c}\includegraphics[width=0.5\textwidth]{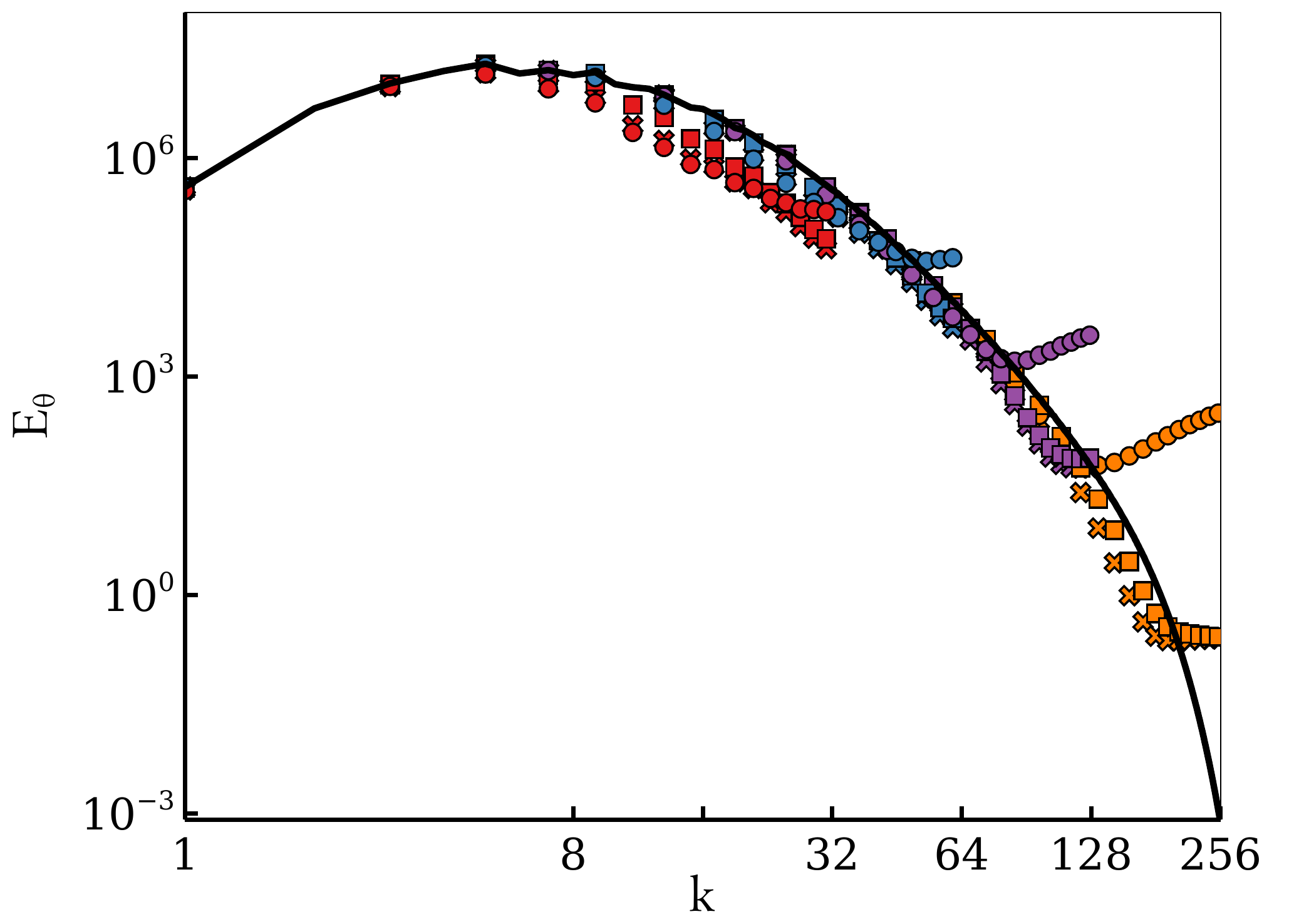}}
\subfloat[Density]{\label{fig:HIT_spectra_WENO_PPM_full_d}\includegraphics[width=0.5\textwidth]{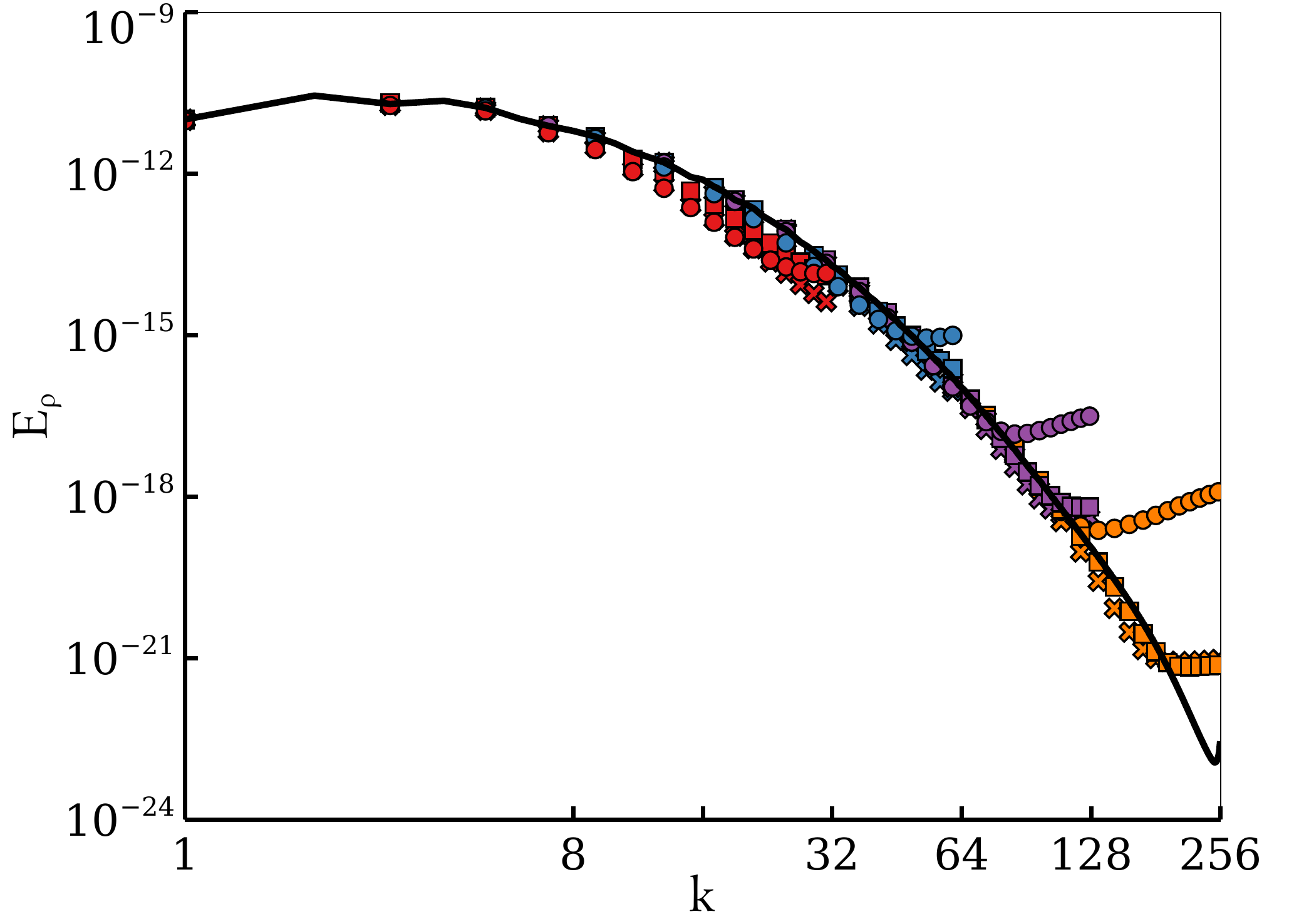}}
\caption{Spectra of selected physical quantities for simulations performed with different mesh resolution and numerical methods. Legend is recalled in the text and in \cref{fig:HIT_legend}.}
\label{fig:HIT_spectra_WENO_PPM_full}
\end{figure}

\begin{figure}[tbhp]
\centering
\includegraphics[width=0.8\textwidth]{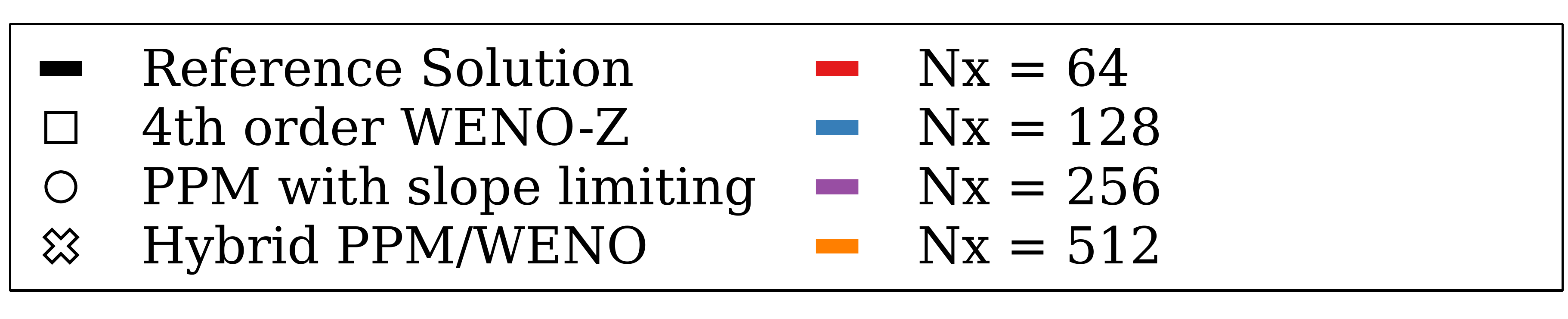}
\caption{Symbols and color legend for \cref{fig:HIT_tseries_WENO_PPM_full,fig:HIT_spectra_WENO_PPM_full,fig:HIT_spectra_WENO_zoom}.}
\label{fig:HIT_legend}
\end{figure}

\begin{figure}[tbhp]
\centering
\subfloat[$N_x=64^3$]{\label{fig:HIT_spectra_WENO_zoom_a}\includegraphics[width=0.5\textwidth]{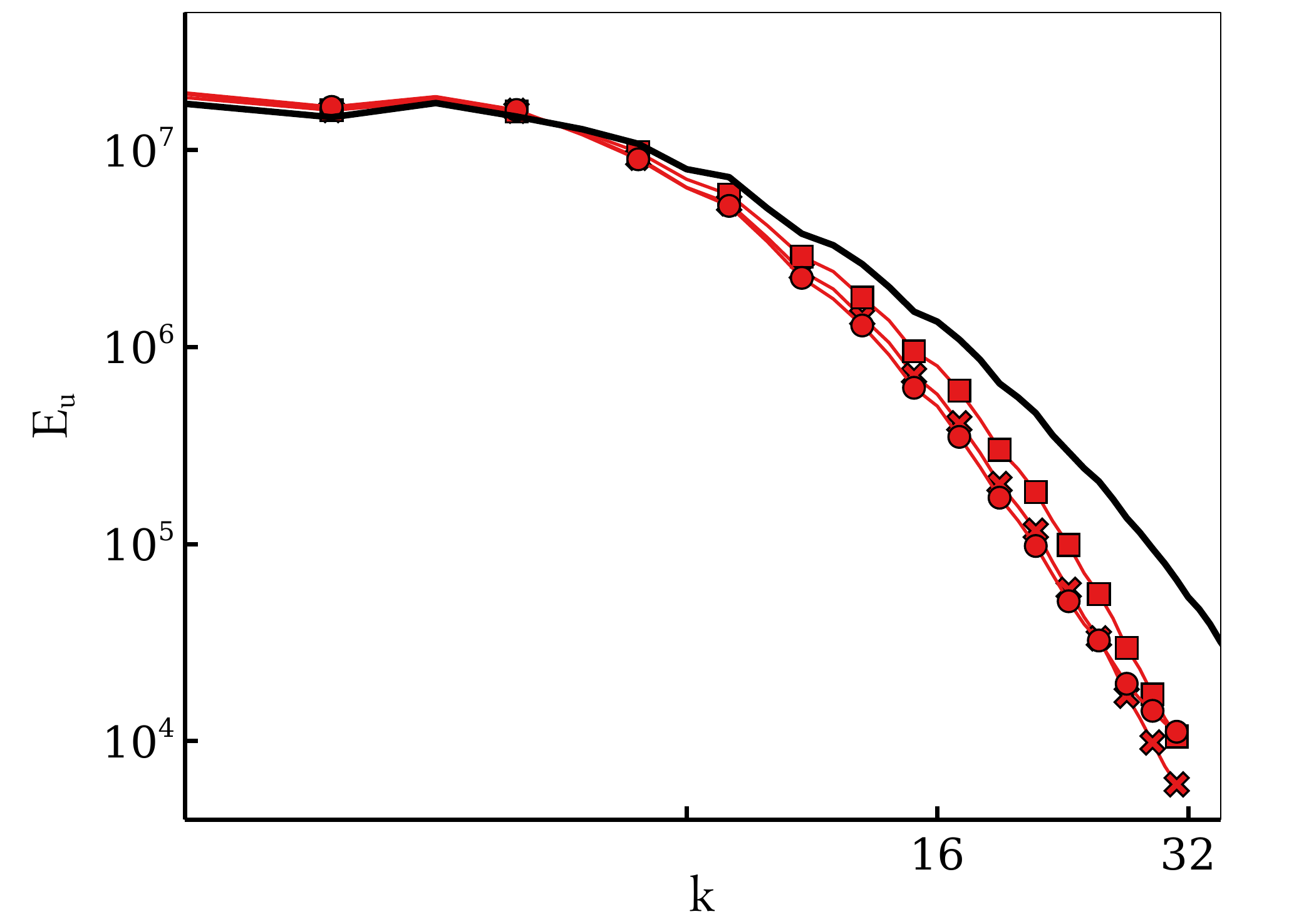}}
\subfloat[$N_x=128^3$]{\label{fig:HIT_spectra_WENO_zoom_b}\includegraphics[width=0.5\textwidth]{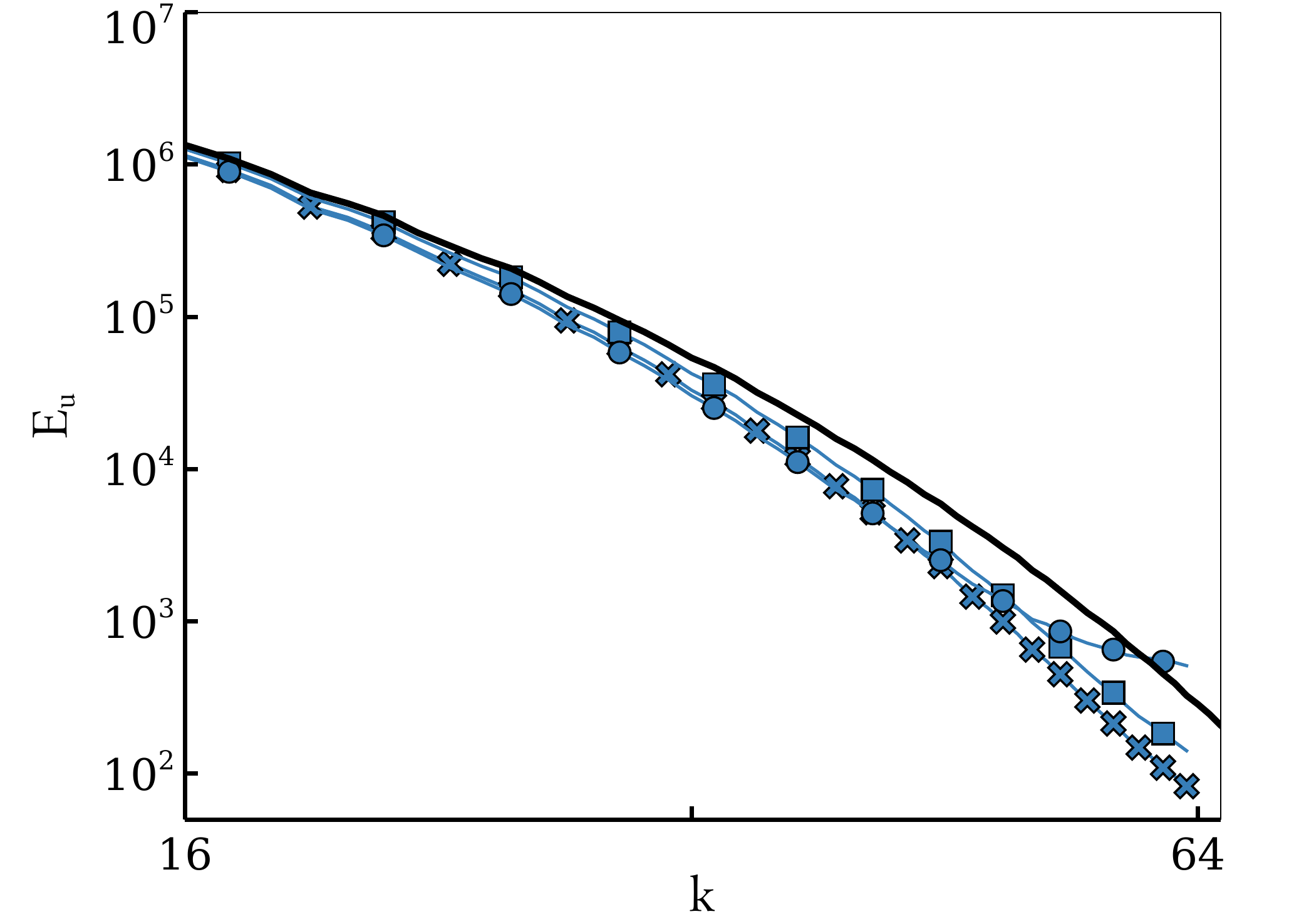}} \\
\subfloat[$N_x=256^3$]{\label{fig:HIT_spectra_WENO_zoom_c}\includegraphics[width=0.5\textwidth]{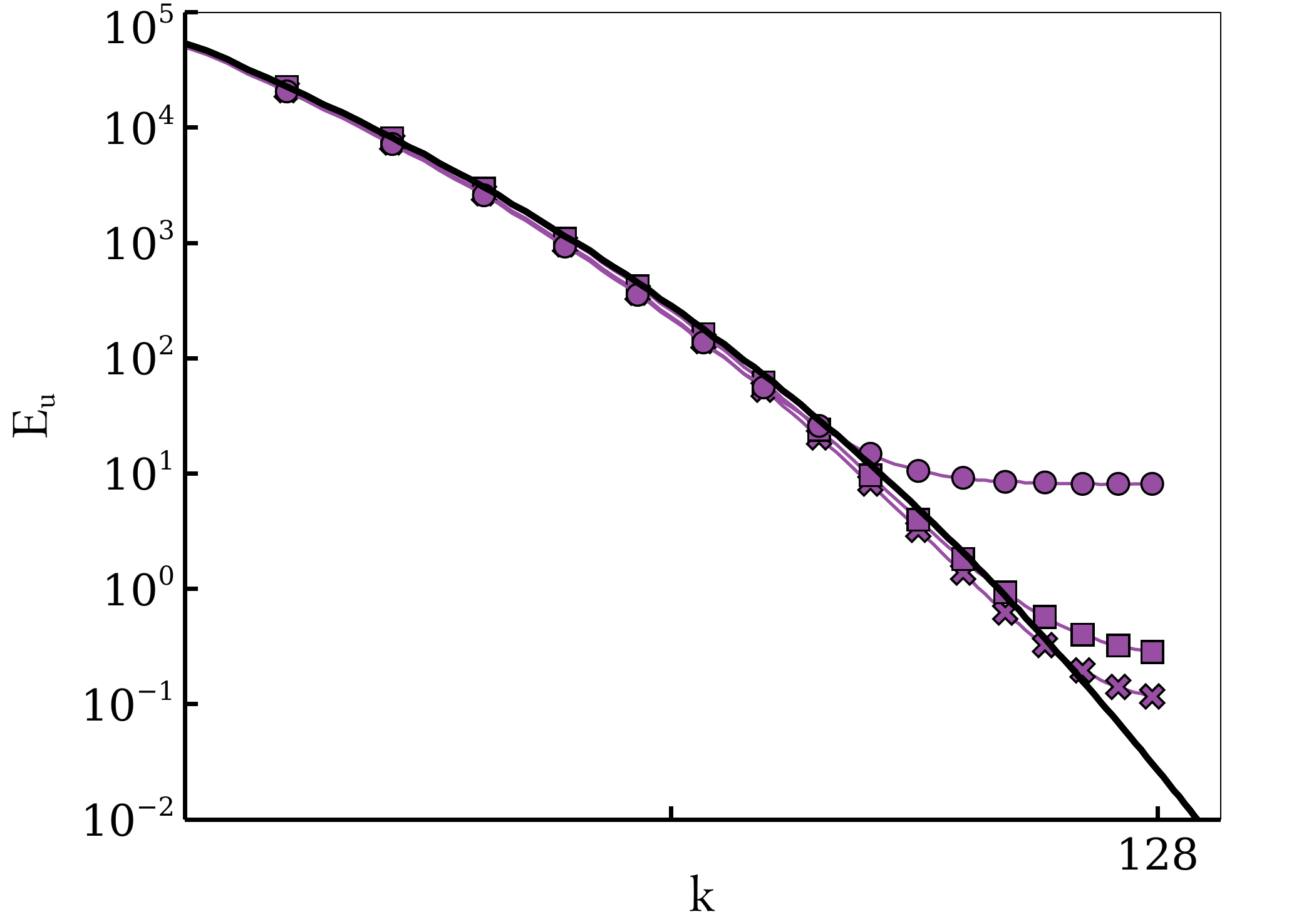}}
\subfloat[$N_x=512^3$]{\label{fig:HIT_spectra_WENO_zoom_d}\includegraphics[width=0.5\textwidth]{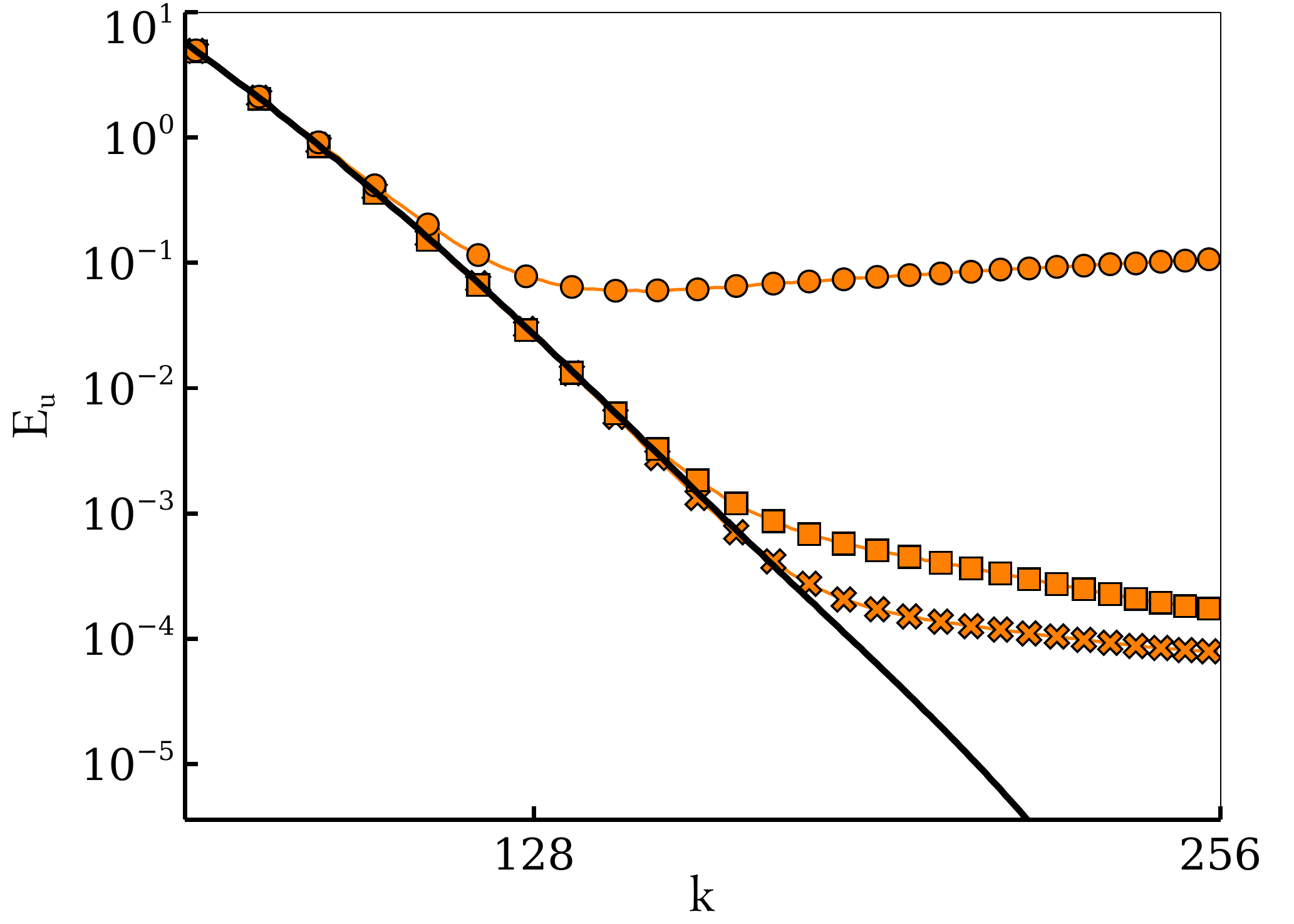}}
\caption{Zoom of the spectra of kinetic energy in \cref{fig:HIT_spectra_WENO_PPM_full} for results computed with different mesh resolutions.}
\label{fig:HIT_spectra_WENO_zoom}
\end{figure}

\Cref{fig:HIT_tseries_WENO_PPM_full_a,fig:HIT_tseries_WENO_PPM_full_b,fig:HIT_tseries_WENO_PPM_full_c,fig:HIT_tseries_WENO_PPM_full_d}
present the temporal evolution of the kinetic energy, the enstrophy, the variance of temperature and the dilatation from $t=0$ to $t/\tau=4$. It can be seen that strong compressibility effects are generated quickly after the beginning of the simulation, suggesting the generation of eddy shocklets in the domain until  $t/\tau\approx 0.5$. After  $t/\tau\approx 1$, compressible shocks are no longer generated and they start to decay in a monotone way. \Cref{fig:HIT_spectra_WENO_PPM_full_a,fig:HIT_spectra_WENO_PPM_full_b,fig:HIT_spectra_WENO_PPM_full_c,fig:HIT_spectra_WENO_PPM_full_d} present the spectra taken at $t/\tau=4$ for the kinetic energy, the vorticity, the dilatation and the density. In these figures, the circle, cross and square symbols represent the second-order Godunov with PPM interpolation and slope-limiting, the second-order Godunov method with the hybrid PPM/WENO procedure, and the fourth-order finite-volume WENO strategy, respectively. The red, blue, purple and orange colors represent simulations performed with  $N_x=64$, $N_x=128$, $N_x=256$ and $N_x=512$, respectively. It is emphasized that these figures contain a significant number of curves. For clarity, the legend is recalled in \cref{fig:HIT_legend} and a zoom on the high-end of the spectra for the kinetic energy is shown in \cref{fig:HIT_spectra_WENO_zoom}
for each mesh resolution. Note that the behavior of the numerical methods highlighted in \cref{fig:HIT_spectra_WENO_zoom} is virtually the same for the spectra of other physical quantities.

From the temporal evolution of physical quantities presented in \cref{fig:HIT_tseries_WENO_PPM_full}, it is clear that the second-order Godunov method, with either the PPM interpolation method with slope limiting or the hybrid PPM/WENO method, gives virtually the same results, with the exception of the very coarse mesh where some slight differences exist. In any case, for the same mesh resolution, the fourth-order finite-volume WENO method provides a better solution.

However, the spectra depicted in \cref{fig:HIT_spectra_WENO_PPM_full} do not follow the same behavior as for the temporal series. Indeed, given a mesh resolution all the numerical methods give virtually the same spectra, but as the refinement of the mesh allows small turbulent structures to be resolved, it turns out that the different numerical methods do not perform equally in the high-frequencies of the spectrum. As it can be seen in \cref{fig:HIT_spectra_WENO_zoom}, whereas all methods present a pile-up of energy in the high-frequency range, the fourth-order finite-volume WENO method resolves the spectra with a monotone decreasing energy, which is not the case for the second-order Godunov method with PPM interpolation and slope limiting. Most interesting, the second-order Godunov method with the hybrid PPM/WENO reconstruction method is able to reproduce virtually the same spectra as the fourth-order finite-volume WENO method, meaning that replacing the slope-limiting procedure by the WENO reconstruction method recovers a monotone spectra close to the reference solution.

\begin{figure}[tbhp]
\centering
\includegraphics[width=0.85\textwidth]{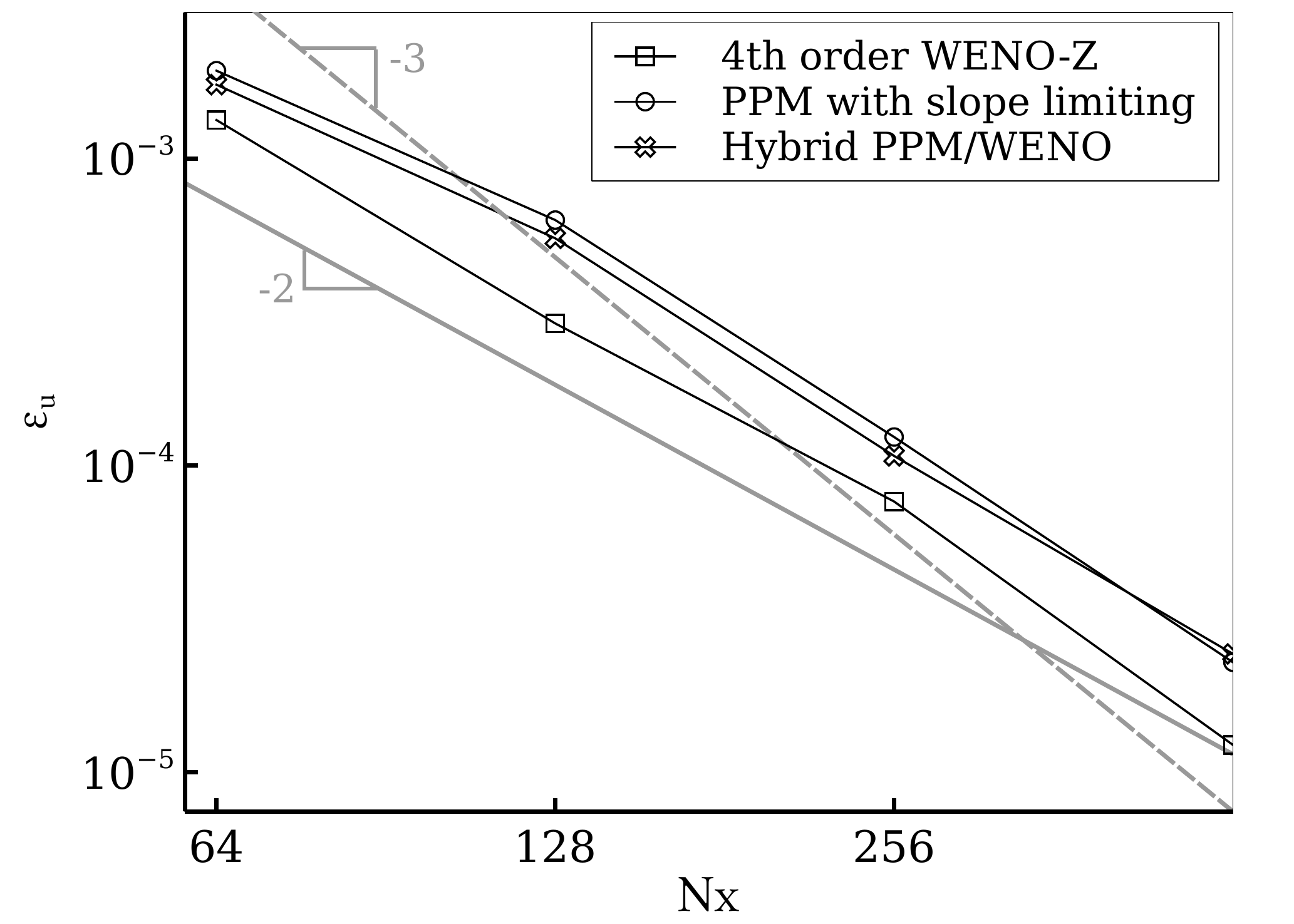}
\caption{HIT test case: $\mathcal{L}^1$-norm of the error on the density.}
\label{fig:HIT_convergence_study}
\end{figure}

Among these general trends, what emerges from all the figures is that for a given mesh resolution, the solutions are very close to each other regardless of the numerical method employed, with the exception of the high-end frequencies at fine mesh resolution. Such observations make sense, because as the turbulent Mach number is $0.6$, the present 3D HIT test case can be seen as a mix between the Shu-Osher test case (see \cref{subsec:Shu_Osher}) where all the methods collapse to first-order, and the smooth solution test case presented at \cref{subsec:COVO} where each numerical method follows its own theoretical order of convergence.  This is highlighted by the study of the convergence rate with the $\mathcal{L}^1$-norm of the error on the $x$-velocity profile. The error $\epsilon_u$ is reported in \cref{fig:HIT_convergence_study} and the convergence rate computed with a best-fitting curve method is reported in \cref{tab:HIT_convergence_rate}. Overall, all the numerical methods present a second-order convergence rate. It is emphasized that this finding is the opposite to the conclusion in \cite{Almgren:2013}, where the fourth-order method always gives better results than the second-order one. This again makes sense, because the decay of turbulence investigated in \cite{Almgren:2013} is simulated in an incompressible regime, leading to a solution always smooth. In that case, the findings of the study in \cite{Almgren:2013} are consistent with the behavior shown in our study in \cref{subsec:COVO}, where a smooth vortex is simulated and where all the numerical methods follow their theoretical order of convergence. Our study highlights that in presence of strong compressibility effects, the theoretical expectations of a numerical method no longer hold because of the interaction with shocks.


\begin{table}[tbhp]
{\footnotesize
\caption{HIT test case: convergence rate of the $\mathcal{L}^1$-norm of the error on the density}\label{tab:HIT_convergence_rate}
\begin{center}
\begin{tabular}{|c|c|} \hline
Method & \bf $\mathcal{O}\left( \epsilon_\rho\right)$  \\ \hline
PPM with slope-limiting & $2.15$  \\
$4$th-order WENO-Z &  $2.22$ \\
Hybrid PPM/WENO & $2.08$  \\  \hline
\end{tabular}
\end{center}
}
\end{table}

All the results presented so far are investigations of the accuracy of the solutions, but another important parameter to take into account is the computational cost of each numerical method. As the \textbf{PeleC} and \textbf{RNS} codes are based on the \textbf{AMReX} framework, the profiling functionality of the library has been used to extract the actual computational cost to evaluate the hyperbolic terms in the set of governing equations. In practice, a timer has been put around the main routine called to compute the terms. \Cref{tab:HIT_cpu_time} presents the average of the computational time for the evaluation of the routines involved in the computation of the hyperbolic convection term, divided by the number of calls during the whole simulation. This nondimensionalization is adopted here because the second-order Godunov procedure requires only one evaluation of the convection term, whereas the finite-volume WENO method is implemented with a Runge-Kutta time integration procedure that requires many calls by time iteration. Also, the simulations are performed with the same mesh resolution of $N_x=256$ and with the same parallelization over $512$ MPI process. It turns out that the fourth-order finite-volume WENO method is about $200$ times more computationally expensive than the second-order Godunov method. For the Godunov method, the new hybrid PPM/WENO method proposed in the present paper has roughly the same computational cost as the original PPM method with slope-limiting.

This significant difference can be explained by the number of interpolation procedures required for each cell and by time-step. If we consider only one component in the system of equations, the PPM method requires only $6$ interpolations in total (one by face), whereas for the high-order finite-volume method, the required  number of interpolations is estimated with the following equation: 
\begin{equation}
2D \left( 2^D -1 \right) \times 2
\end{equation}
where $D$ is the number of dimensions in the computational domain and the factor $2$ in the right hand side stems for the number of Runge-Kutta stages. In three dimensions, achieving fourth-order accuracy requires $14$ times more interpolation procedures by cell than with the PPM algorithm, because data have to be evaluated through Gauss integration points. It is emphasized that this computational burden does not only depend on the interpolation procedures via the WENO schemes, to this count must be added the number of calls to the Riemann solver and all the conversions between conservative and primitive variables.

Overall, from the results presented in this section, it becomes apparent that an accurate representation of a compressible turbulent flow can be achieved faster with a second-order accurate Godunov method, together with the new hybrid PPM/WENO strategy for the reconstruction of physical values at faces that can achieve the same spectra resolution as a more complex and costly high-order method. Because the computational cost of the second-order Godunov method with PPM interpolation is significantly lower than the high-order finite-volume WENO method, it turns out that refining a simulation with the second-order method is still less costly than running a coarse high-order simulation. In this test case, it appears that the use of a fourth-order finite-volume WENO method is unnecessary in practice. The major finding of this study is that for finite-volume methods, the accuracy of the reconstruction of fluxes at cells interface has significantly more impact than the formal order of the method.

\begin{table}[tbhp]
{\footnotesize
\caption{HIT computational time}\label{tab:HIT_cpu_time}
\begin{center}
\begin{tabular}{|c|c|} \hline
Method &  Nondimensional CPU time [s]  \\ \hline
PPM with slope-limiting & $5.06 \times 10^{-3}$  \\
$4$th-order WENO-Z &  $1.1149$ \\
Hybrid PPM/WENO & $5.03 \times 10^{-3}$  \\  \hline
\end{tabular}
\end{center}
}
\end{table}

\section{Conclusions}
\label{sec:conclusions}

A comparison between low-order and high-order finite-volume methods has been performed on a series of test cases: the convection of a smooth 2D vortex, the Shu-Osher problem, and the decay of 3D homogeneous isotropic turbulence. The choice to assess the performance of finite-volume methods is justified by the fact that they are more robust and flexible to use in the context of simulations of industrial applications. The study focus on the second-order Godunov method, as well as the fourth-order finite-volume WENO method. Results show that while on a smooth problem the high-order method perform better than the second-order one, when the solution contains a shock all the methods collapse to first-order accuracy. The study of the decay of compressible homogeneous isotropic turbulence with shocklets shows that the actual overall order of accuracy of the methods reduces to near second-order, despite the use of fifth-order reconstruction schemes. Most important, results in terms of turbulent spectra are similar regardless of the numerical methods employed, except for the higher end of the frequencies. Because our results show that the original PPM method with slope limiting fails to provide an accurate representation in the high-frequency range of the spectra, a novel hybrid PPM/WENO method is proposed. It is demonstrated that such hybrid PPM/WENO method has the ability to capture the turbulent spectra with the accuracy of a formally high-order method, but at the cost of the second-order Godunov method. Moreover, this study highlights that for finite-volume methods, the accuracy of the reconstruction of fluxes at cells interface has significantly more impact than the formal order of the method. Overall, the present study demonstrates the importance of evaluating the accuracy of a numerical method in terms of its actual spectral dissipation and dispersion properties on mixed smooth/shock cases, rather than by the theoretical formal order of convergence rate.

\appendix
\section{Slope-flattening procedure}
\label{sec:appendix_flattening}

In \cref{subsubsec:interpolation_limiting} a flattening limiter is imposed at \cref{eqn:flattening_eq_1,eqn:flattening_eq_2} through a flattening coefficient $\chi_i$. The coefficient $\chi_i \in \left[0,1 \right]$, where $\chi_i=1$ indicates that no additional limiting take place, whereas $\chi_i=0$ means that the Godunov method is dropped to first-order accuracy. The computation of $\chi_i$ is performed as follows:
\begin{enumerate}
\item First, a dimensionless measure of the shock resolution is computed with
\begin{equation}
\varsigma_i = \frac{p_{i+1}-p_{i-1}}{\max\left(p_{\rm small},|p_{i+2}-p_{i-2}| \right)}
\end{equation}
where $p$ is the pressure and $p_{\rm small}$ is a very small value to avoid a division by zero.
\item Then the parameter $\tilde{\chi}_i$ is defined as
\begin{equation}
\tilde{\chi}_i = \min\{1,\max \left[0,a\left(\varsigma_i -b \right) \right] \}
\end{equation}
where $a=10$ and $b=0.75$ are parameters set by the user. In order to confine $\tilde{\chi}_i$ in the range $\left[ 0,1\right]$, $\tilde{\chi}_i=0$ if either $u_{i+1}-u_{i-1} < 0$ or
\begin{equation}
\frac{p_{i+1}-p_{i-1}}{\min\left(p_{i+1},p_{i-1} \right)} \leqslant c
\end{equation}
with $c$ a parameter set by the user, which take the value of $c=1/3$ here.
\item Finally $\chi_i$ is computed as follows:
\begin{equation}
\chi_i = \begin{cases}
        1-\max\left(\tilde{\chi}_i,\tilde{\chi}_{i-1} \right), & \text{if}\;p_{i+1}-p_{i-1} > 0,  \\
        1-\max\left(\tilde{\chi}_i,\tilde{\chi}_{i+1} \right), & \text{otherwise}.
    \end{cases}
\end{equation}
\end{enumerate}

\section{Reference solution with the very high-order SMC code for the decay of homogeneous isotropic turbulence}
\label{sec:reference_SMC}

\begin{figure}[h!]
\centering
\subfloat[Kinetic Energy]{\label{fig:tseries_SMC_a}\includegraphics[width=0.5\textwidth]{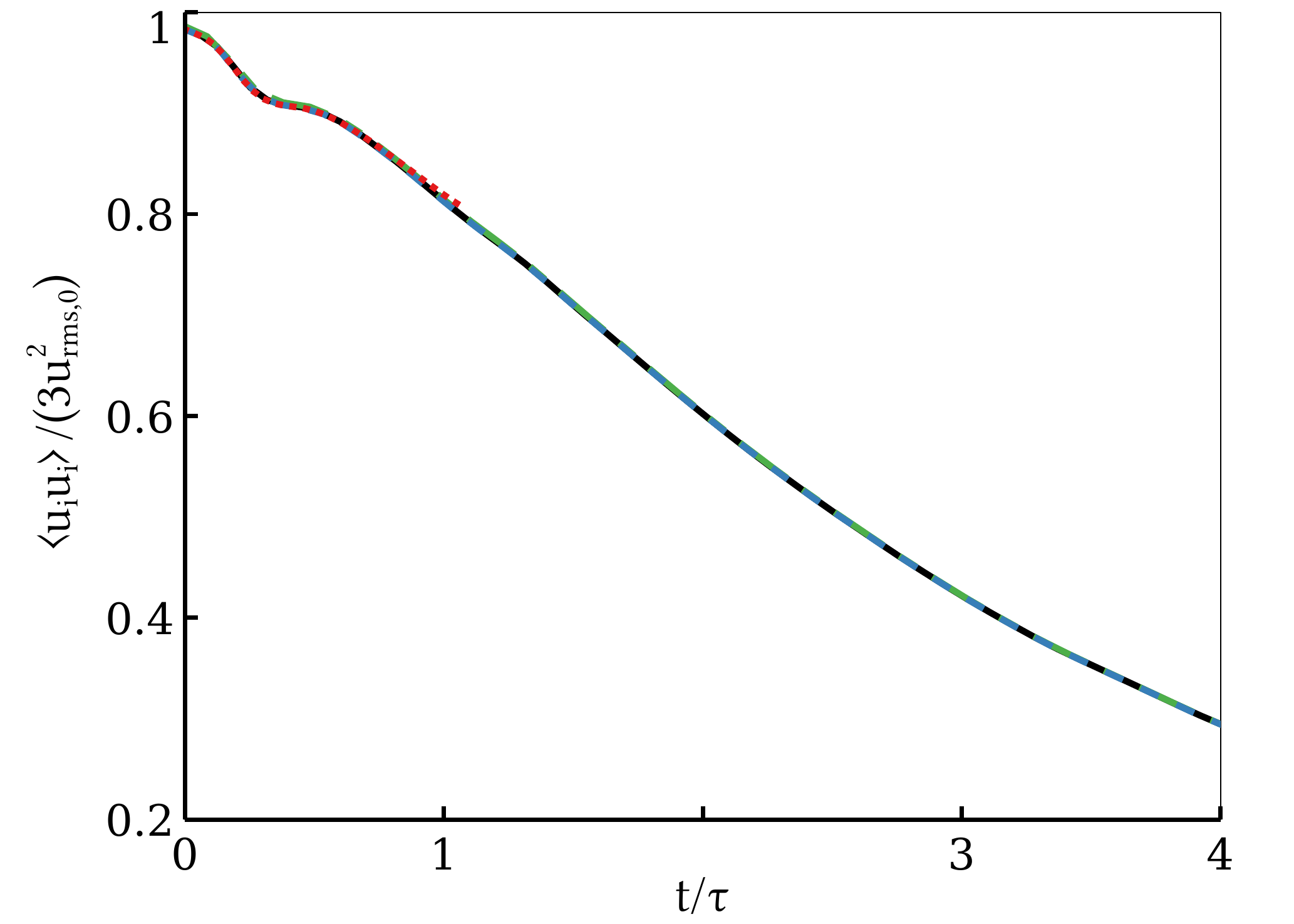}}
\subfloat[Enstrophy]{\label{fig:tseries_SMC_b}\includegraphics[width=0.5\textwidth]{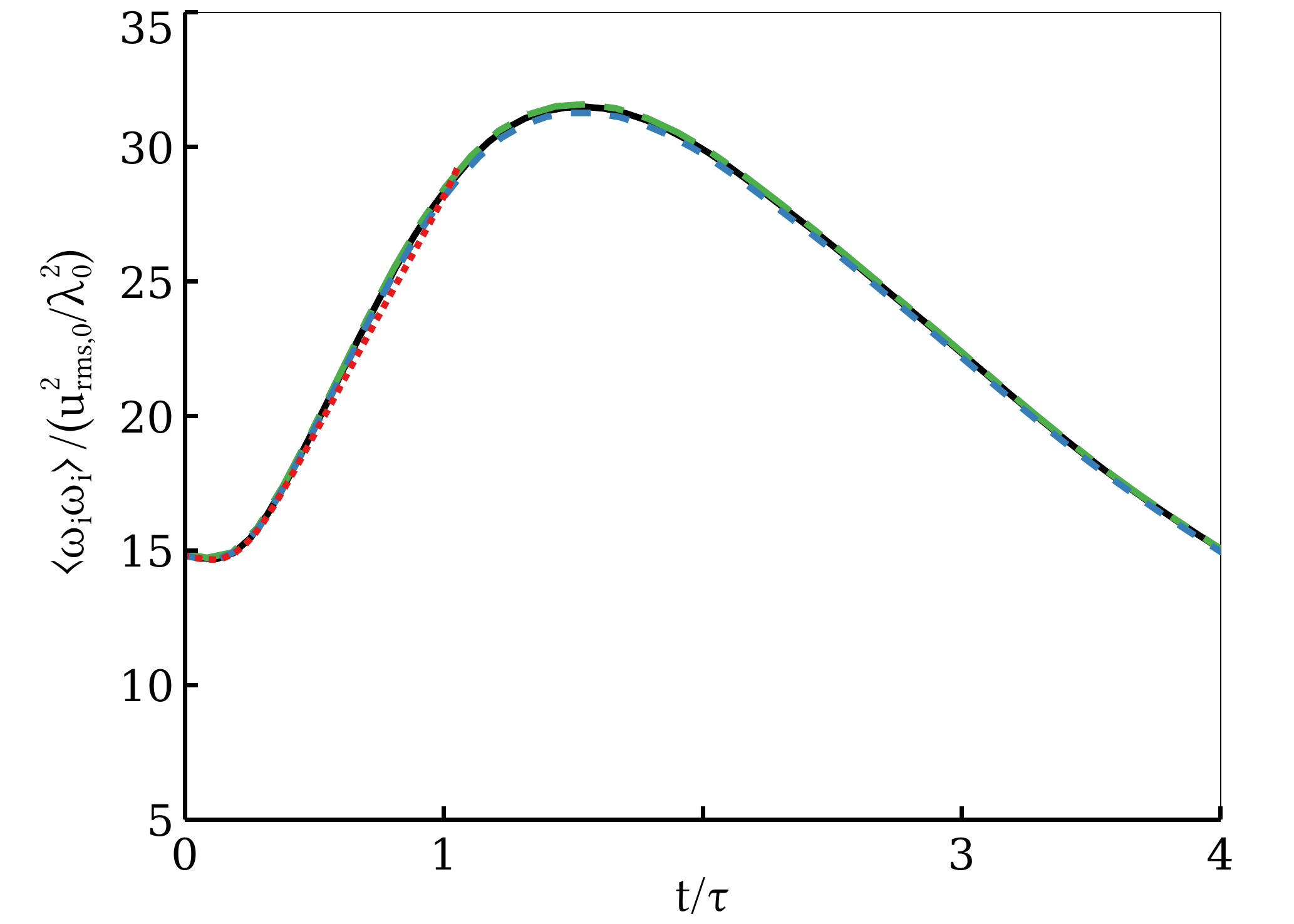}} \\
\subfloat[Temperature]{\label{fig:tseries_SMC_c}\includegraphics[width=0.5\textwidth]{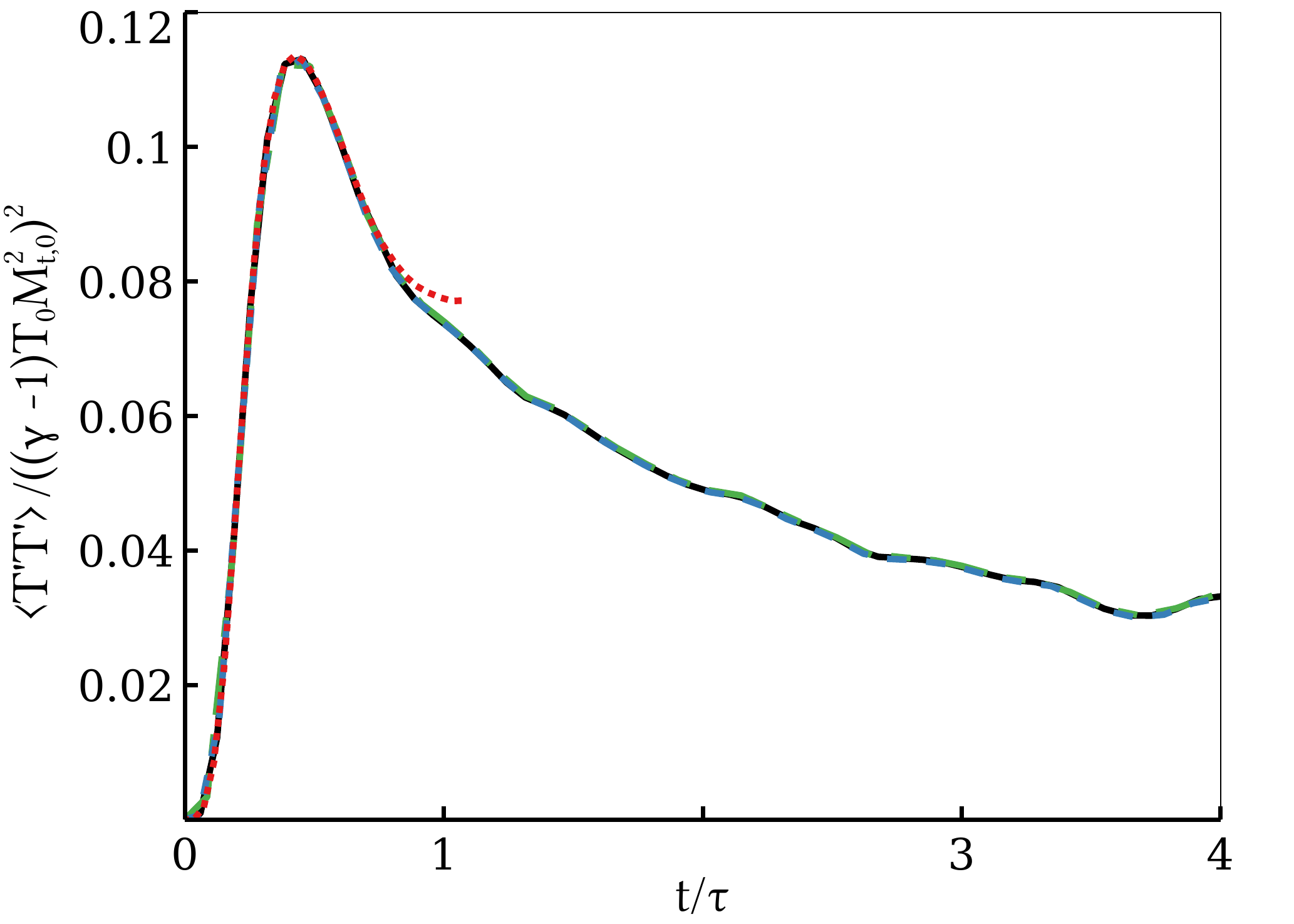}}
\subfloat[Dilatation, $\theta = \partial_j u_j$]{\label{fig:tseries_SMC_d}\includegraphics[width=0.5\textwidth]{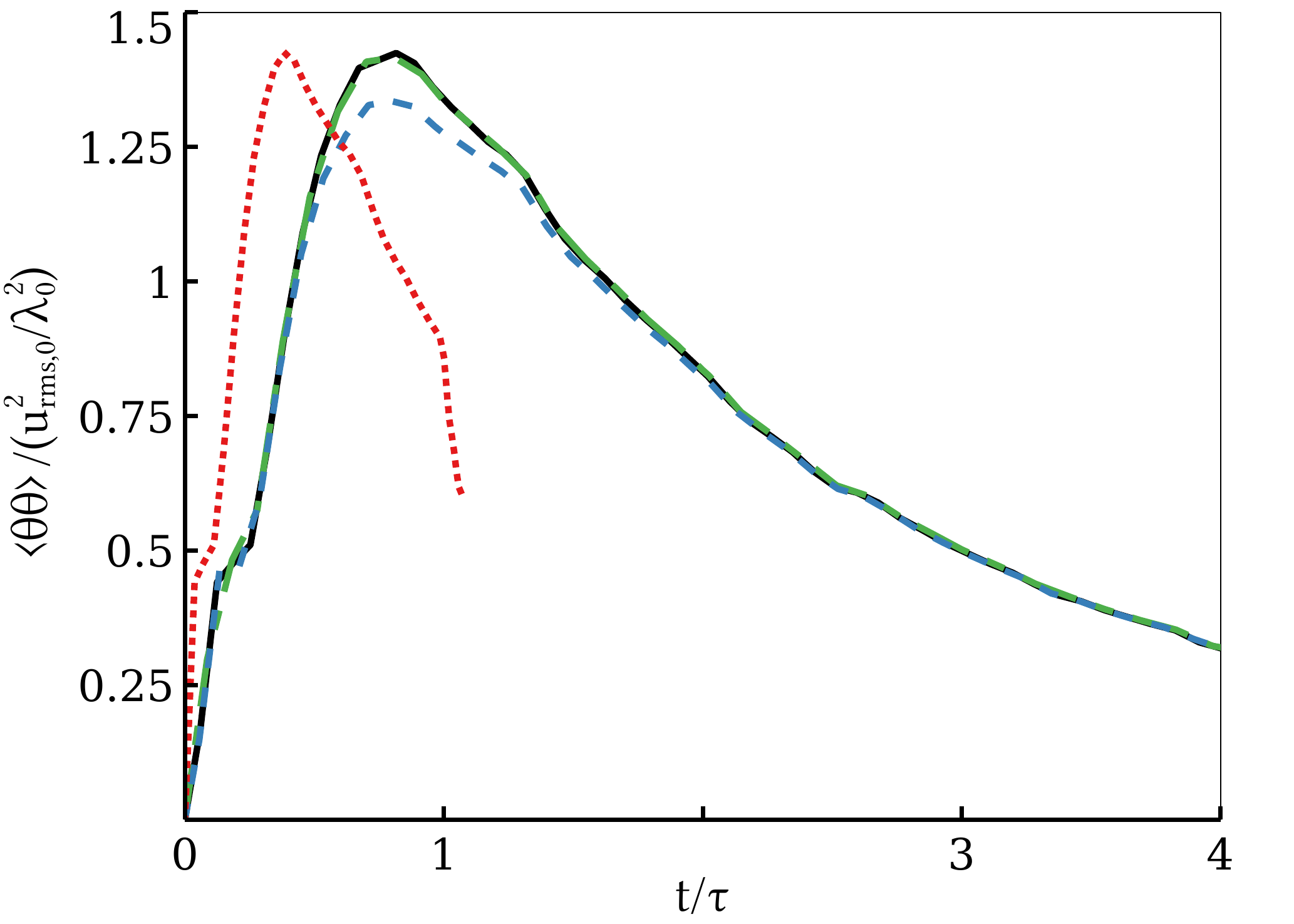}}
\caption{Temporal evolution of selected physical quantities for \textbf{SMC} simulations with different mesh resolution. The red dotted line, the blue dashed line, the green dashed-dotted line and the black line represent the solutions computed on a mesh grid discretized with $N_x=64$, $N_x=128$, $N_x=256$ and $N_x=512$.}
\label{fig:tseries_SMC}
\end{figure}

\begin{figure}[h!]
\centering
\subfloat[Kinetic Energy]{\label{fig:spectra_SMC_a}\includegraphics[width=0.5\textwidth]{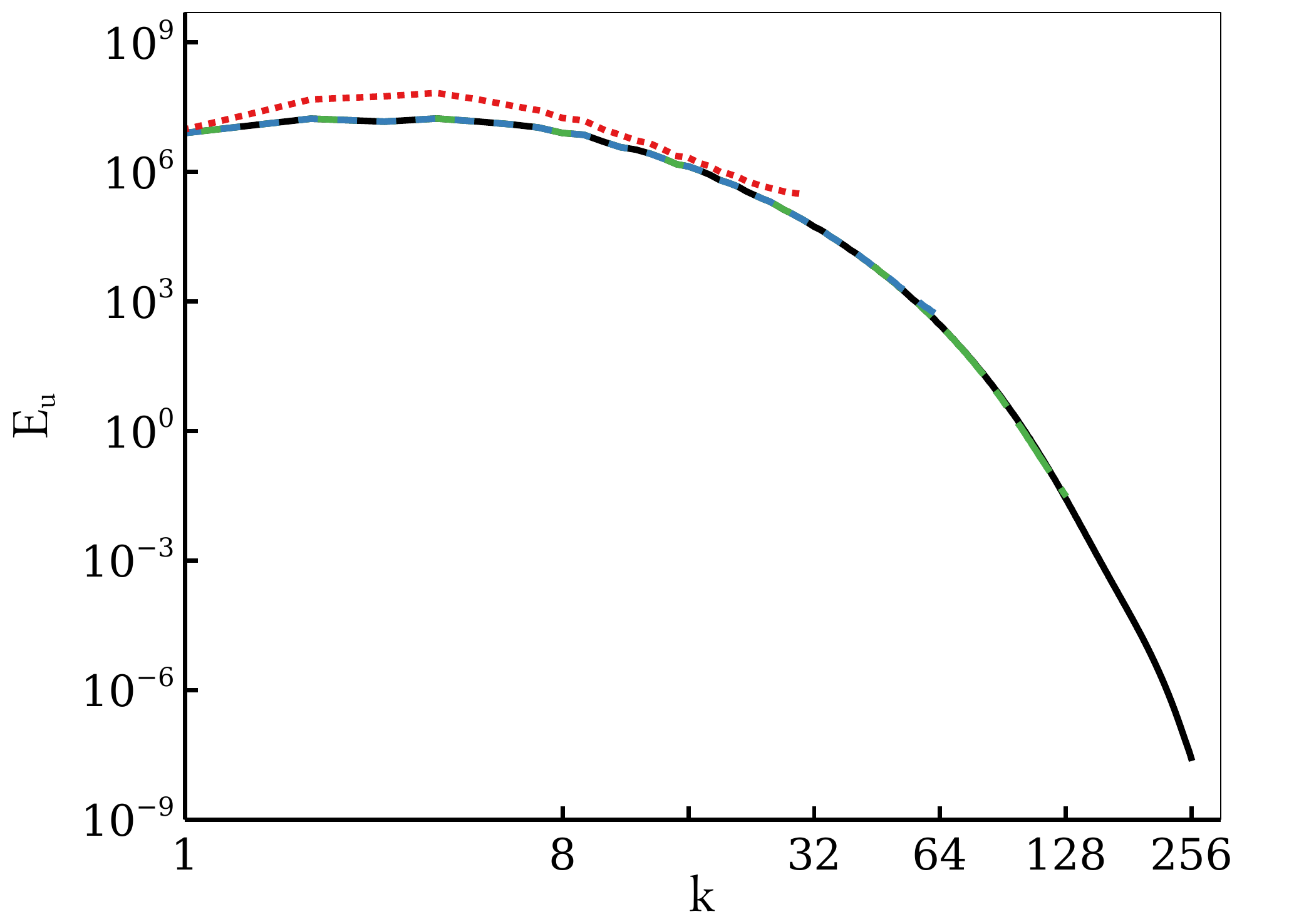}}
\subfloat[Vorticity]{\label{fig:spectra_SMC_b}\includegraphics[width=0.5\textwidth]{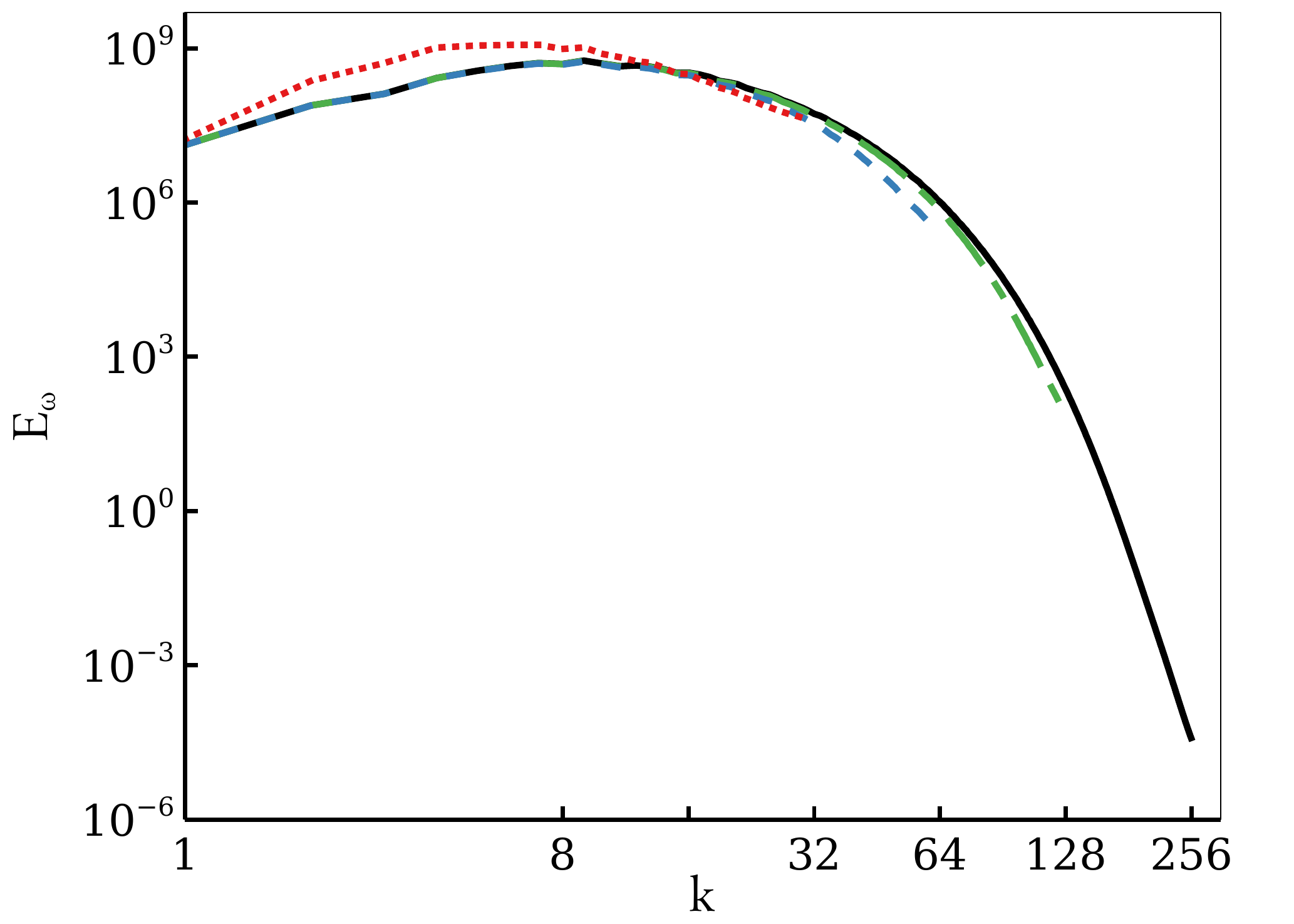}} \\
\subfloat[Dilatation]{\label{fig:spectra_SMC_c}\includegraphics[width=0.5\textwidth]{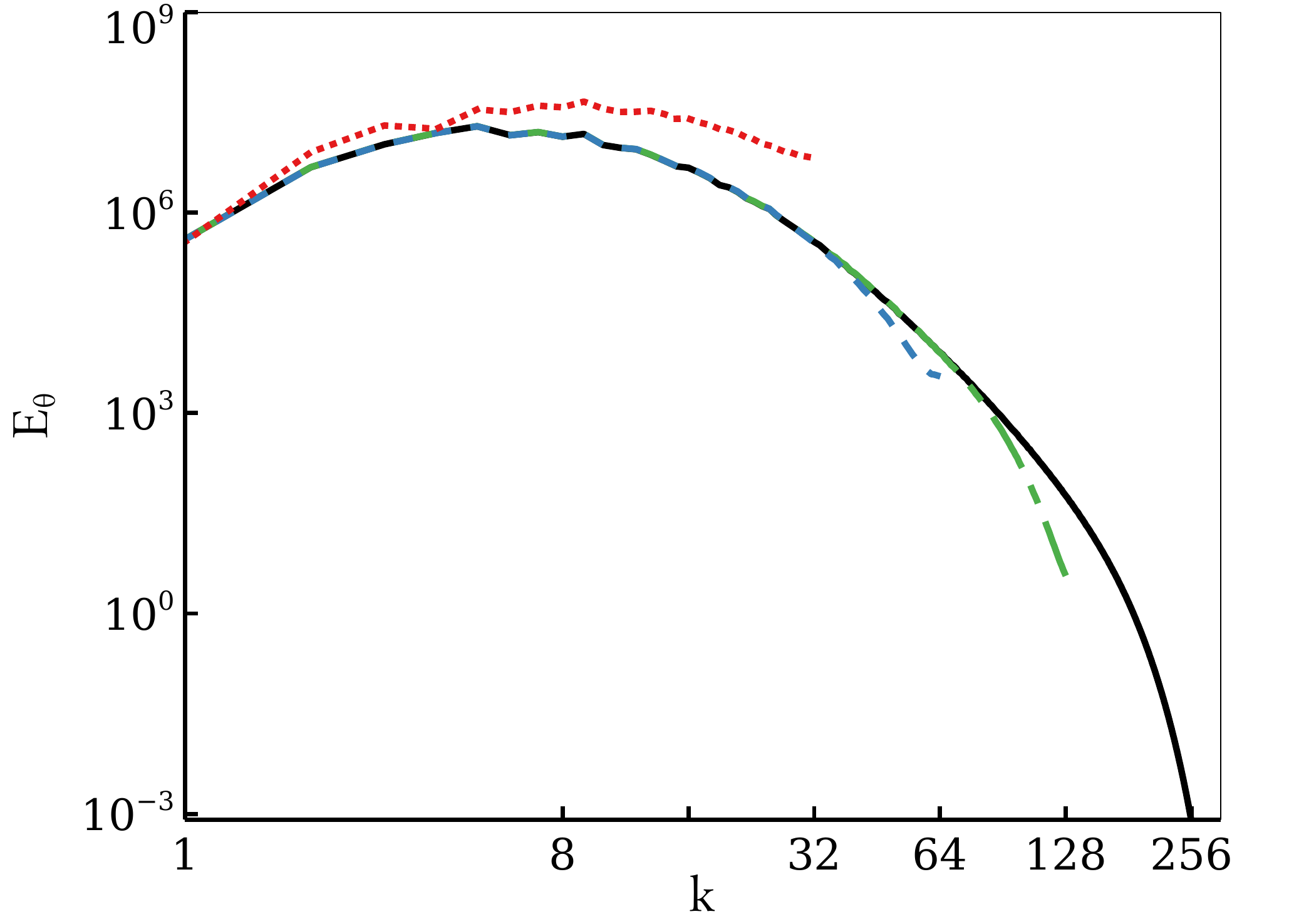}}
\subfloat[Density]{\label{fig:spectra_SMC_d}\includegraphics[width=0.5\textwidth]{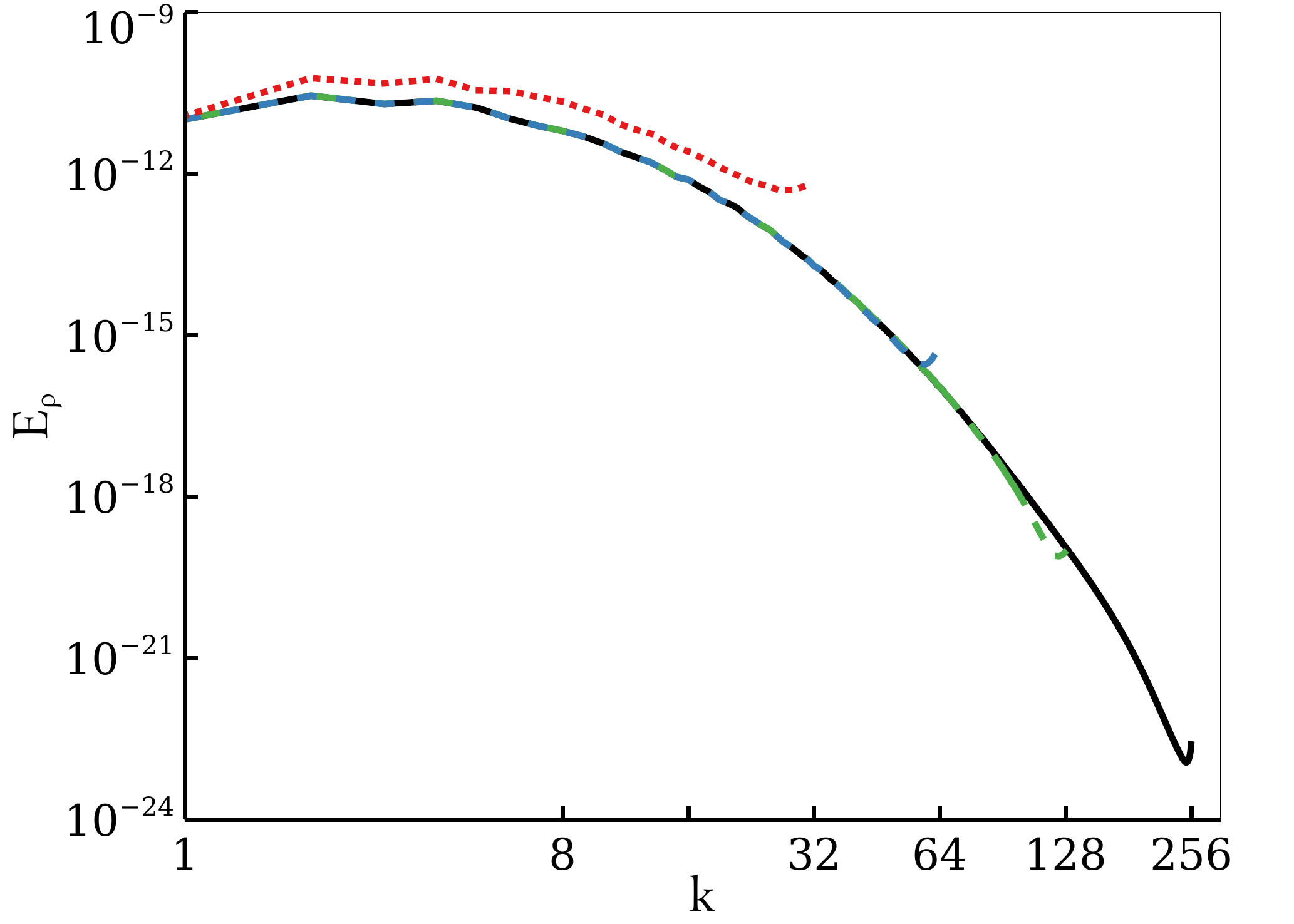}}
\caption{Spectra of selected physical quantities for \textbf{SMC} simulations with different mesh resolution. The red dotted line, the blue dashed line, the green dashed-dotted line and the black line represent the solutions computed on a mesh grid discretized with $N_x=64$, $N_x=128$, $N_x=256$ and $N_x=512$.}
\label{fig:spectra_SMC}
\end{figure}

In order to generate a reference solution, simulations are performed with the very high-order code \textbf{SMC} \cite{Emmett:2014}, which employs eighth-order accurate centered finite-difference schemes for the spatial discretization, and a fourth-order Runge-Kutta algorithm for the time advancement. \Cref{fig:tseries_SMC_a,fig:tseries_SMC_b,fig:tseries_SMC_c,fig:tseries_SMC_d} present the temporal evolution of the kinetic energy, the enstrophy, the variance of temperature and the dilatation from $t=0$ to $t/\tau=4$. \Cref{fig:spectra_SMC_a,fig:spectra_SMC_b,fig:spectra_SMC_c,fig:spectra_SMC_d} present the spectra taken at $t/\tau=4$ for the kinetic energy, the vorticity, the dilatation and the density. In these figures, the red dotted line, the blue dashed line, the green dashed-dotted line and the solid black line represent the solutions computed on a mesh grid discretized with $N_x=64$, $N_x=128$, $N_x=256$ and $N_x=512$, respectively. As can be seen in \cref{fig:tseries_SMC}, the simulation computed with $N_x=64$ is unable to complete and crashes at approximately $t/\tau=1$, because the mesh is too coarse to resolve the diffusion up to the Kolmogorov scale. The solution computed with $N_x=512$ (solid black line) differs slightly from the one computed with $N_x=256$, and is considered converged and will be used at the reference solution.

\section{WENO comparisons}
\label{sec:appendix_weno_comparisons}

As recalled in \cref{subsec:WENO_Z}, a significant amount of schemes for the reconstruction of data at interfaces are based on the WENO paradigm. Indeed, the classical WENO-JS scheme is not optimal and is often considered too dissipative in smooth regions. Many variants have been developed to overcome such isue. Among all of these variants, we have chosen to focus on the most popular ones to assess their robustness and performance on the test cases investigated in the present paper. A complete description of the variations introduced by these schemes is beyond the scope of the paper. So far, the WENO variants tested in this study are: WENO-JS \cite{WENO_JS}, WENO-M \cite{Henrick:2005}, WENO-Z \cite{Borges:2008}, WENO-MDCD \cite{Martin:2006}  and TENO \cite{Fu:2016}.

\subsection{Shu-Osher test case}
\label{subsec:apprendix_shu_osher}

\begin{figure}[tbhp]
\centering
\subfloat[Full domain]{\label{fig:comparisons_Shu_Osher_WENO_PPM_256_a}\includegraphics[width=0.5\textwidth]{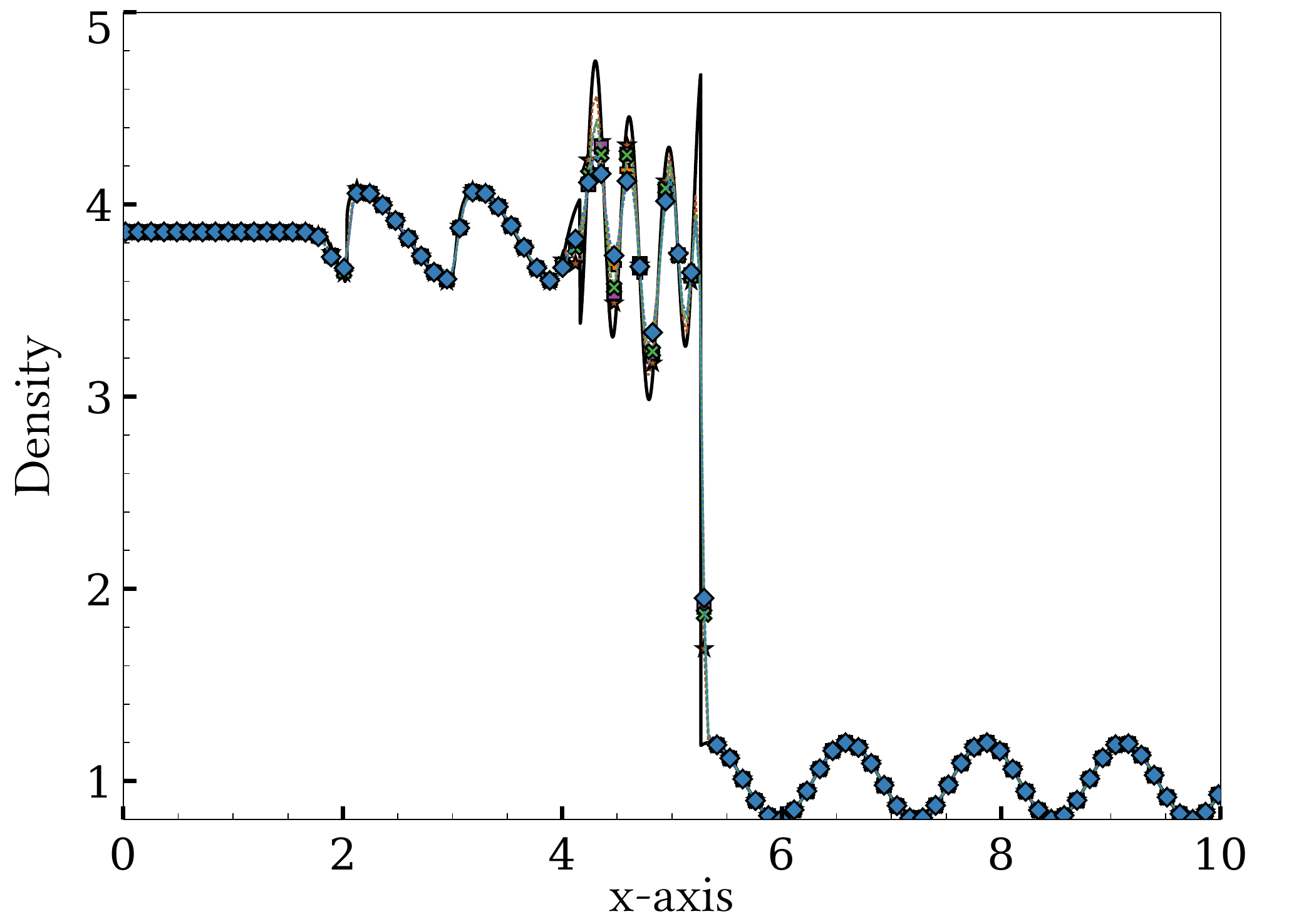}}
\subfloat[Zoom]{\label{fig:comparisons_Shu_Osher_WENO_PPM_256_b}\includegraphics[width=0.5\textwidth]{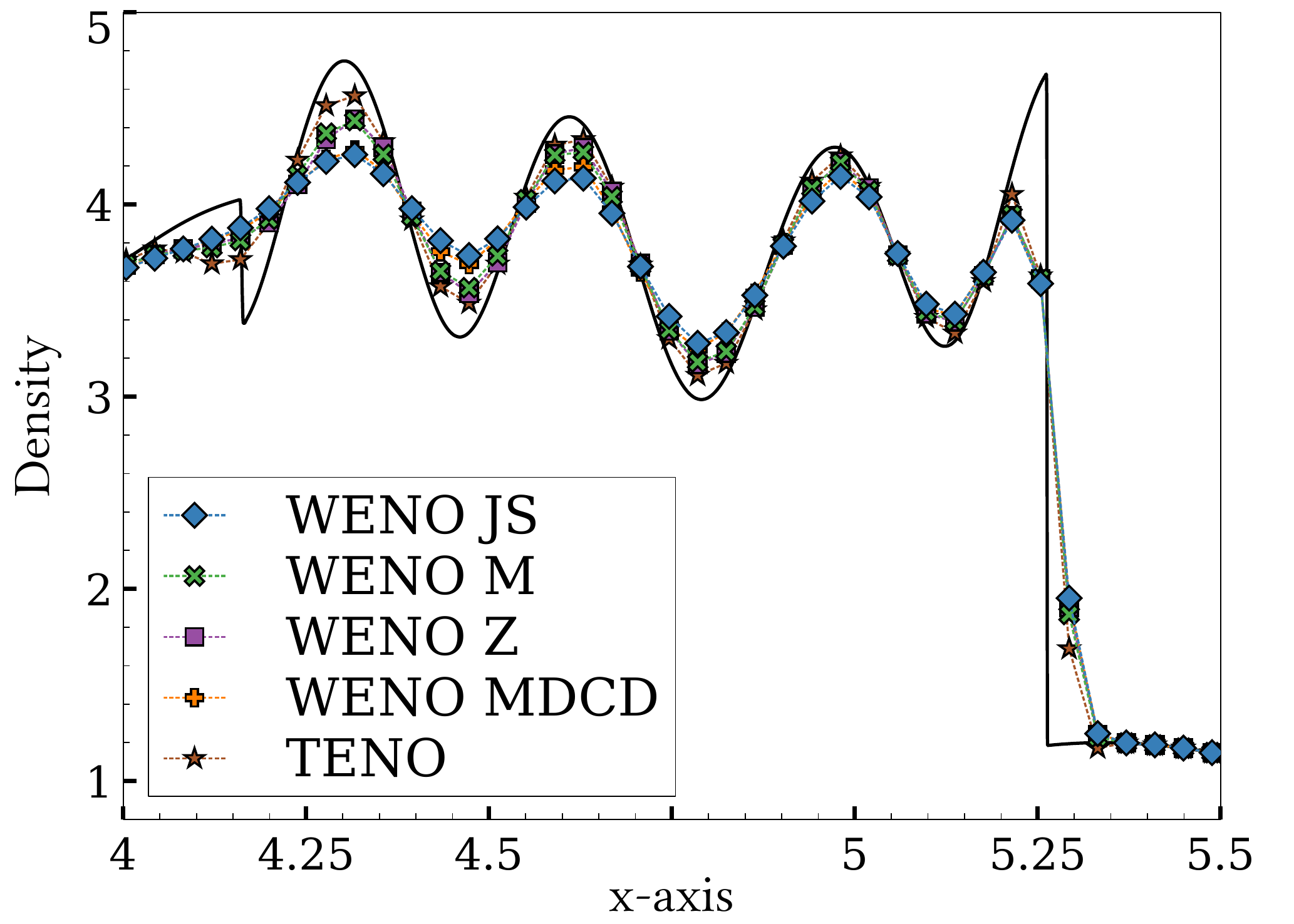}}
\caption{Shu-Osher test case: profile of density with PPM and WENO methods for $N_x=256$.}
\label{fig:comparisons_Shu_Osher_WENO_PPM_256}
\end{figure}

\begin{figure}[tbhp]
\centering
\subfloat[Full domain]{\label{fig:comparisons_Shu_Osher_WENO_PPM_512_a}\includegraphics[width=0.5\textwidth]{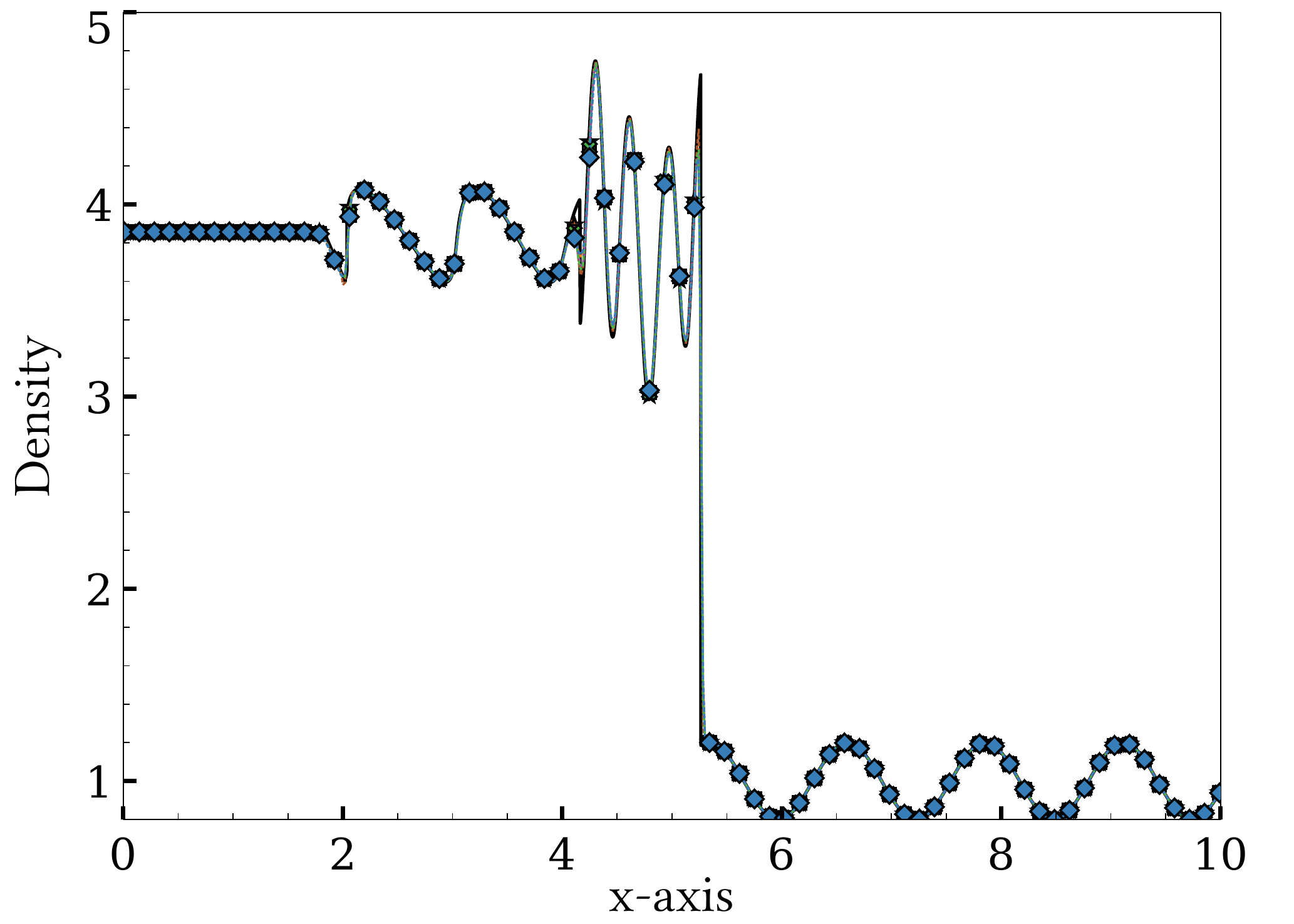}}
\subfloat[Zoom]{\label{fig:comparisons_Shu_Osher_WENO_PPM_512_b}\includegraphics[width=0.5\textwidth]{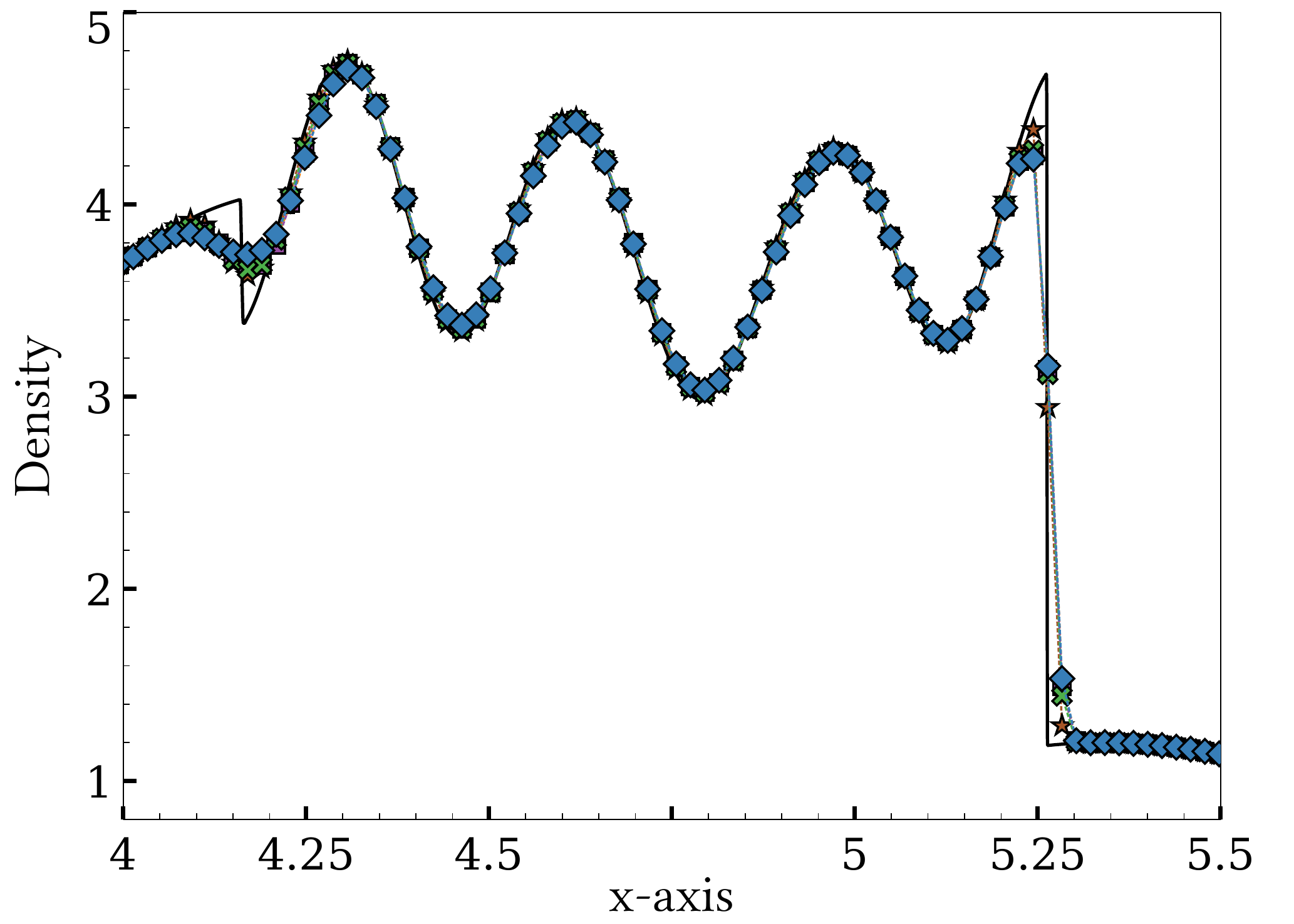}}
\caption{Shu-Osher test case: profile of density with PPM and WENO methods for $N_x=512$.}
\label{fig:comparisons_Shu_Osher_WENO_PPM_512}
\end{figure}

\begin{figure}[tbhp]
\centering
\subfloat[Full domain]{\label{fig:comparisons_Shu_Osher_WENO_PPM_1024_a}\includegraphics[width=0.5\textwidth]{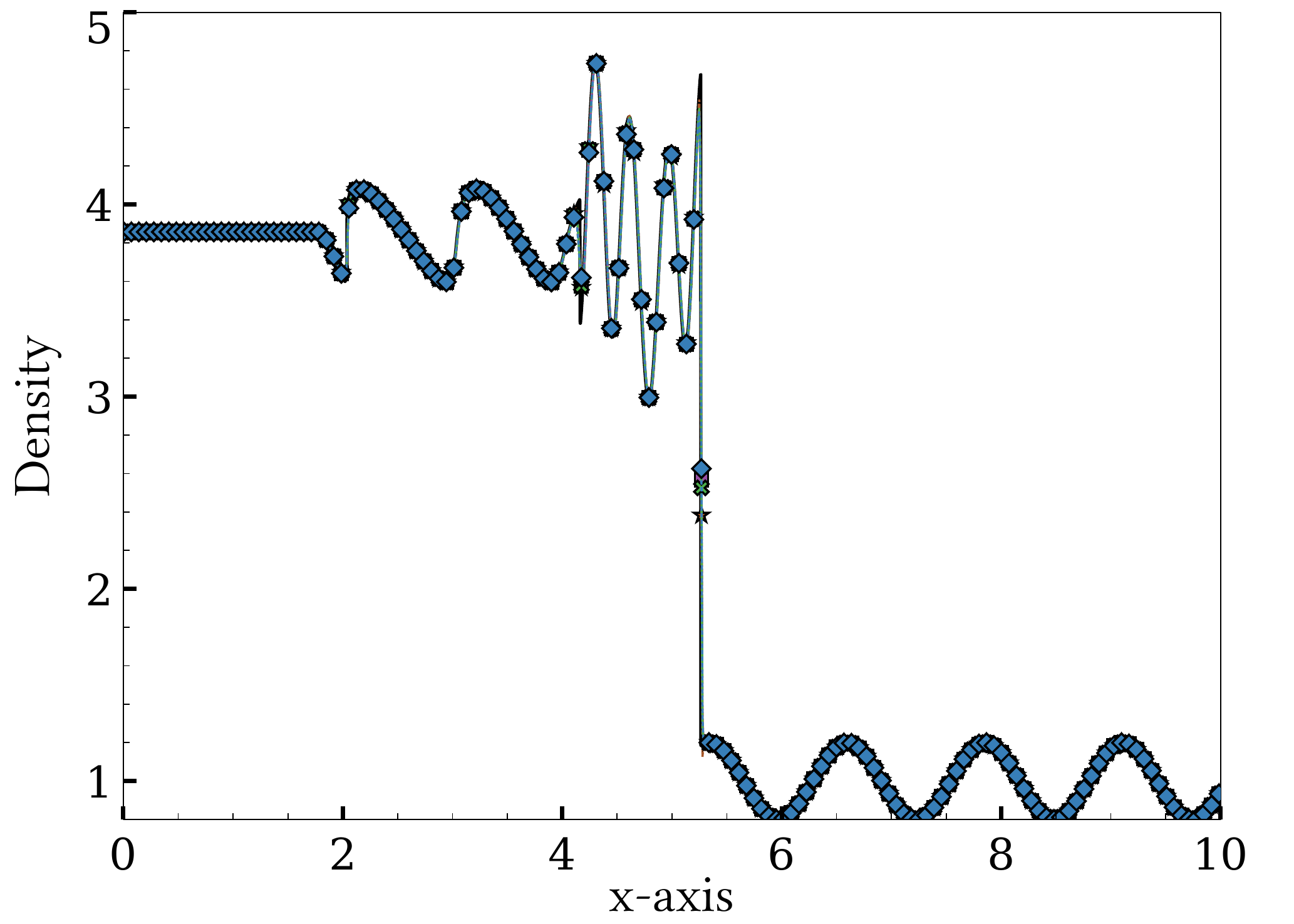}}
\subfloat[Zoom]{\label{fig:comparisons_Shu_Osher_WENO_PPM_1024_b}\includegraphics[width=0.5\textwidth]{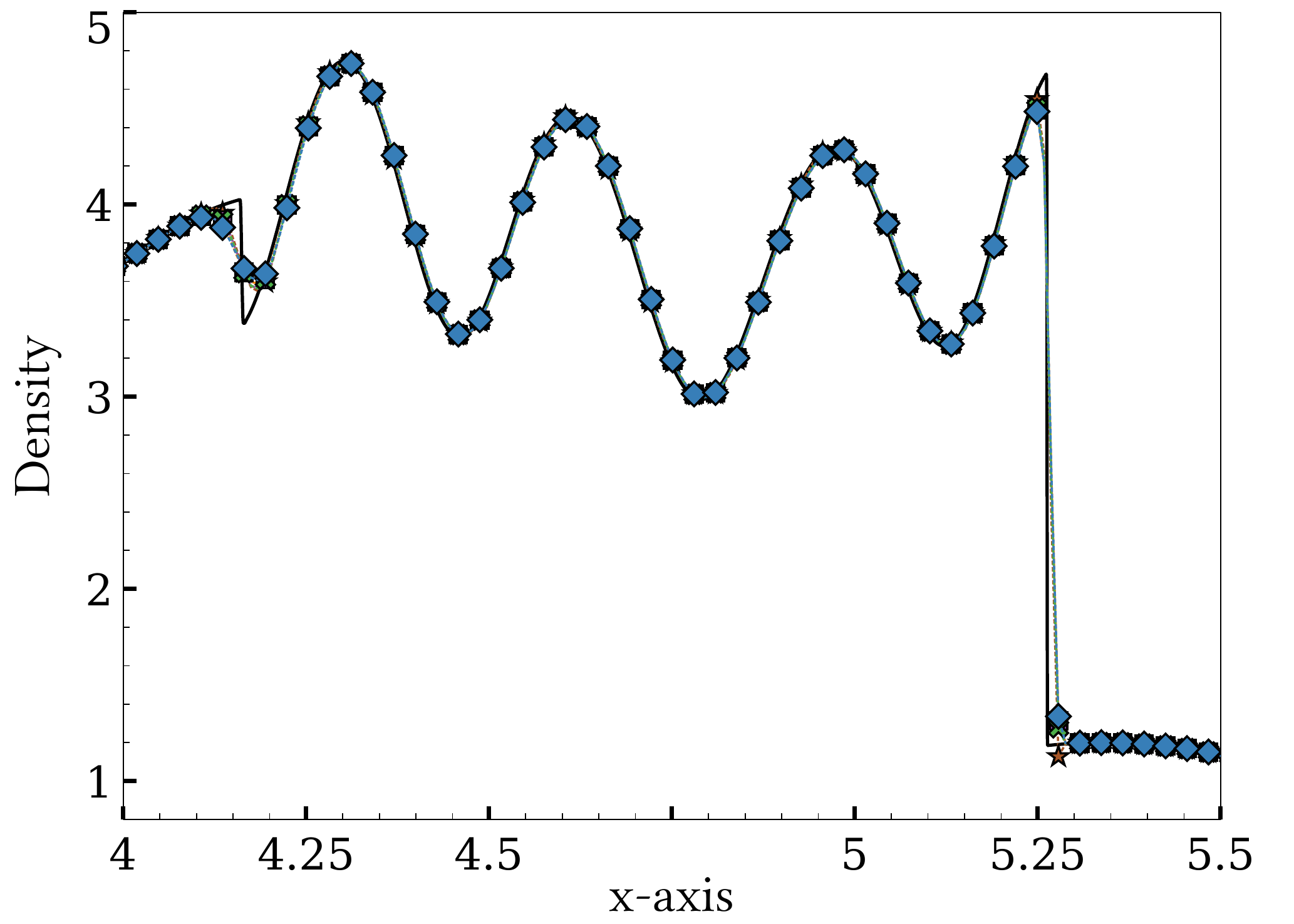}}
\caption{Shu-Osher test case: profile of density with PPM and WENO methods for $N_x=1024$.}
\label{fig:comparisons_Shu_Osher_WENO_PPM_1024}
\end{figure}

\begin{figure}[tbhp]
\centering
\includegraphics[width=0.9\textwidth]{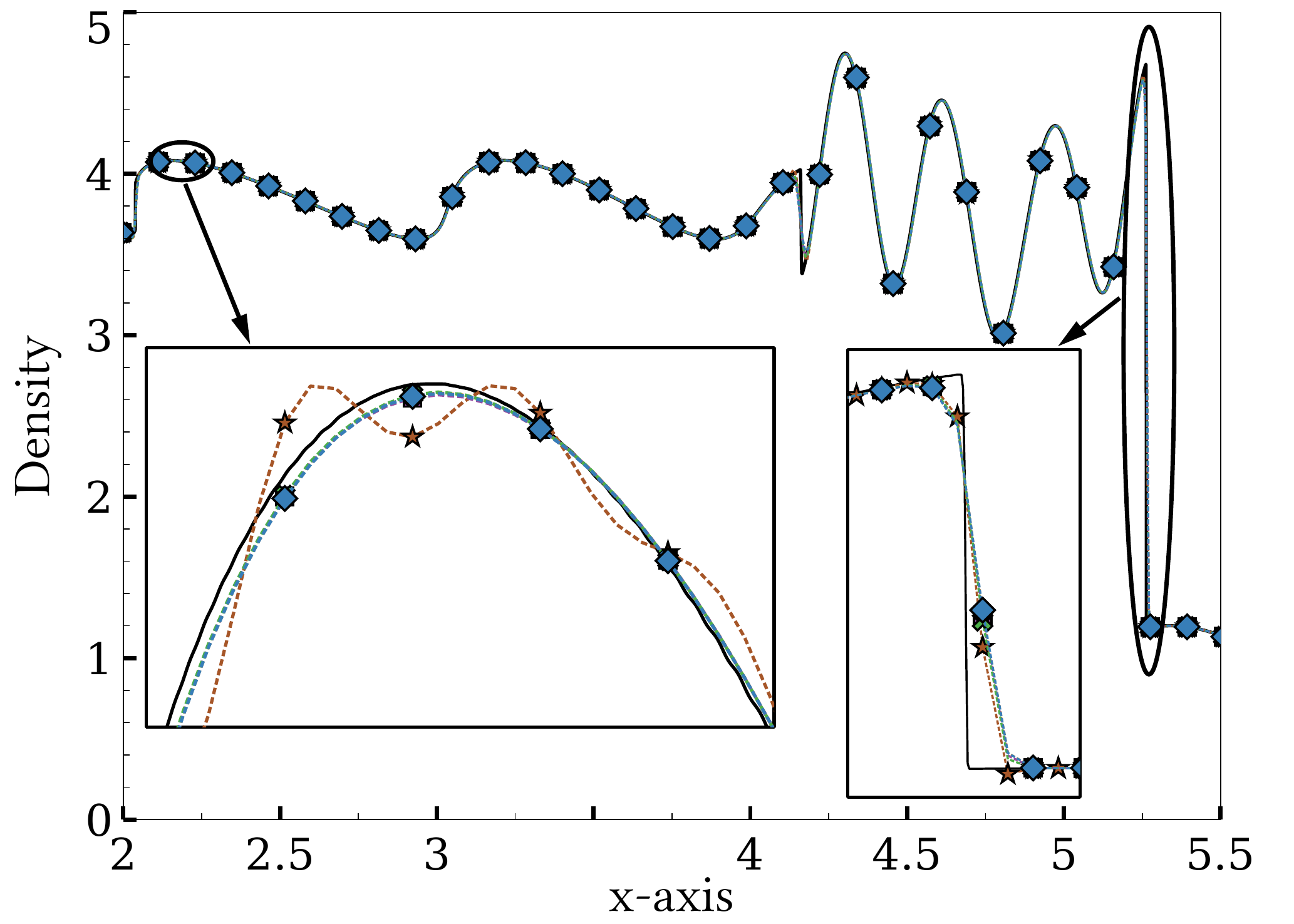}
\caption{Shu-Osher test case: profile of density with PPM and WENO methods for $N_x=2048$.}
\label{fig:comparisons_Shu_Osher_WENO_PPM_2048}
\end{figure}

The density at $t=1.2$ computed with $N_x = 256, 512, 1024$, and $2048$ is shown in \cref{fig:comparisons_Shu_Osher_WENO_PPM_256,fig:comparisons_Shu_Osher_WENO_PPM_512,fig:comparisons_Shu_Osher_WENO_PPM_1024,fig:comparisons_Shu_Osher_WENO_PPM_2048}, respectively. In these figures, the blue diamond, green cross, purple square, orange plus and maroon star symbols represent the WENO-JS, WENO-M, WENO-Z, WENO-MDCD and TENO methods, respectively (see legend in \cref{fig:comparisons_Shu_Osher_WENO_PPM_256_b}). Note also that the panels (a) and (b) in \cref{fig:comparisons_Shu_Osher_WENO_PPM_256}, \cref{fig:comparisons_Shu_Osher_WENO_PPM_512} and  \cref{fig:comparisons_Shu_Osher_WENO_PPM_1024} present the full domain and a zoom in the domain, respectively, while \cref{fig:comparisons_Shu_Osher_WENO_PPM_2048} is only a zoom in the domain.

For the coarse mesh, a close look at \cref{fig:comparisons_Shu_Osher_WENO_PPM_256} reveals that all the WENO variants reproduce the correct phase of the oscillation. The WENO-M, WENO-Z and TENO methods give virtually similar results in terms of estimation of the amplitudes of the waves, while the WENO-JS and WENO-MDCD methods are equivalently the least accurate of the WENO variants. As shown in \cref{fig:comparisons_Shu_Osher_WENO_PPM_512},  an increase of the mesh resolution by a factor $2$ leads all the WENO variants to virtually collapse to the same curve.

As shown in \cref{fig:comparisons_Shu_Osher_WENO_PPM_1024}, with an another increase of the mesh resolution by a factor $2$, all the numerical methods investigated in the present study are virtually equivalent and very close to the reference solution computed on a very fine mesh. However, as can be seen at $x \approx 5.25$ after another increase of the mesh resolution by a factor $2$ in \cref{fig:comparisons_Shu_Osher_WENO_PPM_1024_b}, the TENO method provides an incorrect representation of the discontinuity. As shown in \cref{fig:comparisons_Shu_Osher_WENO_PPM_2048}, this trend becomes worse when the mesh is refined again by a factor $2$. As can be seen in the detailed zoom, the solution computed with the TENO scheme shows large oscillations in the smooth regions. All other WENO variants are, however, robust. 

This present study shows that when the mesh is small enough, it allows high-frequency waves to be resolved but small oscillations around discontinuities can appear and propagate, because the mesh is no longer coarse enough to filter them out. The most surprising result is the fact that TENO variant, which appears to be a good choice on a coarse mesh, becomes the worst on a fine mesh. This can be attributed to the fact that the method fails to properly avoid the application of the central linear scheme in the region of large gradients. Furthermore, consistently with the convergence rate analysis performed at \cref{subsec:Shu_Osher}, it should be noted that all the WENO variants provide the same rate of convergence of the error, which is approximately $\mathcal{O}(0.9)$ here for this test case.

\subsection{Decay of compressible isotropic turbulence}

\begin{figure}[tbhp]
\centering
\subfloat[Kinetic Energy]{\label{fig:comparisons_HIT_tseries_WENO_PPM_full_a}\includegraphics[width=0.5\textwidth]{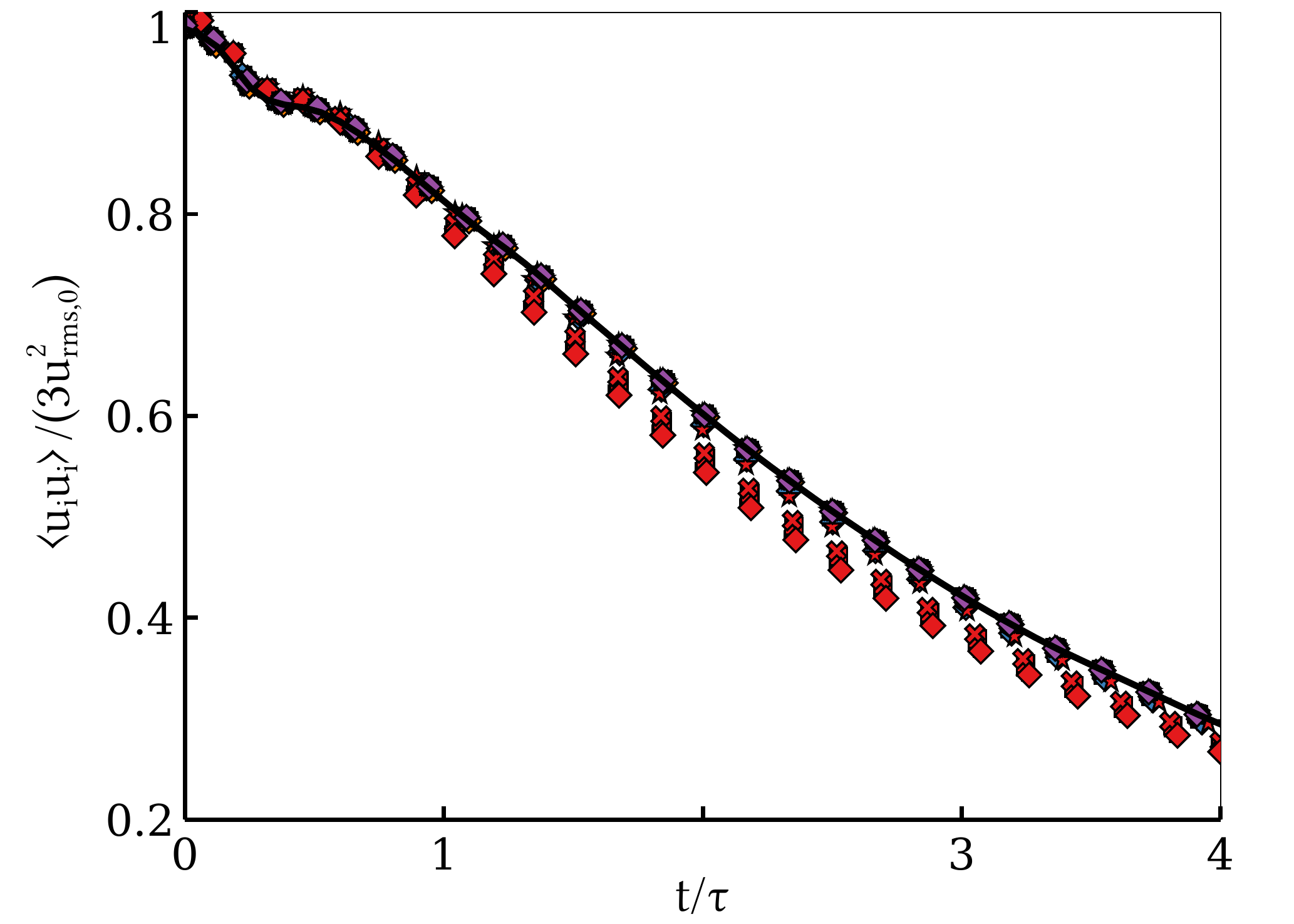}}
\subfloat[Enstrophy]{\label{fig:comparisons_HIT_tseries_WENO_PPM_full_b}\includegraphics[width=0.5\textwidth]{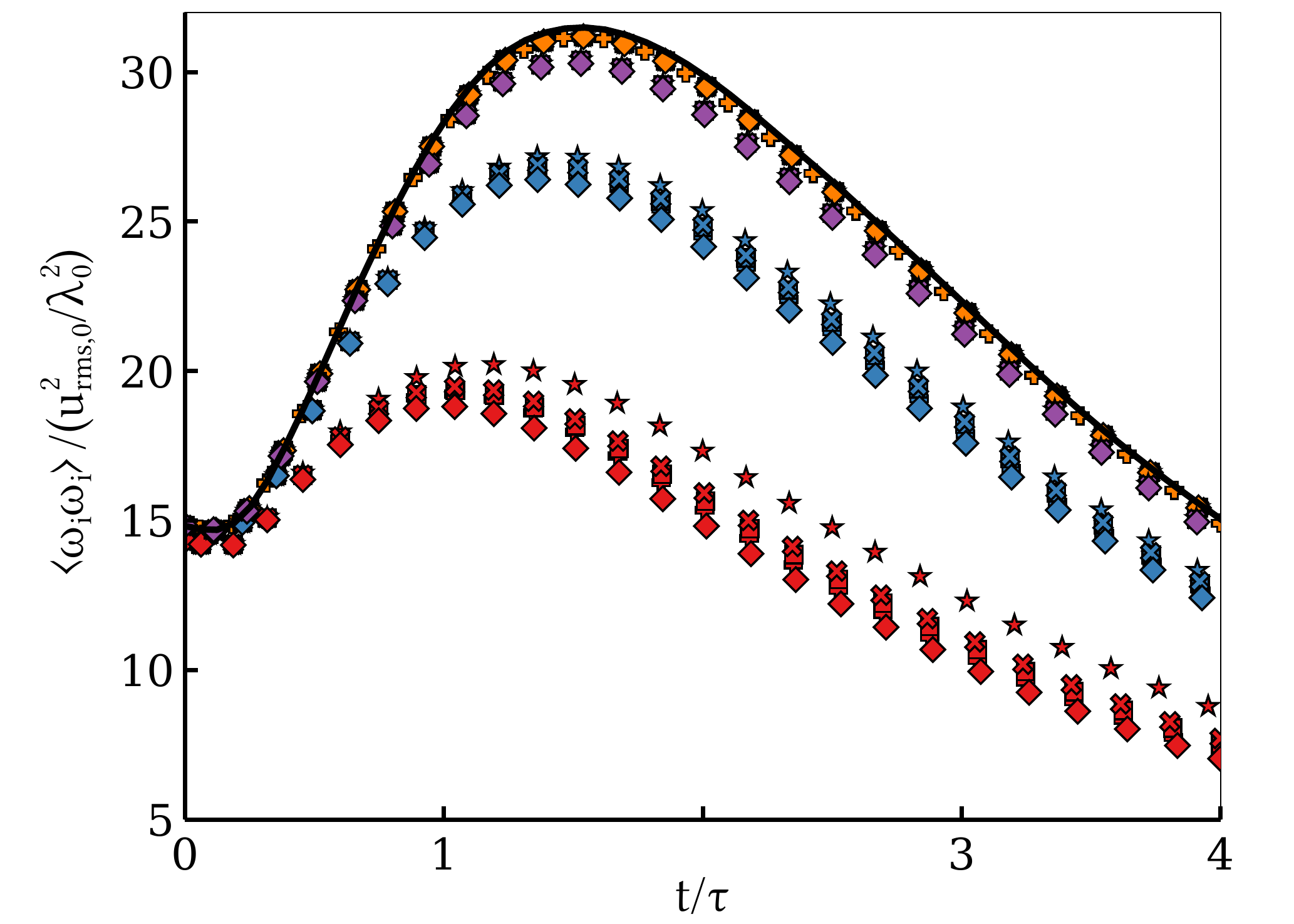}} \\
\subfloat[Temperature]{\label{fig:comparisons_HIT_tseries_WENO_PPM_full_c}\includegraphics[width=0.5\textwidth]{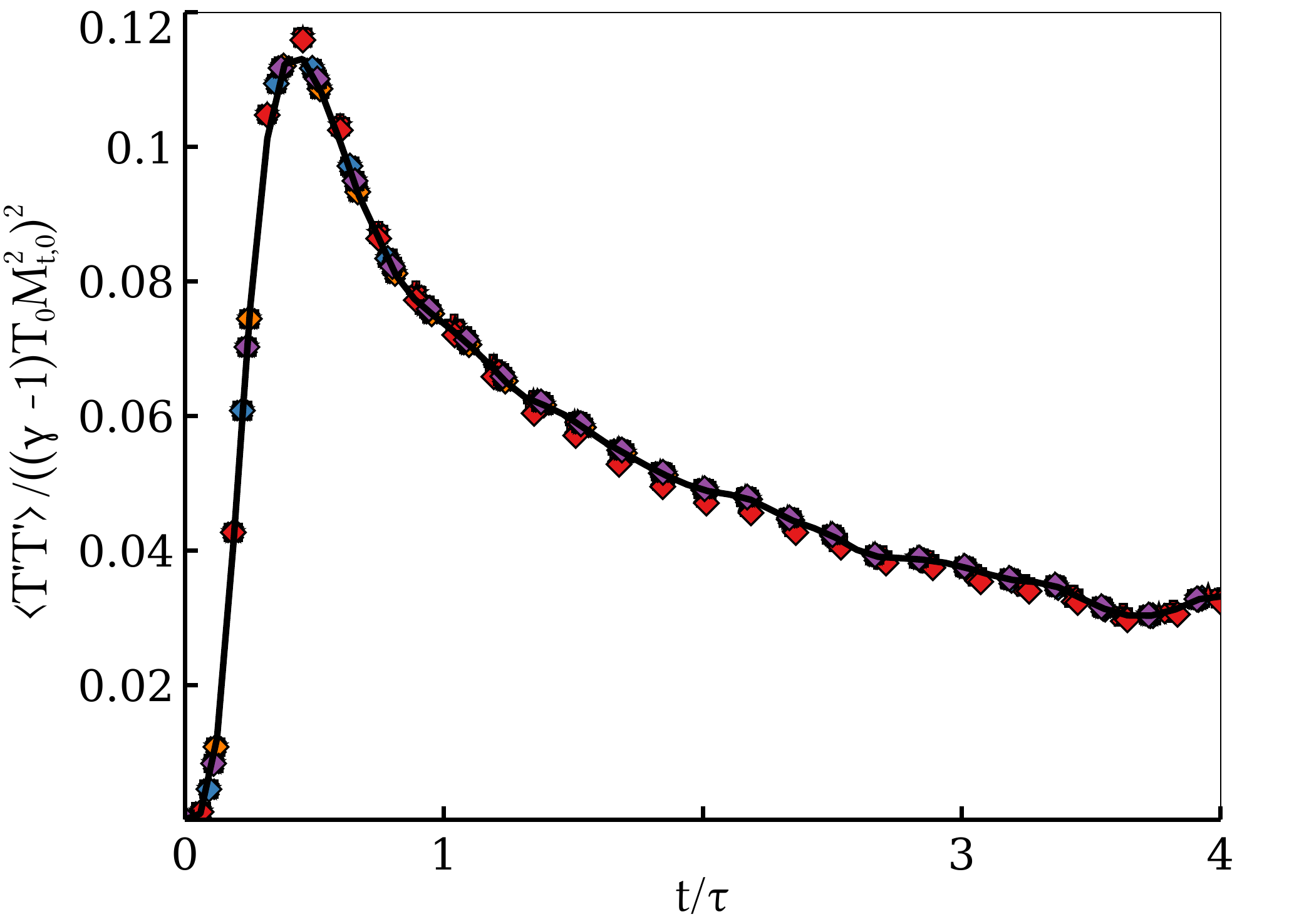}}
\subfloat[Dilatation, $\theta = \partial_j u_j$]{\label{fig:comparisons_HIT_tseries_WENO_PPM_full_d}\includegraphics[width=0.5\textwidth]{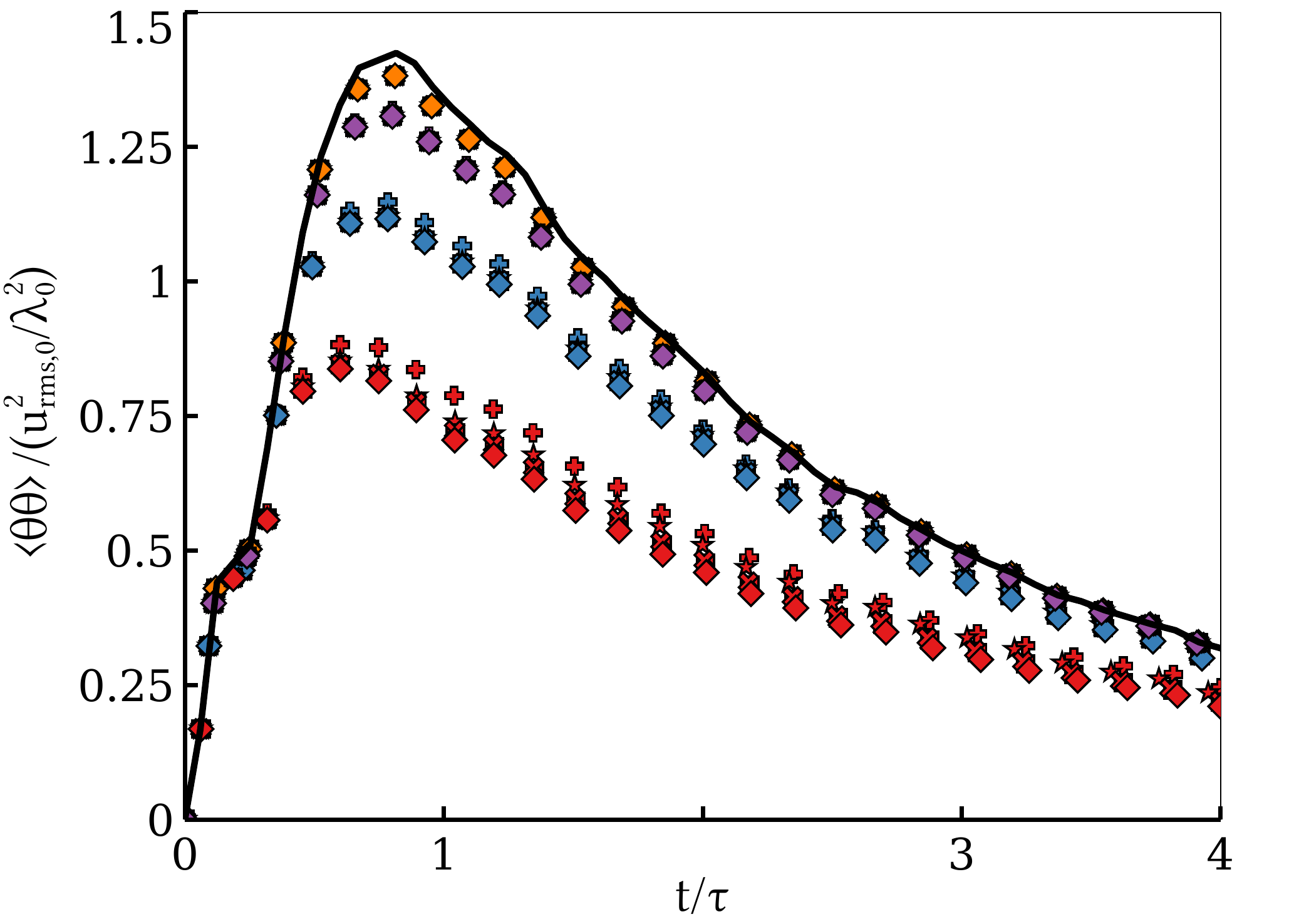}}
\caption{Time series of selected physical quantities for simulations performed with different WENO reconstruction schemes and with different mesh resolution.  The diamond, cross, square, plus and star symbols represent the WENO-JS, WENO-M, WENO-Z, WENO-MDCD and TENO methods, respectively. The red, blue, purple and orange colors represent simulations performed with  $N_x=64$, $N_x=128$, $N_x=256$ and $N_x=512$, respectively.}
\label{fig:comparisons_HIT_tseries_WENO_PPM_full}
\end{figure}

\begin{figure}[tbhp]
\centering
\subfloat[Kinetic Energy]{\label{fig:comparisons_HIT_spectra_WENO_PPM_full_a}\includegraphics[width=0.5\textwidth]{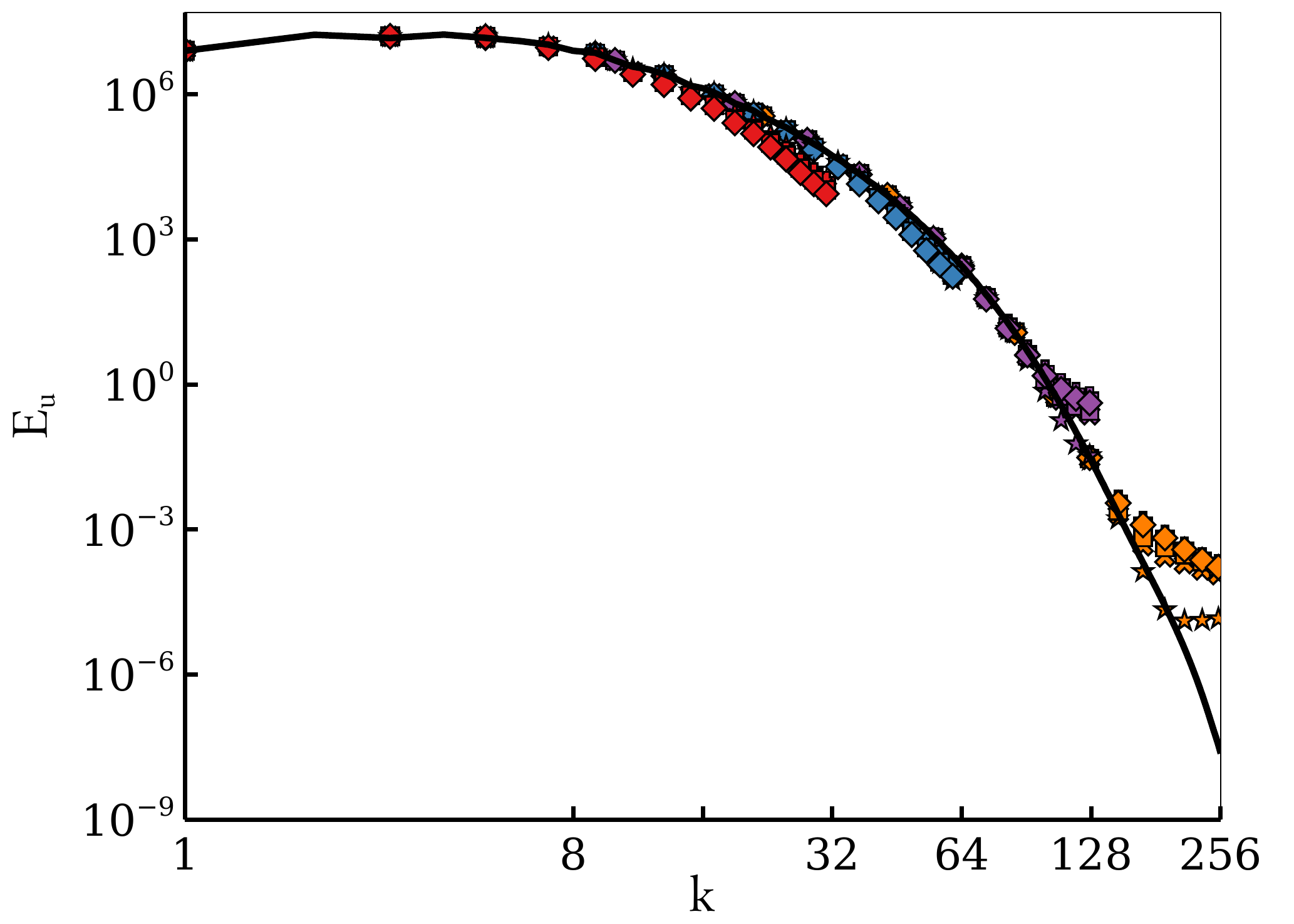}}
\subfloat[Vorticity]{\label{fig:comparisons_HIT_spectra_WENO_PPM_full_b}\includegraphics[width=0.5\textwidth]{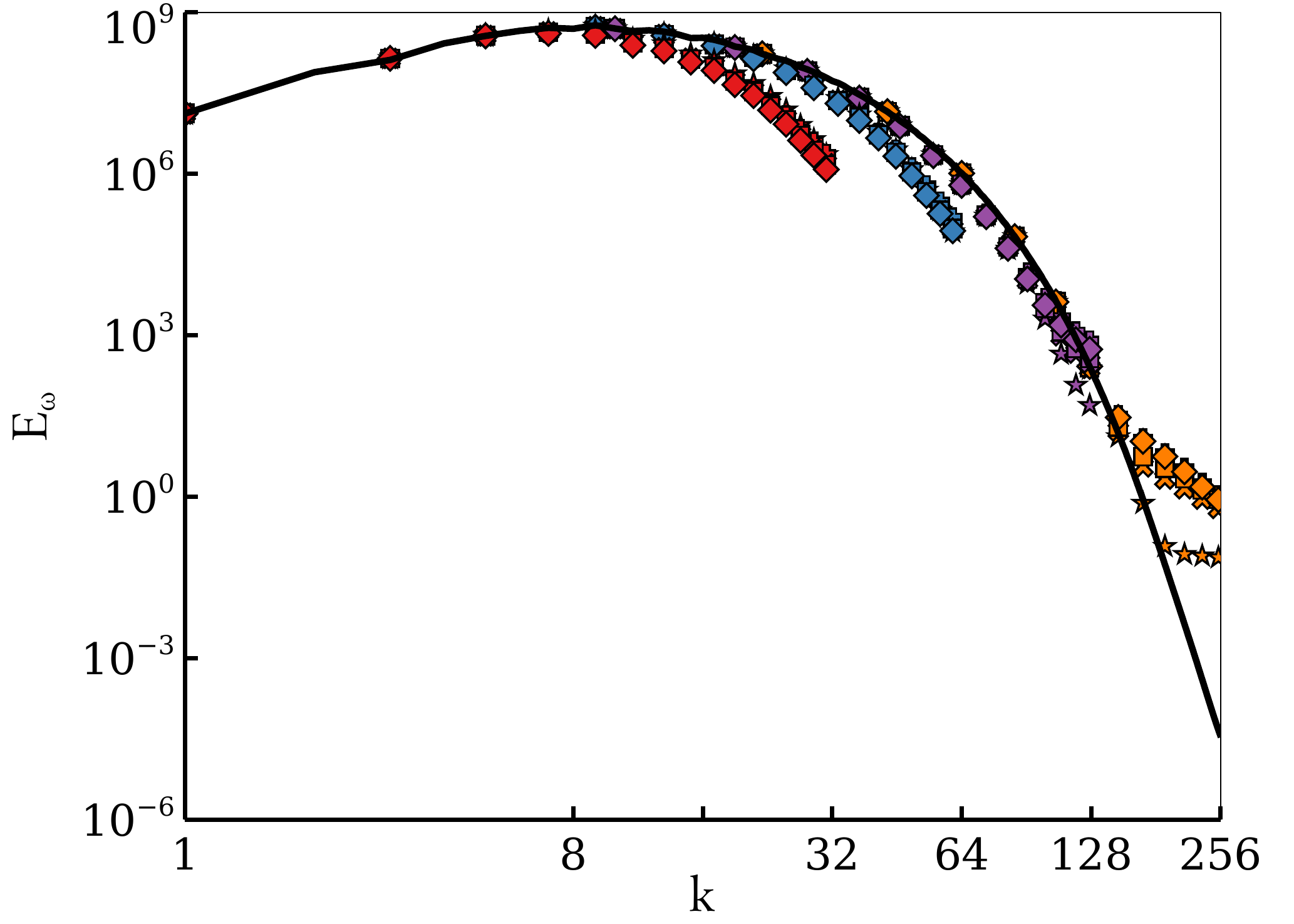}} \\
\subfloat[Dilatation]{\label{fig:comparisons_HIT_spectra_WENO_PPM_full_c}\includegraphics[width=0.5\textwidth]{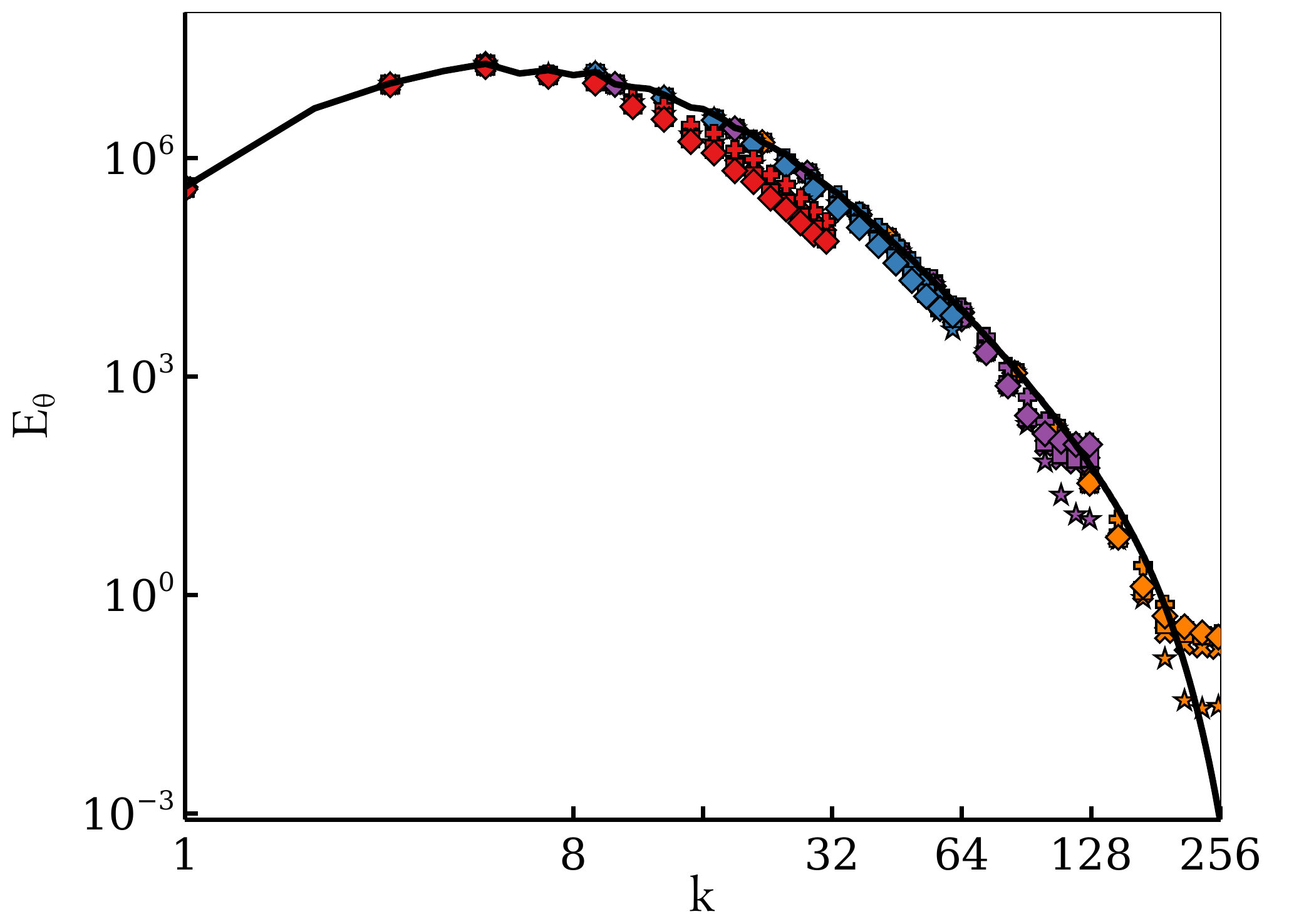}}
\subfloat[Density]{\label{fig:comparisons_HIT_spectra_WENO_PPM_full_d}\includegraphics[width=0.5\textwidth]{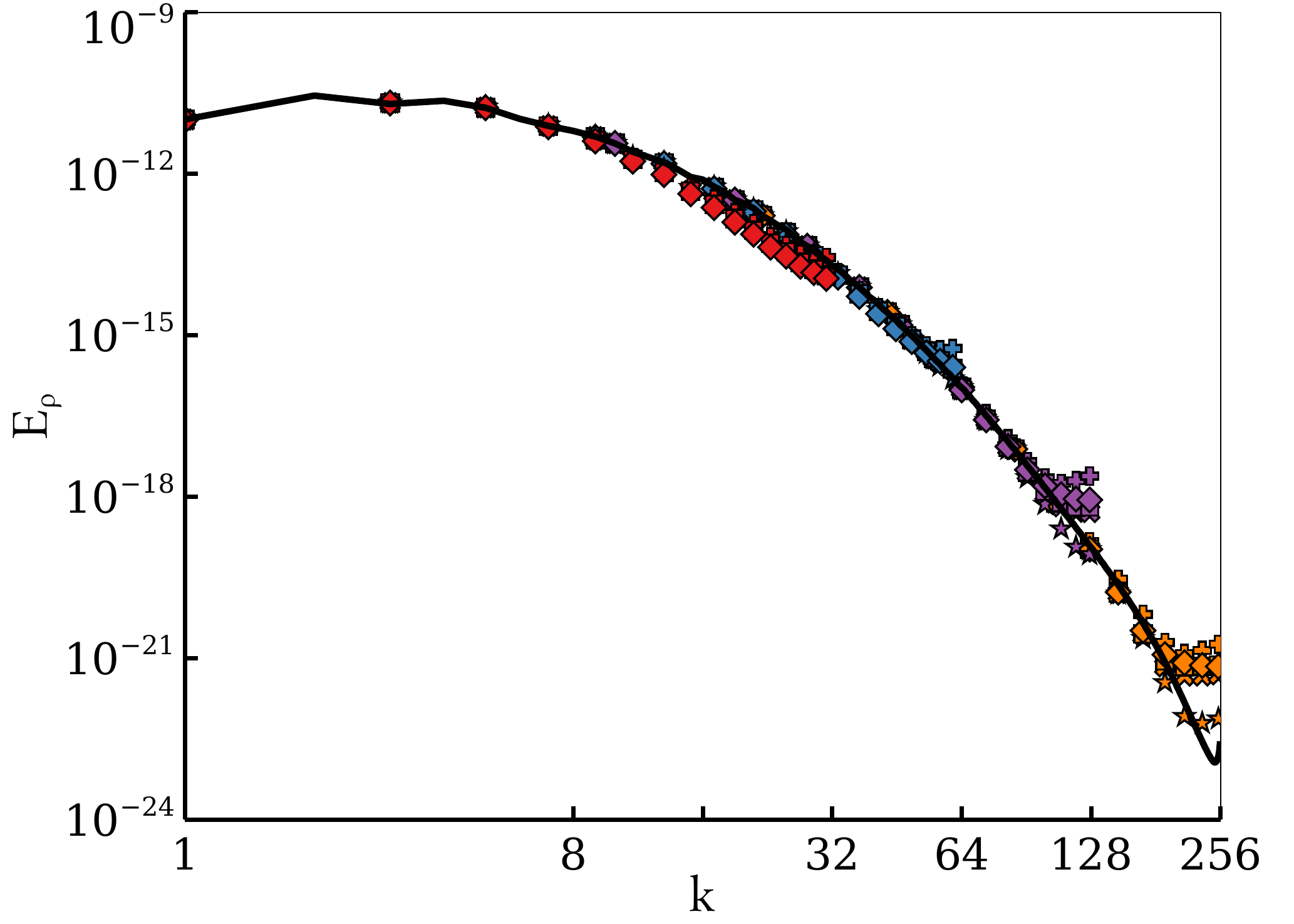}}
\caption{Spectra of selected physical quantities for simulations performed with different WENO reconstruction schemes and with different mesh resolution. The diamond, cross, square, plus and star symbols represent the WENO-JS, WENO-M, WENO-Z, WENO-MDCD and TENO methods, respectively. The red, blue, purple and orange colors represent simulations performed with  $N_x=64$, $N_x=128$, $N_x=256$ and $N_x=512$, respectively.}
\label{fig:comparisons_HIT_spectra_WENO_PPM_full}
\end{figure}

\begin{figure}[tbhp]
\centering
\subfloat[$N_x=64^3$]{\label{fig:comparisons_HIT_spectra_WENO_PPM_a}\includegraphics[width=0.5\textwidth]{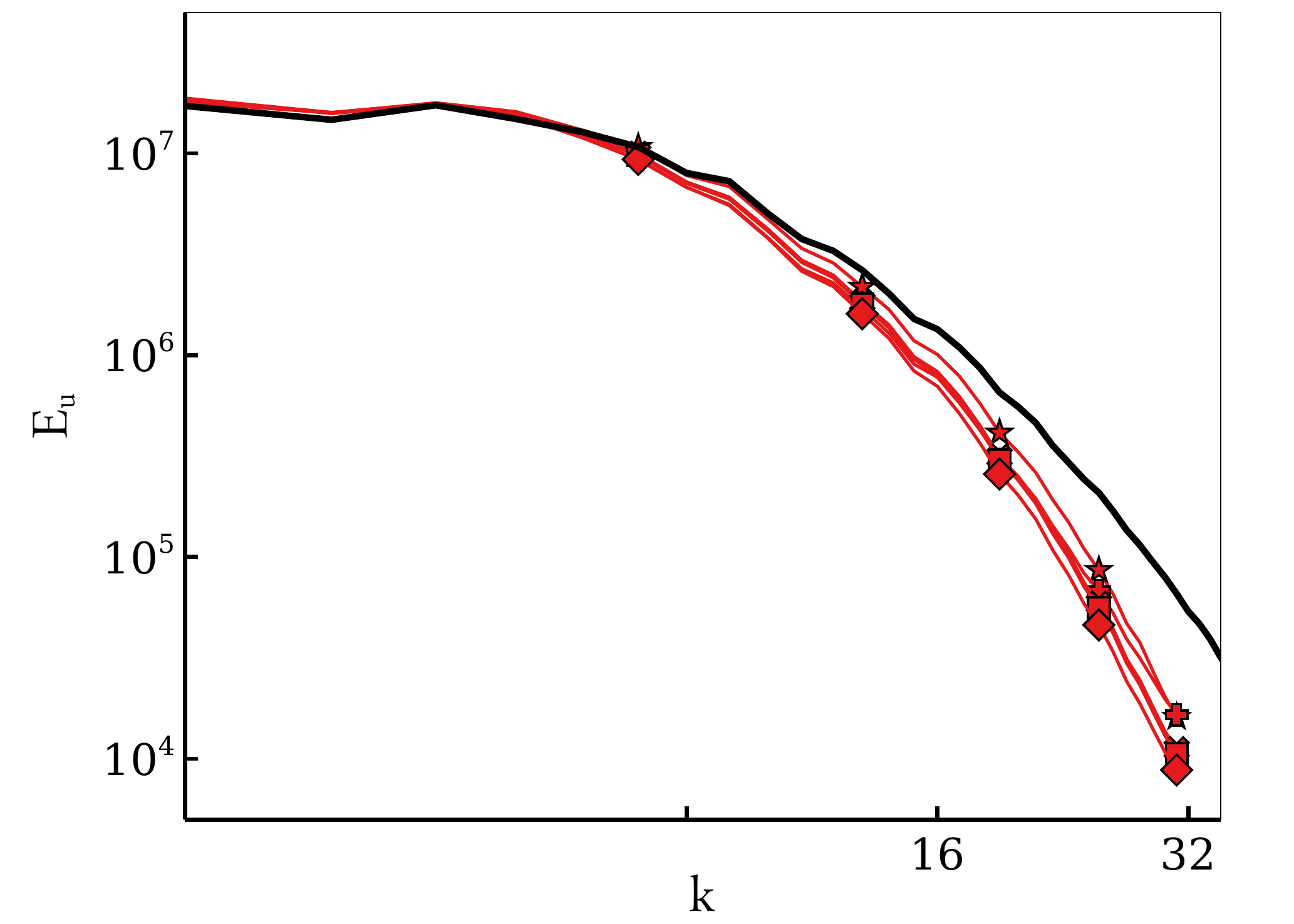}}
\subfloat[$N_x=128^3$]{\label{fig:comparisons_HIT_spectra_WENO_PPM_b}\includegraphics[width=0.5\textwidth]{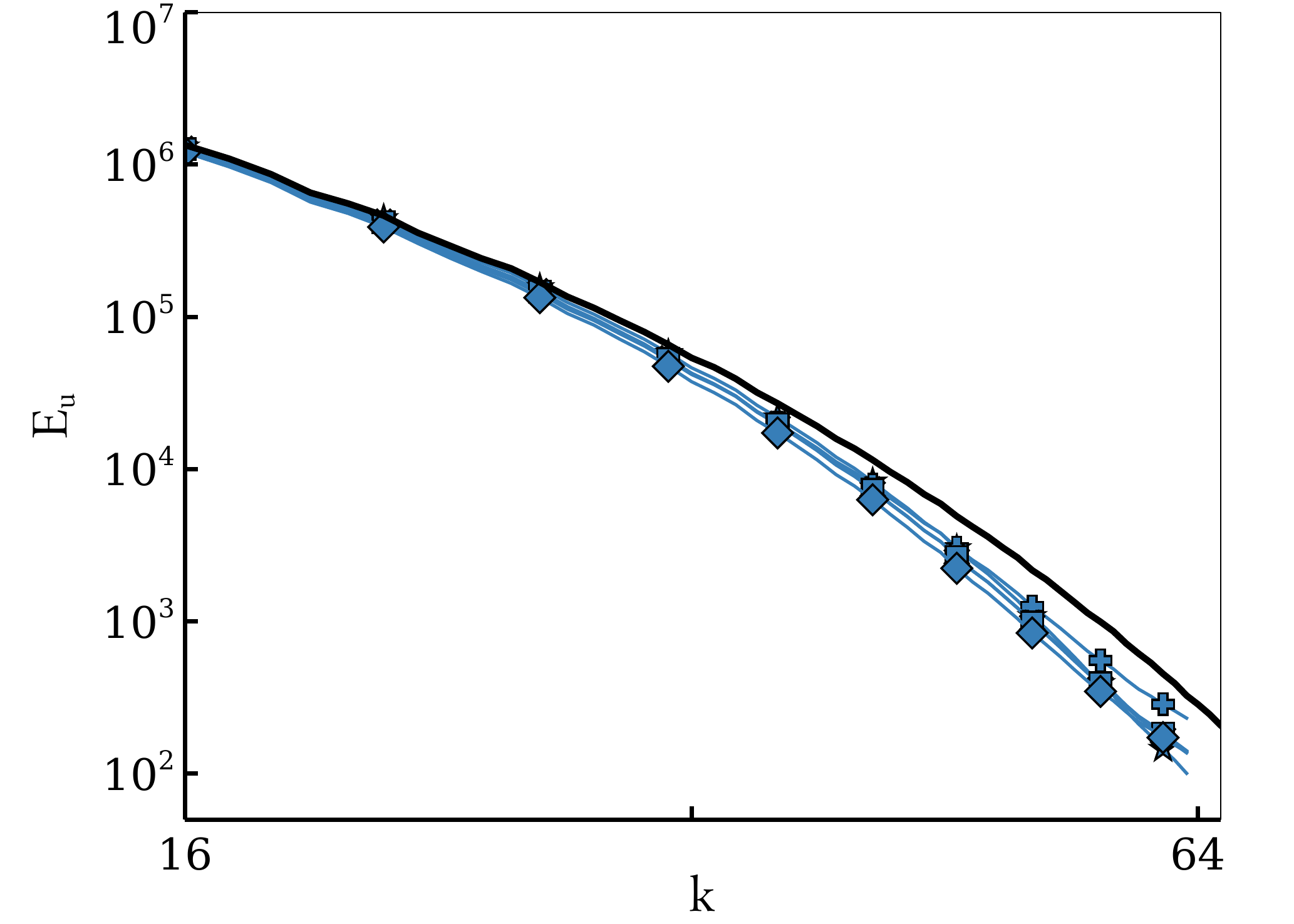}} \\
\subfloat[$N_x=256^3$]{\label{fig:comparisons_HIT_spectra_WENO_PPM_c}\includegraphics[width=0.5\textwidth]{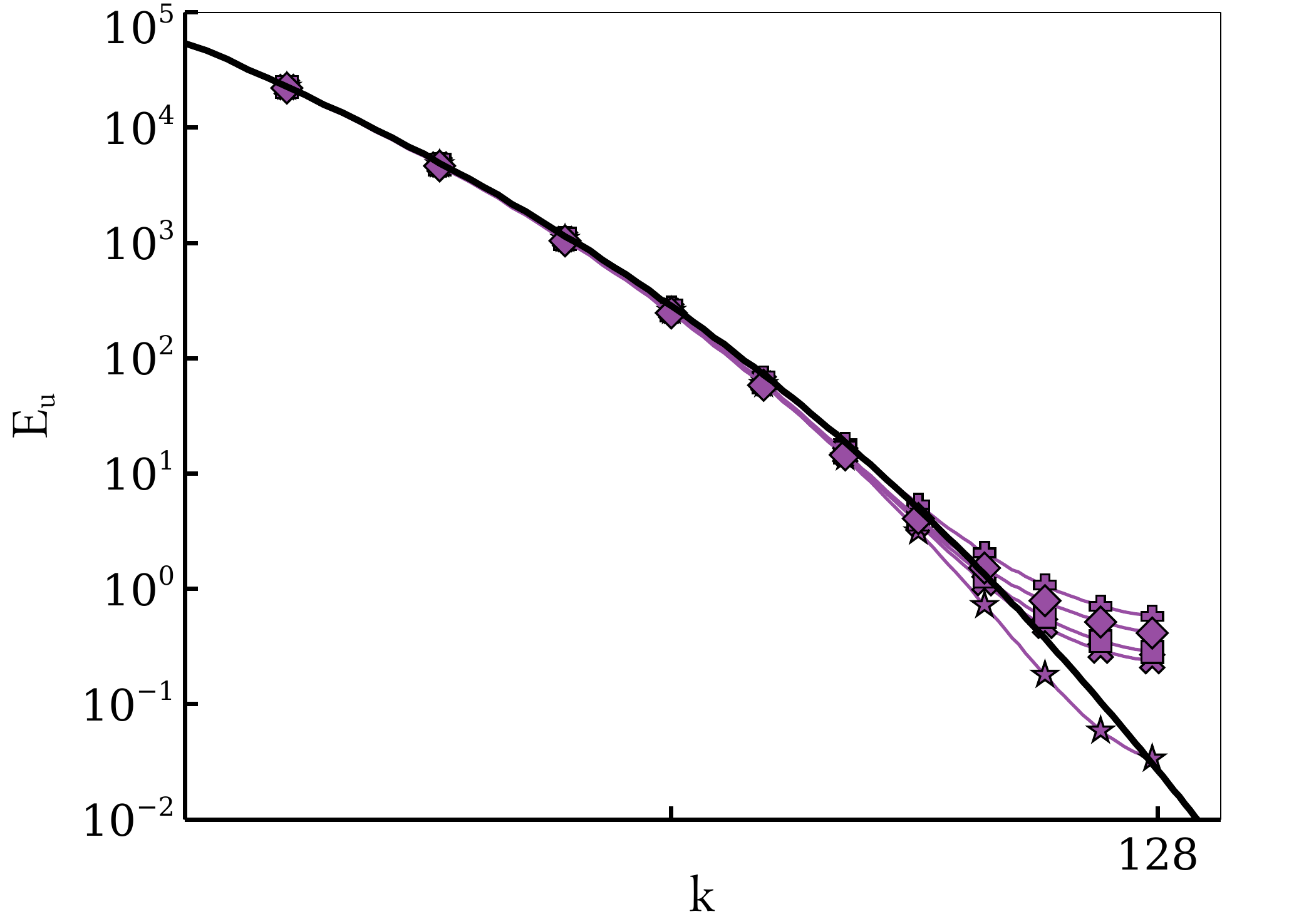}}
\subfloat[$N_x=512^3$]{\label{fig:comparisons_HIT_spectra_WENO_PPM_d}\includegraphics[width=0.5\textwidth]{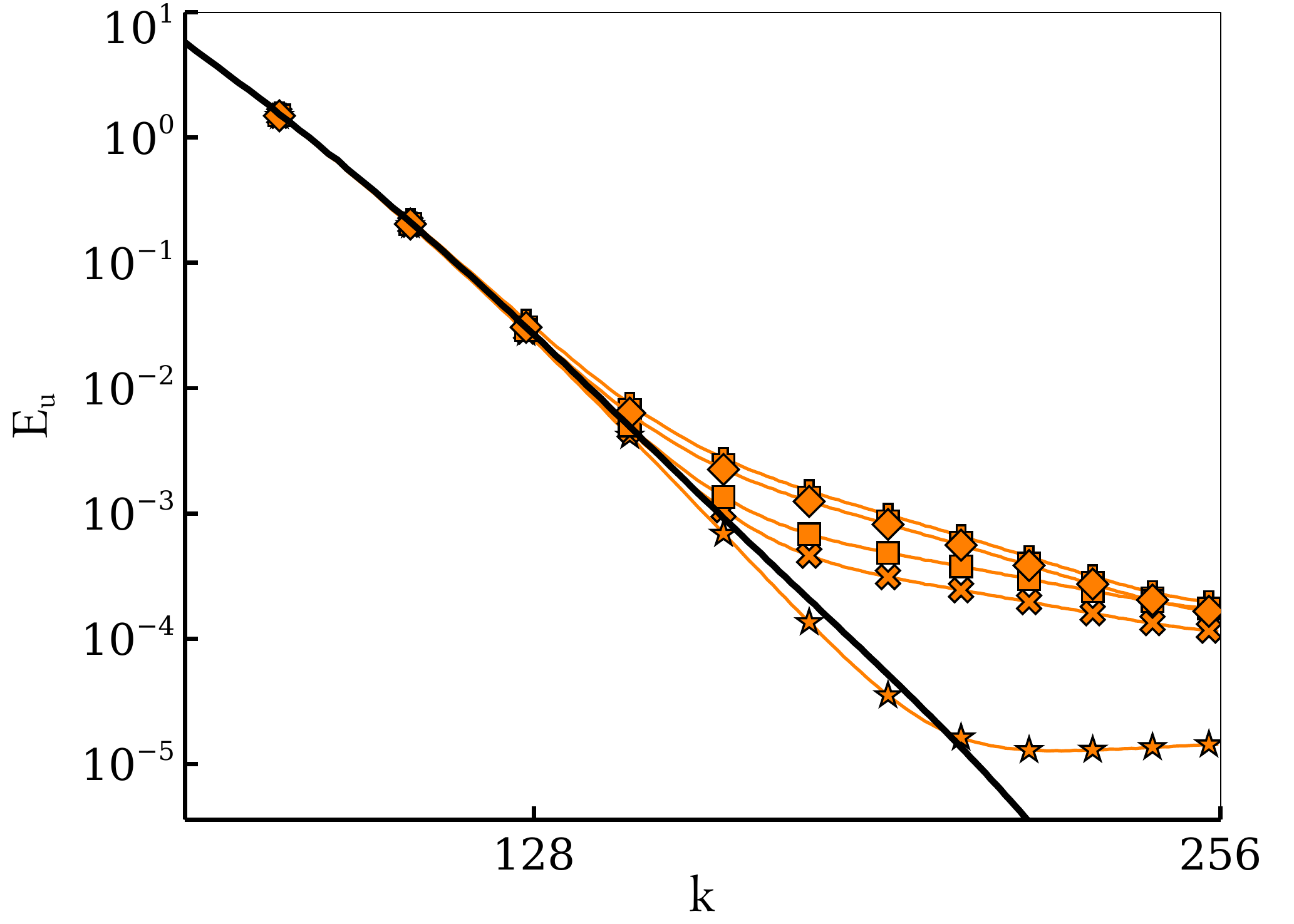}}
\caption{Zoom of the spectra of kinetic energy in \cref{fig:comparisons_HIT_spectra_WENO_PPM_full} for results computed with different mesh resolutions.}
\label{fig:comparisons_HIT_spectra_zoom}
\end{figure}

The decay of compressible isotropic turbulence is now simulated. \Cref{fig:comparisons_HIT_tseries_WENO_PPM_full_a,fig:comparisons_HIT_tseries_WENO_PPM_full_b,fig:comparisons_HIT_tseries_WENO_PPM_full_c,fig:comparisons_HIT_tseries_WENO_PPM_full_d}
present the temporal evolution of the kinetic energy, the enstrophy, the variance of temperature and the dilatation from $t=0$ to $t/\tau=4$.  \Cref{fig:comparisons_HIT_spectra_WENO_PPM_full_a,fig:comparisons_HIT_spectra_WENO_PPM_full_b,fig:comparisons_HIT_spectra_WENO_PPM_full_c,fig:comparisons_HIT_spectra_WENO_PPM_full_d} present the spectra taken at $t/\tau=4$ for the kinetic energy, the vorticity, the dilatation and the density. In these figures, the diamond, cross, square, plus and star symbols represent the WENO-JS, WENO-M, WENO-Z, WENO-MDCD and TENO methods, respectively. The red, blue, purple and orange colors represent simulations performed with  $N_x=64$, $N_x=128$, $N_x=256$ and $N_x=512$, respectively. It is emphasized that these figures contain a significant number of curves. For clarity, a zoom on the high-end of the spectra of kinetic energy is shown in \cref{fig:comparisons_HIT_spectra_zoom}
for each mesh resolution.

\begin{figure}[tbhp]
\centering
\includegraphics[width=0.85\textwidth]{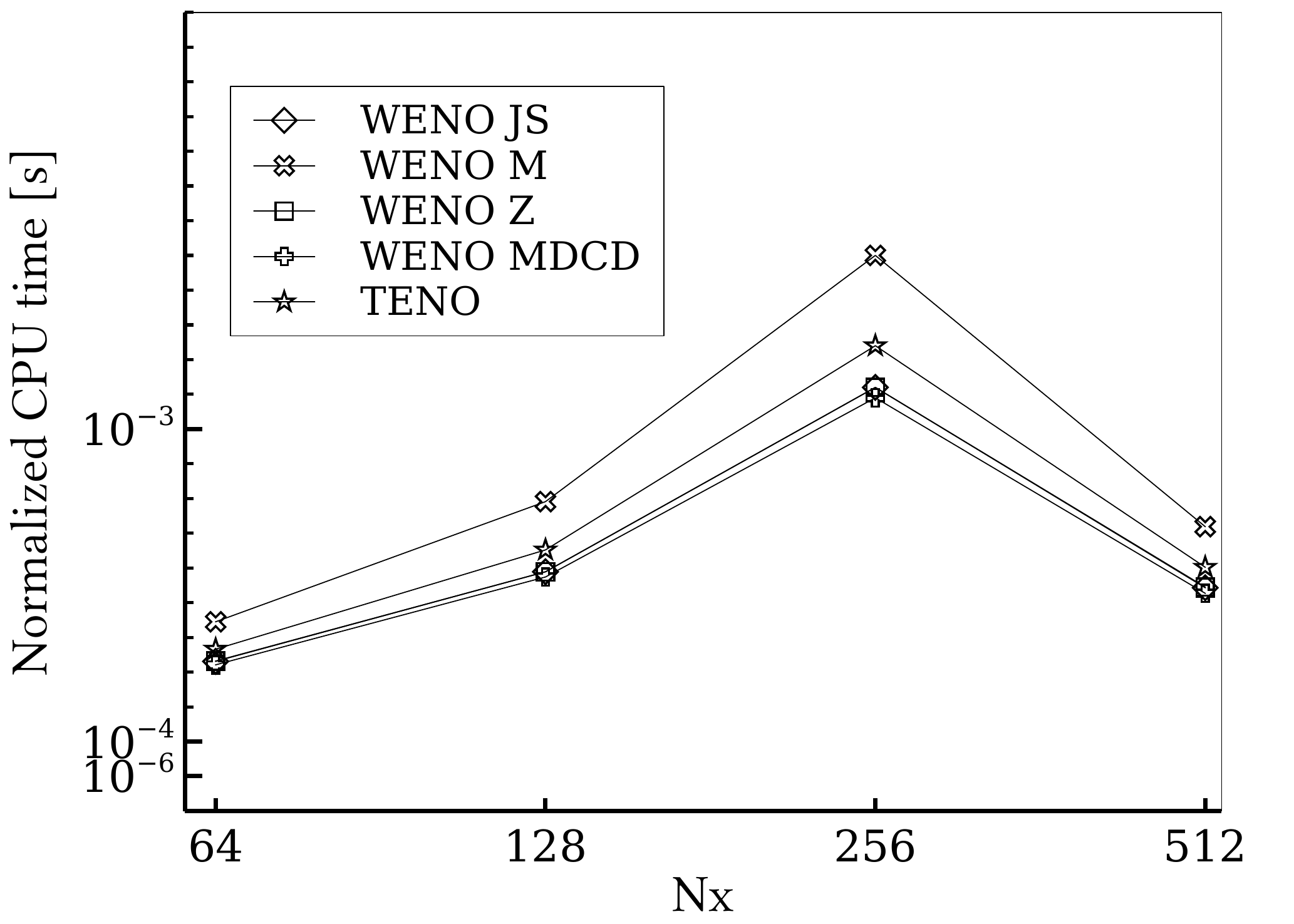}
\caption{CPU time for only the convection term with different methods.}
\label{fig:comparisons_HIT_cpu_time_PPM1_WENO}
\end{figure}

Overall, two general trends can be seen in  \cref{fig:comparisons_HIT_tseries_WENO_PPM_full,fig:comparisons_HIT_spectra_WENO_PPM_full}. For the temporal evolution of physical quantities, the different WENO variants investigated present significant differences when the mesh is coarse. However, when the mesh is refined, they quickly collapse to give similar results. However, as shown in \cref{fig:comparisons_HIT_spectra_WENO_PPM_full},  all the different WENO variants give virtually the same spectra, at the exception of the very high-frequencies of the spectrum when the mesh in refined enough to allow small turbulent structures to be resolved (see \cref{fig:comparisons_HIT_spectra_WENO_PPM_d}). Similarly to the previous section, a convergence rate analysis has been performed and all the different WENO variants exhibit a second-order convergence rate. Furthermore, \cref{fig:comparisons_HIT_cpu_time_PPM1_WENO} presents the computational time of each WENO variant. Recall here that the normalized CPU time is defined as the averaged wall-clock time spent in the routines required for the computation of the hyperbolic terms, divided by the number of iterations performed during the simulation and the number of CPUs employed.  It can be seen that the computational cost of the WENO-M variant is higher, followed by the TENO, whereas the WENO-JS, WENO-Z and WENO MDCD are virtually the same.

Based on these results, the present comparison study reveals that the TENO is not robust enough to avoid instabilities near strong shocks (see \cref{subsec:apprendix_shu_osher}). Other WENO variants are found to be robust, but the WENO-M is more costly. Among the remaining variants, the WENO-Z scheme presents slightly better results, and is thus adopted as the best WENO reconstruction scheme for the whole study presented in this paper.

\section*{Acknowledgments}

The work here was supported by the U.S. Department of Energy, Office of Science, Office of Advanced Scientific Computing Research, Applied Mathematics program under contract number DE-AC02005CH11231. John Wakefield has been funded by The National Science Foundation (NSF) Division of Mathematical Sciences (DMS) Mathematical Sciences Graduate Internship Program, administered by the Oak Ridge Institute for Science and Education (ORISE) through an interagency agreement between the U.S. Department of Energy (DOE) and NSF. ORISE is managed by ORAU under DOE contract number DE-SC0014664. All opinions expressed in this paper are the author’s and do not necessarily reflect the policies and views of NSF, DOE or ORAU/ORISE.

\bibliographystyle{amsplain}
\bibliography{bib_manu_2018}
\end{document}

%% file: ex_shared.tex
\usepackage{lipsum}
\usepackage{amsfonts}
\usepackage{graphicx}
\usepackage{epstopdf}
\usepackage{algorithmic}

\ifpdf
  \DeclareGraphicsExtensions{.eps,.pdf,.png,.jpg}
\else
  \DeclareGraphicsExtensions{.eps}
\fi
\usepackage[caption=false]{subfig} 
\usepackage{amssymb}

\usepackage{ifpdf}
\usepackage{hyperref}
\usepackage[noabbrev]{cleveref}
 \usepackage[foot]{amsaddr}


\title[Capturing shocks and turbulence spectra in compressible flows]{Investigation of finite-volume methods to capture shocks and turbulence spectra in compressible flows}

\author{Emmanuel Motheau and John Wakefield}

\usepackage{amsopn}

\renewcommand{\vec}[1]{\mbox{\boldmath$#1$}}
\newcommand{\tensor}[1]{\mbox{\boldmath{\ensuremath{#1}}}}

\usepackage{color}
